\documentclass[%
 reprint,
 amsmath,amssymb,
 aps,
]{revtex4-2} 

\usepackage{aasmacros}
\usepackage{enumitem}
\usepackage{natbib}
\usepackage{amsmath}
\usepackage{graphicx}
\usepackage{dcolumn}
\usepackage{bm}
\usepackage{xcolor}

\usepackage{apjfonts} 
\usepackage{graphicx}

\usepackage{caption}
\usepackage{subcaption}
\graphicspath{{./}{figures/}}

\newcommand{\reply}[1]{\textcolor{black}{{#1}}}

\begin{document}

\title{Inferring the neutron star maximum mass and lower mass gap in neutron star--black hole systems with spin}

\author{Christine Ye}
\affiliation{Eastlake High School, 400 228th Ave NE, Sammamish, WA 98074, USA}

\author{Maya Fishbach}
\altaffiliation{NASA Hubble Fellowship Program Einstein Postdoctoral Fellow}
\affiliation{Center for Interdisciplinary Exploration and Research in Astrophysics (CIERA) and Department of Physics and Astronomy,
Northwestern University, 1800 Sherman Ave, Evanston, IL 60201, USA}

\begin{abstract}
Gravitational-wave (GW) detections of merging neutron star--black hole (NSBH) systems probe astrophysical neutron star (NS) and black hole (BH) mass distributions, especially at the transition between NS and BH masses. Of particular interest are the maximum NS mass, minimum BH mass, and potential mass gap between them. While previous GW population analyses assumed all NSs obey the same maximum mass, if rapidly spinning NSs exist, they can extend to larger maximum masses than nonspinning NSs. In fact, several authors have proposed that the $\sim2.6\,M_\odot$ object in the event GW190814 -- either the most massive NS or least massive BH observed to date -- is a rapidly spinning NS. We therefore infer the NSBH mass distribution jointly with the NS spin distribution, modeling the NS maximum mass as a function of spin. Using 4 LIGO--Virgo NSBH events including GW190814, if we assume that the NS spin distribution is uniformly distributed up to the maximum (breakup) spin, we infer the maximum non-spinning NS mass is $2.7^{+0.5}_{-0.4}\,M_\odot$ (90\% credibility), while assuming only nonspinning NSs, the NS maximum mass must be $>2.53 M_\odot$ (90\% credibility). The data support the mass gap's existence, with a minimum BH mass at $5.4^{+0.7}_{-1.0} M_\odot$. With future observations, under simplified assumptions, 150 NSBH events may constrain the maximum nonspinning NS mass to $\pm0.02\,M_\odot$, and we may even measure the relation between the NS spin and maximum mass entirely from GW data. If rapidly rotating NSs exist, their spins and masses must be modeled simultaneously to avoid biasing the NS maximum mass.

\end{abstract}

\maketitle
\section{Introduction}
\label{sec:intro}

The transition between neutron star (NS) and black hole (BH) masses is key to our understanding of stellar evolution, supernova physics, and nuclear physics. In particular, the maximum mass that a NS can support before collapsing to a black hole (BH), known as the Tolman–Oppenheimer–Volkoff (TOV) mass $M_\mathrm{TOV}$ for a nonspinning NS,  is governed by the unknown high-density nuclear EOS~\citep{1996A&A...305..871B,1996ApJ...470L..61K,2012ARNPS..62..485L}.
Constraints on the maximum NS mass can therefore inform the nuclear EOS, together with astrophysical observations such as X-ray timing of pulsar hotspots~\citep{2019ApJ...887L..25B}, gravitational-wave (GW) tidal effects from mergers involving NSs~\citep{2018PhRvL.121p1101A,2019EPJA...55..209L,2020PhRvD.101l3007L,2020Sci...370.1450D}, and electromagnetic observations of binary neutron star (BNS) merger remnants~\citep{2017ApJ...850L..19M,2018ApJ...852L..25R}, as well as lab experiments~\citep[e.g.][]{2021PhRvL.126q2502A}. Recent theoretical and observational constraints on the EOS have placed $M_\mathrm{TOV} = 2.2$--$2.5\,M_\odot$~\citep[e.g.][]{2021PhRvD.104f3003L}. If astrophysical NSs exist up to the maximum possible NS mass, $M_\mathrm{TOV}$ can be measured by fitting the NS mass distribution to Galactic NS observations~\citep{2011MNRAS.414.1427V,2012ApJ...757...55O,2018MNRAS.478.1377A,2019ApJ...876...18F,2020RNAAS...4...65F}. A recent fit to Galactic \reply{neutron stars} finds a maximum mass of $2.22^{+0.85}_{-0.23}\,M_\odot$~\citep{2020RNAAS...4...65F}. In particular, observations of massive pulsars~\citep{2013Sci...340..448A,2020NatAs...4...72C} set a lower limit of $M_\mathrm{TOV} \gtrsim 2\,M_\odot$.

Meanwhile, the minimum BH mass and the question of a mass gap between NSs and BHs is of importance to supernova physics~\citep{2001ApJ...554..548F,2012ApJ...749...91F,2012ApJ...757...91B,2021ApJ...908..106L}. Observations of BHs in X-ray binaries first suggested a mass gap between the heaviest NSs (limited by $M_\mathrm{TOV}$) and the lightest BHs ($\sim5\,M_\odot$; \citealt{2010ApJ...725.1918O,2011ApJ...741..103F}), although recent observations suggest that the mass gap may not be empty~\citep{2019Sci...366..637T,2020ApJ...896L..44A}.

Over the last few years, the GW observatories Advanced LIGO~\citep{2015CQGra..32g4001L} and Virgo~\citep{2015CQGra..32b4001A} have revealed a new astrophysical population of NSs and BHs in merging binary black holes (BBHs)~\citep{2016PhRvL.116f1102A}, BNS~\citep{2017PhRvL.119p1101A,2020ApJ...892L...3A}, neutron-star black hole (NSBH) systems~\citep{2021ApJ...915L...5A}. These observations can be used to infer the NS mass distribution in merging binaries and constrain the maximum NS mass~\citep{2020PhRvD.102f4063C,2021ApJ...909L..19G,2021arXiv210704559L,2021arXiv210806986L,2021arXiv211202605Z,2021arXiv211103634T}. Furthermore, jointly fitting the NS and BH mass distribution using GW data probes the existence of the mass gap~\citep{2017MNRAS.465.3254M,2020ApJ...899L...8F,2021arXiv211103498F}. Recent fits of the BNS, BBH and NSBH mass spectrum finds a relative lack of objects between $2.6$--$6\,M_\odot$~\citep{2021ApJ...913L...7A,2021arXiv211103498F,2021arXiv211103634T}.

Gravitational-wave NSBH detections can uniquely explore both the maximum NS mass and the minimum BH mass simultaneously with the same system. In particular, the NS and BH masses in the first NSBH detections~\citep{2021ApJ...915L...5A} seem to straddle either side of the proposed mass gap~\citep{2021arXiv211103498F}, especially when assuming astrophysically-motivated BH spins~\citep{2021arXiv210914759M}. However, our understanding of the NS maximum mass and the mass gap from GWs is challenged by one discovery: GW190814~\citep{2020ApJ...896L..44A}. The secondary mass of GW190814 is tightly measured at $2.6\,M_\odot$, making it exceptionally lighter than BHs in BBH systems~\citep{2021arXiv210900418E} but heavier than most estimates of $M_\mathrm{TOV}$~\citep{2020ApJ...896L..44A,2020ApJ...904...80E}. As a possible explanation, several authors have proposed that GW190814 is a \emph{spinning} NS~\citep{2020MNRAS.499L..82M}. While $M_\mathrm{TOV}$ limits the mass of nonspinning NSs, NSs with substantial spins can support $\sim 20\%$ more mass~\citep{1994ApJ...424..823C}. Unfortunately, it is difficult to test the spinning NS hypothesis for a single system, because the spin of the secondary $2.6\,M_\odot$ object in GW190814 is virtually unconstrained from the GW signal.

In this paper, we show that by studying a \emph{population} of NSBH events, we may measure the NS maximum mass as a function of spin. \reply{We build upon the work of~\citet{2021arXiv211202605Z,2021arXiv211103498F,2021arXiv211103634T}, who studied the population statistics of NSBH masses and BH spins, but allow the NS mass distribution to depend on NS spin for the first time.} This method will not only enable more accurate classifications for NSBH versus BBH events in cases like GW190814, but will also prevent biases that would result from measuring $M_\mathrm{TOV}$ while neglecting the dependence of the maximum NS mass on spin. \reply{As~\citet{2022MNRAS.511.4350B} previously showed, mismodeling the NS spin distribution can bias the inferred mass distribution even in cases where the NS mass distribution does not vary with spin, simply because masses and spins are correlated in the GW parameter estimation of individual events.} The rest of this paper is structured as follows. Section \ref{Methods} describes population-level spin and mass models, our hierarchical Bayesian framework, the current GW data, and our procedure for simulating future NSBH events. Results from analyzing the LIGO--Virgo NSBH mergers are presented in Section \ref{ligodata}; results from simulating future GW NSBH observations are presented in Section \ref{Projections}. We conclude in Section~\ref{sec:conclusion}.


\section{Methods} \label{Methods}

\subsection{Population Models} \label{models}
We use the following phenomenological models to describe the astrophysical spin (Section~\ref{spin models}) and mass (Section~\ref{ns models}--\ref{bh mass}) distribution of NSBH systems.

\subsubsection{Spin Models} \label{spin models}

It remains unclear whether NSs, specifically those in merging BNS and NSBH systems, can have significant spins. The most rapidly spinning NS in a (nonmerging) double NS system is the Pulsar J1807-2500B with a period of 4.2 ms or dimensionless spin magnitude $a = 0.12$~\citep{2012ApJ...745..109L}. Among recycled pulsars, the fastest spinning is Pulsar J1748-2446ad with a period of $\sim1.4$ ms~\citep{2006Sci...311.1901H}. However, rapidly spinning NSs in which spin down is inefficient (due to e.g. weak magnetic fields) may have avoided electromagnetic discovery for the same reasons. In NSBH systems, it may also be possible for the NS spin to grow through accretion if the NS is born before the BH~\citep{2021MNRAS.504.3682C}, or through tidal synchronization as has been studied in BBH systems ~\citep{2018A&A...616A..28Q}.

We remain agnostic about NS spin magnitudes, modeling their distribution as a power law,
\begin{equation}
\label{eq:NSspin}
    p(a_2 | a_\mathrm{max}, \beta_{{s}}) \propto  \begin{cases}  
      {(1-a_2)}^{\beta_{s}} & 0 < a_2 < a_\mathrm{max}\\
      0 & \text{otherwise},
      \end{cases}
\end{equation}
where $a_\mathrm{max}$ sets an upper limit on possible values of $a_2$ and $\beta_s$ controls the slope. For $\beta_s = 0$, the secondary spin magnitude follows a uniform distribution; for $\beta_s > 0$, the secondary spin distribution prefers low spin. The maximum value of $a_\mathrm{max}$ is the breakup spin $a_\text{Kep}$, which is around $a_\text{Kep} \approx 0.7$ for most EOSs.

We do not explicitly model NS spin tilts (the angle between the spin vector and the orbital angular momentum axis), but consider a few different assumptions and explore how they affect our inference. By default, we consider a NS spin tilt distribution that is isotropic, or flat in $-1< \cos(\mathrm{tilt}_2) < 1$. We also explore a restricted model in which NS spins are perfectly aligned with the orbit, $\cos(\mathrm{tilt}_2) = 1$.
For the distribution of BH spins, by default we assume that BHs are nonspinning~\citep[$a_1 = 0$;][]{2019ApJ...881L...1F,2021arXiv210914759M}. We alternatively assume that the BH spin distribution is uniform in spin magnitude with isotropic spin tilts.

\begin{figure*}
     \centering
     \begin{subfigure}[b]{0.3\textwidth}
         \centering
         \includegraphics[width=\textwidth]{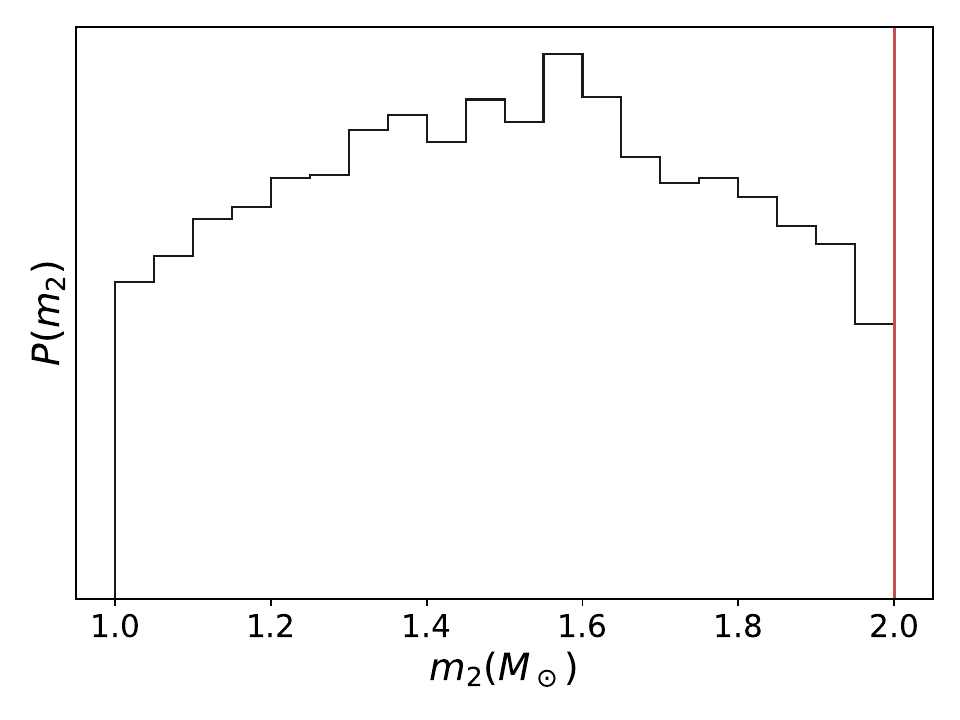}
         \caption{1 component, no spin ($a_\mathrm{max}/a_{\text{Kep}}= 0$)}
     \end{subfigure}
     \hfill
     \begin{subfigure}[b]{0.3\textwidth}
         \centering
         \includegraphics[width=\textwidth]{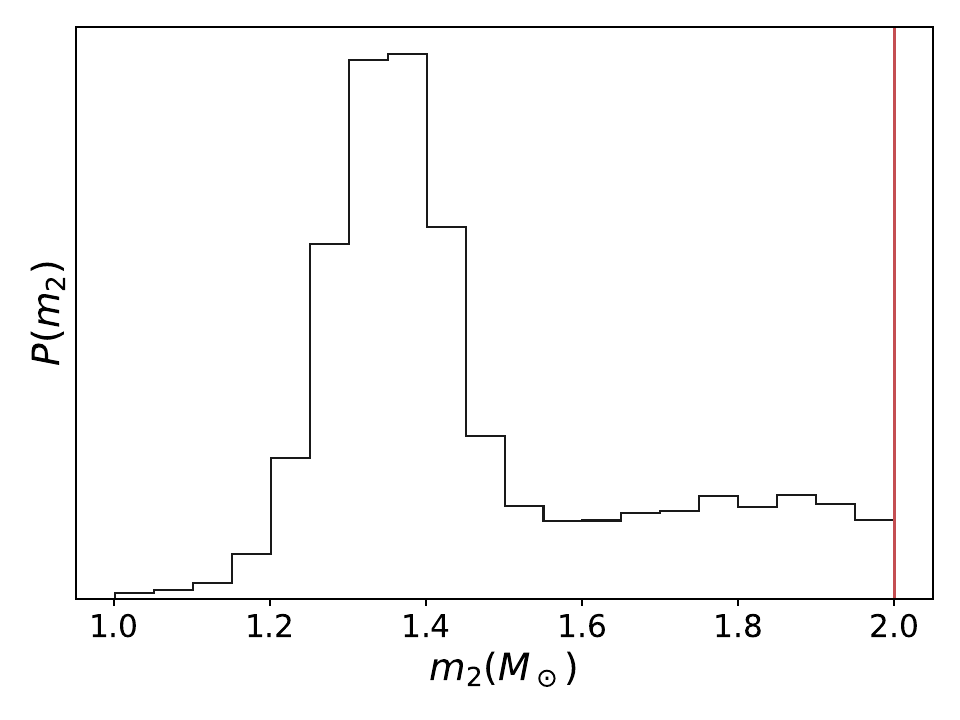}
         \caption{2 components, no spin ($a_\mathrm{max}/a_{\text{Kep}}= 0$)}
     \end{subfigure}
     \hfill
     \begin{subfigure}[b]{0.3\textwidth}
         \centering
         \includegraphics[width=\textwidth]{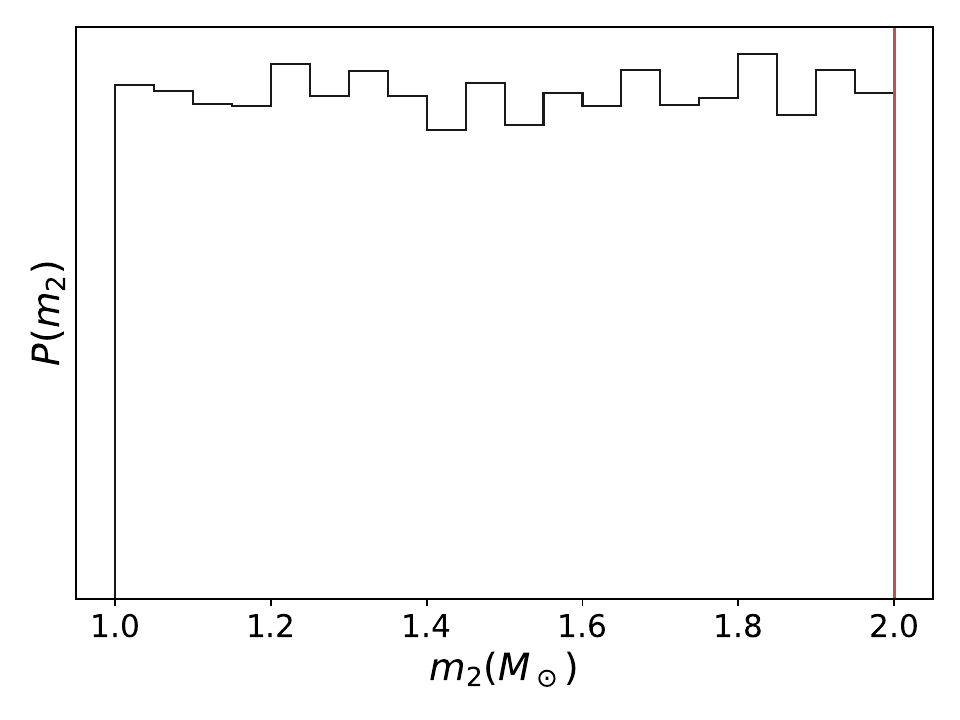}
         \caption{Uniform distribution, no spin ($a_\mathrm{max}/a_{\text{Kep}}= 0$)}
     \end{subfigure}
     \bigskip
     \begin{subfigure}[b]{0.3\textwidth}
         \centering
         \includegraphics[width=\textwidth]{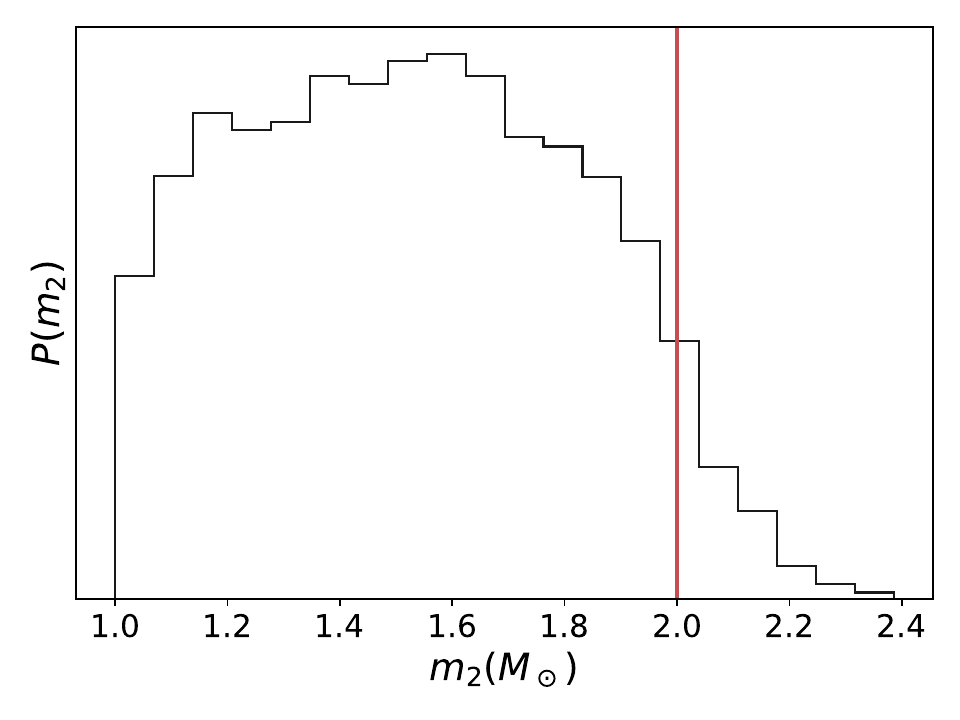}
         \caption{1 component, uniform spin ($\beta_s = 0$)  with $a_\mathrm{max}/a_\mathrm{Kep} = 1.0$.}
     \end{subfigure}
     \hfill
     \begin{subfigure}[b]{0.3\textwidth}
         \centering
         \includegraphics[width=\textwidth]{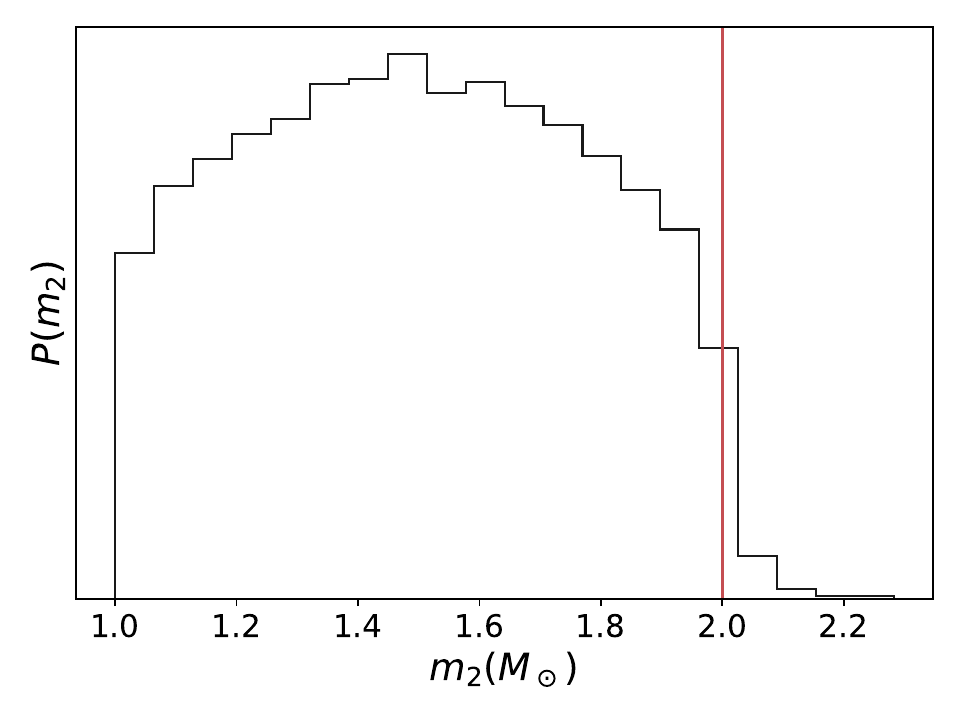}
         \caption{1 component, uniform spin ($\beta_s = 0$)  with $a_\mathrm{max}/a_\mathrm{Kep} = 0.5$.}
     \end{subfigure}
     \hfill
     \begin{subfigure}[b]{0.3\textwidth}
         \centering
         \includegraphics[width=\textwidth]{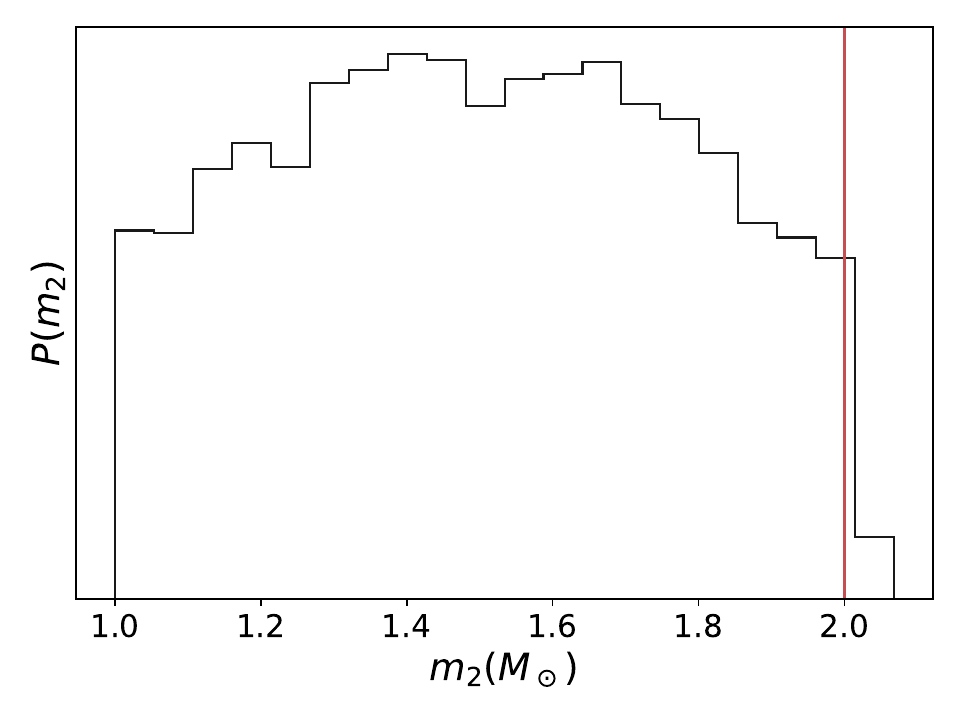}
         \caption{1 component, power-law spin ($\beta_s = 2$)  with $a_\mathrm{max}/a_\mathrm{Kep} = 1.0$.}
     \end{subfigure}
     \hfill
    \caption{Simulated astrophysical NS mass distributions $p(m_2)$, with $M_{\text{TOV}} = 2\,M_\odot$, marked in red. The NS spin distribution follows Eq.~\ref{eq:NSspin}, with $\beta_s$ and $a_\mathrm{max}$ specified in the subcaptions.}
    \label{fig:distributions}
\end{figure*}

In summary, we consider the following spin models:
\begin{enumerate}
    \item \emph{Zero spin BH} (``ZS", default spin model): Primary BH is nonspinning ($a_1  = 0$). Secondary NS spin is isotropic in spin tilt (flat in $-1 < \cos(\mathrm{tilt}_2) < 1$) and follows a power law in the spin magnitude $a_2$ (Eq.~\ref{eq:NSspin}).
    
    \item \emph{Zero spin BH + aligned spin NS} (``ZS + AS"): 
    Same as \emph{Default}, but with $\cos(\mathrm{tilt}_2) = 1$.
    
    \item \emph{Uniform and isotropic} (``U+I"): Same as \emph{Default}, but primary BH spin is flat in magnitude $a_1$ and $\cos(\mathrm{tilt}_1)$ rather than nonspinning.
    
\end{enumerate} 

\subsubsection{NS Mass Models} \label{ns models}

Like the case with spins, we consider a few different mass models to check the robustness of our conclusions. We consider three models for NS masses, which describe the distribution of NSBH secondary masses $m_2$ (see Fig.~\ref{fig:distributions}):
\begin{enumerate}
    \item \emph{Default}: Single Gaussian distribution (panels a, d--f of Figure \ref{fig:distributions})
    \begin{multline*}
    p(m_2 | \mu, \sigma, M_{min}, M_{max}) = \\ \begin{cases}  
      \mathcal{N}_T(m_2 | \mu, \sigma) & M_{min} \leq m_2 \leq M_{max} \\
      0 & \mathrm{otherwise},
      \end{cases}
    \end{multline*}
    where $\mathcal{N}_T(x \mid \mu, \sigma)$ denotes a truncated Gaussian distribution with mean $\mu$ and standard deviation $\sigma$.
    
    \item Two-component (bimodal) Gaussian distribution (``2C"), as in the Galactic NS distribution~\citep[][panel b of Figure\ref{fig:distributions}]{2018MNRAS.478.1377A} 
    \begin{multline}
    p(m_2 | \mathcal{A}, \mu_1, \sigma_1, \mu_2, \sigma_2, M_\mathrm{min}, M_\mathrm{max}) = \\
    \begin{cases}  
       \mathcal{A}\mathcal{N}_T(m_2 | \mu_1, \sigma_1) + \\ (1- \mathcal{A})\mathcal{N}_T(m_2 | \mu_2, \sigma_2) & M_\mathrm{min} \leq m_2 \leq M_\mathrm{max} \\
       0 & \mathrm{otherwise}.
      \end{cases}
    \end{multline}
    
        \item Uniform distribution (``U") with sharp cutoffs at the minimum and maximum NS mass [$M_{\rm min}, M_{\rm max}$] (panel c of Figure~\ref{fig:distributions}) 
    \begin{multline*}
    p(m_2 | M_\mathrm{min}, M_\mathrm{max}) = \\
    \begin{cases}  
      \frac{1}{M_\mathrm{max} - M_\mathrm{min}}  & M_\mathrm{min} \leq m_2 \leq M_{\rm max} \\
      0 & \mathrm{otherwise}.
      \end{cases}
    \end{multline*}
\end{enumerate}
All normal distributions ($\mathcal{N}_T$) are truncated sharply and normalized to integrate to 1 between $M_\text{min} = 1 M_\odot$ and $M_\text{max}$. In this work, we focus on inferring the maximum NS mass. While the minimum NS mass can also be inferred with GWs~\citep{2020PhRvD.102f4063C}, we fix the minimum NS mass to $1\,M_\odot$ in our models. If binary stellar evolution can produce NSs with extreme masses, then $M_\mathrm{min}$ and $M_\mathrm{max}$ correspond to the minimum and maximum \emph{allowable} masses set by nuclear physics.  

Crucially, we allow NSs to have significant spin. Rapid uniform rotation may provide additional support to the NS, allowing it to reach masses greater than the non-spinning maximum mass $M_{TOV}$. 
We model the dependence of $M_\mathrm{max}$ on NS spin $a_2$ using the universal relationship from \citet{2020MNRAS.499L..82M}:
\begin{multline*} 
    M_{\text{max}}(a_2, a_{\text{Kep}}, M_{\text{TOV}}) = \\
    M_{\text{TOV}}\left(1 + A_2\left(\frac{a_2}{a_{\text{Kep}}}\right)^2+A_4\left(\frac{a_2}{a_{\text{Kep}}}\right)^4\right)
\end{multline*}
with $A_2 = 0.132$, $A_4 = 0.0071$, where $a_{\text{Kep}}$ corresponds to the dimensionless spin at the mass-shedding limit. For concreteness, we assume $a_{\mathrm{Kep}} = 0.7$, which is true for most EOS. For a neutron star with spin $a_{\text{Kep}}$, the maximum possible mass is around $1.2\times$ the (non-spinning) TOV limit. To measure this relation directly from gravitational-wave data, we also optionally measure a free, linear dependence between maximum spin and critical mass (see Section \ref{sec:spinmass}):

\begin{equation}
    M_{\text{max}}(a_2, a_{\text{Kep}}, M_{\text{TOV}}) = M_{\text{TOV}} A_1\left(\frac{a_2}{a_{\text{Kep}}}\right)
\end{equation}

The extent to which the NS mass distribution can extend above $M_\text{TOV}$ depends on the spin distribution. 
The NS mass distributions $p(m_2)$ above include a dependence on spin, and can be written as $p(m_2 | M_{\rm max}(a_2), \theta)$ where $\theta$ includes all other parameters.
Figure \ref{fig:distributions}d--f shows the NS mass distribution under three variations of the spin distributions outlined in \ref{spin models}.

\subsubsection{BH Mass Models and Pairings}
\label{bh mass}
We model the primary (BH) mass distribution $p(m_1)$ as a power law with slope $-\alpha$, and a minimum mass cutoff at $M_{\rm BH}$:
\begin{equation}
\label{eq:pm1}
p(m_1 | \alpha, M_{\rm BH}) \propto  \begin{cases}  
      0 & x < M_{\rm BH}\\
      {m_1}^{-\alpha} & \text{otherwise}.
      \end{cases}
\end{equation}
We fix $\alpha > 0$ such that the probability density decreases for increasing BH mass. The minimum BH mass represents the upper boundary of the mass gap. In order to restrict the range of $m_1$ to reasonable values, we optionally include a maximum BH mass of $30\,M_\odot$  in Eq.~\ref{eq:pm1}. However, for most of our NSBH models, high-mass BHs are rare due to a relatively steep slope $\alpha$ and/or a pairing function that disfavors extreme mass ratio pairings, and we do not explicitly model the BH maximum mass.

We assume that the pairing function between $m_1$ and $m_2$ NSBH systems follows a power law in the mass ratio $m_2/m_1 = q < 1$~\citep{2020ApJ...891L..27F}:
\begin{equation}
    p(q) \propto q^{\beta},
\end{equation}
where by default we assume $\beta =0$~\citep{2021arXiv211103498F}. We alternatively consider the case $\beta = 3$, which favors equal-mass pairings. Depending on the width of the mass gap, NSBHs may necessarily have unequal masses, but on a population level, higher $q$ may still be relatively preferred.

Putting the mass and spin distributions together, we model the distribution of NSBH masses and spins $\theta \equiv (m_1, m_2, a_1, a_2)$ given population hyperparameters $\Lambda$ and model $H$ as:

\begin{equation}
    \label{eq:fullpopmodel}
    \begin{split}
    \pi(\theta | \Lambda, H) \propto p(m_1 \mid \alpha, M_\mathrm{BH}, H)p(m_2 | \Lambda_\mathrm{NS}, a_2, H) \\
    p(a_1 \mid H)     p(a_2 | a_\mathrm{max}, \beta_s, H)p(q \mid \beta, H),
    \end{split}
\end{equation}
where $H$ refers to the choice of model as described in the earlier subsections. For the extrinsic source parameters not in $\theta$, we assume isotropic distributions in sky position, inclination and orientation, and the local-Universe approximation to a uniform-in-volume distribution $p(d_L) \propto d_L^2$, where $d_L$ is the luminosity distance.

\subsection{Hierarchical Inference}
\subsubsection{Likelihood}
We infer properties of the overall NSBH population with a hierarchical Bayesian approach~\citep{2004AIPC..735..195L,2019MNRAS.486.1086M}. This allows us to marginalize over the uncertainties in individual events' masses and spins (grouped together in the set $\theta_i$ for event $i$) in order to estimate the hyperparameters $\Lambda$ describing the NS and BH mass and spin distributions. For $N_{det}$ GW detections producing data $d$, the likelihood of the data is described by an inhomogeneous Poisson process:

\begin{equation}
    \mathcal{L}({d}|\Lambda, N) = N^{N_{\text{det}}} e^{-N\xi(\Lambda)}\prod_{i=1}^{N_{\text{det}}} \int \mathcal{L}({d_i}|\theta_i)\pi(\theta_i|\Lambda)\:d\theta_i
\end{equation}
where $N$ is the total number of NSBH mergers in the Universe within some observing time, $\xi(\Lambda)$ is the fraction of detectable events in the population described by hyperparameters $\Lambda$ (see Section~\ref{sec:selection}), $\mathcal{L}({d_i|\theta_i})$ is the likelihood for event $i$ given its masses and spins $\theta_i$, and $\pi(\theta|\Lambda)$ describes the NSBH mass and spin distribution given population hyperparameters $\Lambda$ (Eq.~\ref{eq:fullpopmodel}. As we do not attempt to calculate event rates, we marginalize over $N$ with a log-uniform prior and calculate the population likelihood as~\citep{2019MNRAS.486.1086M,2018ApJ...863L..41F}:
\begin{equation}
    \mathcal{L}({d}|\Lambda) \propto \prod_{i=1}^{N_{\text{det}}} \frac{\int \mathcal{L}({d_i}|\theta_i)\pi(\theta_i|\Lambda)\:d\theta_i}{\xi(\Lambda)}
\end{equation}
We evaluate the single-event likelihood $\mathcal{L}(d \mid \theta)$ via importance sampling over $N_\text{samp}$ parameter estimation samples $\theta_\mathrm{PE}$ for each event:
\begin{equation}
    \int \mathcal{L}({d}|\theta)\pi(\theta|\Lambda)\:d\theta \simeq \frac{1}{N_\text{samp}} \sum_{j=1}^{N_\text{samp}} \frac{\pi(\theta_{\mathrm{PE},j}|\Lambda)}{\pi_{\mathrm{PE}}(\theta_{\mathrm{PE},j})},
\end{equation}
where $\pi_\mathrm{PE}(\theta)$ is the original prior that was used in LIGO parameter estimation.
We calculate the posterior on the population parameters, $p(\Lambda \mid d)$, from the likelihood $\mathcal{L}(d \mid \Lambda)$, under Bayes theorem, using broad, flat priors on the parameters $\Lambda$. For prior ranges, see Table~\ref{table:prior}.

\begin{table}[h!]
\centering
\begin{tabular}{ |c|c|} 
 \hline
 $\mathcal{A}$ & [0.0, 1.0] \\
 $\mu \text{ or } \mu_1, \mu_2$ & [1.0, 3.0] \\
 $\sigma \text{ or } \sigma_1, \sigma_2$ & [0.01, 1.5] \\
 $M_{TOV}$ & [1.5, 3.5] \\ 
 $M_{BH}$ & [1.5, 10] \\
 $\alpha$ & [0, 10] \\
 max $a/a_\text{Kep}$ & [0.1, 1.0] \\
 $\beta_\text{s}$ & [0.0, 5.0] \\
 $A_1 \text{(optional)}$ & [-0.5, 0.5] \\
 \hline
\end{tabular}
\caption{Priors ranges for population parameters.}
\label{table:prior}
\end{table}

\subsubsection{Selection Effects}
\label{sec:selection}
While we model and measure the astrophysical source distributions, GW detectors observe only sources loud enough to be detected, i.e. sources that produce data above some threshold $d > \mathrm{thresh}$. We account for this selection effect by including the term $\xi(\Lambda)$, the fraction of detectable binaries from a population described by parameters $\Lambda$.

\begin{align}
    \xi(\Lambda) &= \int_{d > \mathrm{thresh}} \mathcal{L}({d}|\theta)\pi(\theta|\Lambda)  \:dd\: d\theta \nonumber \\
    &\equiv \int P_\mathrm{det}(\theta) \pi(\theta \mid \Lambda) d\theta
\end{align}
To evaluate $\xi(\Lambda)$, we calculate the detection probability $P_\mathrm{det}(\theta)$ as a function of masses and cosmological redshift following the semi-analytic approach outlined in \citet{2017ApJ...851L..25F}. We assume the detection threshold is a simple single-detector signal-to-noise ratio (SNR) threshold $\rho_\mathrm{thresh} = 8$. We neglect the effect of spin on detectability; although systems with large aligned spins experience orbital hang-up that increases their SNR compared to small or anti-aligned spins, the effect is small compared to current statistical uncertainties~\citep{2018PhRvD..98h3007N}. 

Given masses and redshift of a potential source, we calculate its detectability as follow. We first calculate the optimal matched-filter SNR $\rho_\text{opt}$ using noise power spectral density (PSD) curves corresponding to aLIGO at O3 sensitivity, Design sensitivity, or A+ sensitivity~\citep{2020LRR....23....3A}; the optimal SNR corresponds to a face-on, directly-overhead source. We then calculate the SNR $\rho$ for a random sky position and orientation by generating angular factors $0 < w < 1$ from a single-detector antenna pattern~\citep{1993PhRvD..47.2198F} and set $\rho = w\rho_\text{opt}$. If $\rho > \rho_\mathrm{thresh}$ for a given detector noise curve, we consider the simulated source to be detected.

Finally, we estimate $\xi(\Lambda)$ with a Monte Carlo integral over simulated sources. We draw simulated sources with $m_1, m_2, z$ according to $p_\mathrm{draw}(\theta)$ until injection sets of ~10,000 events are created. BH ($m_1$) are drawn from a power law with $M_\text{BH} = 1.5 M_\odot$. NS ($m_2$) are drawn from a uniform distribution between $1$ and $3.5 M_\odot$. Redshifts $z$ are drawn uniform in comoving volume and source-frame time. Each simulated system is labeled as detected or not based on its SNR, described above. We then approximate the integral $\xi(\Lambda)$ as a sum over $M_\mathrm{det}$ detected simulated systems:
\begin{equation}
    \xi(\Lambda) \simeq \frac{1}{N_\text{draw}} \sum_{j=1}^{M_\text{det}} \frac{\pi(m_{1,j}, m_{2,j}, z_j|\Lambda)}{p_\text{draw}(m_{1,j}, m_{2,j}, z_j)}
\end{equation}

\subsection{Gravitational Wave Data and Simulations}
\subsubsection{Well-Measured Parameters}
While the population distributions in \ref{models} are defined in terms of $m_1$, $m_2$, $a_1$, and $a_2$, gravitational-wave detectors are most sensitive to degenerate combinations of these parameters. These include the gravitational chirp mass 

\begin{equation}
    \mathcal{M} = \frac{(m_1 m_2)^{3/5}}{(m_1 + m_2)^{1/5}},
\end{equation}

the symmetric mass ratio 

\begin{equation}
    \nu = \frac{q}{(1+q)^2},
\end{equation}

and $\chi_{\text{eff}}$, a mass-weighted sum of the component spins that is approximately conserved during the inspiral

\begin{equation}
    \chi_{\text{eff}} = \frac{m_1 a_{1,z} + m_2 a_{2,z}}{m_1 + m_2}
\end{equation}

where $a_{1,z}$ and $a_{2,z}$ are the components of the primary and secondary spin that are aligned with the orbital angular momentum axis. If the primary is nonspinning, $\chi_{\text{eff}}$ reduces to $\frac{m_2 a_{2,z}}{m_1 + m_2} = a_{2,z} \frac{q}{1+q}$.

\begin{figure}
    \centering
    \includegraphics[width=0.5\textwidth]{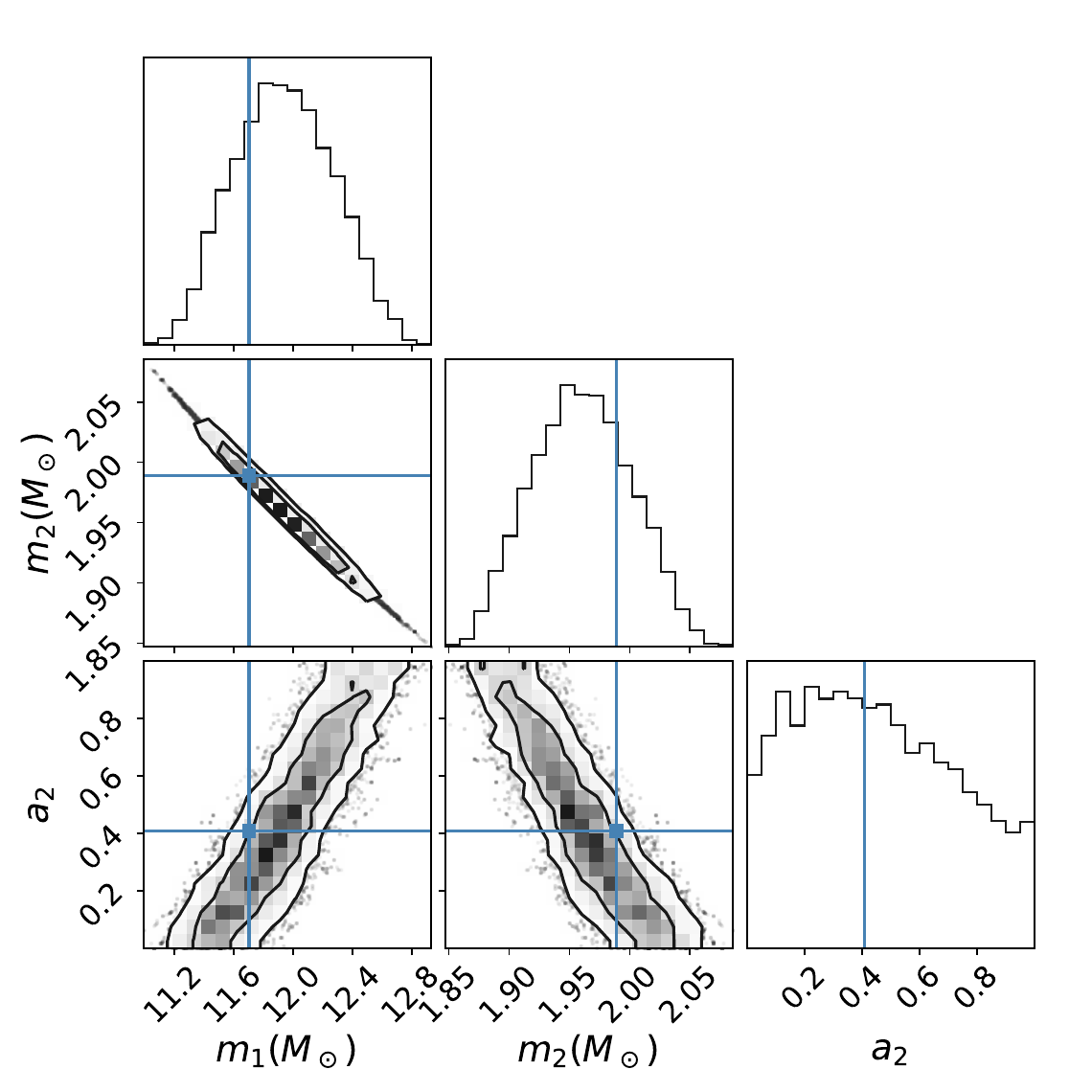}
    \caption{Sample parameter estimation posterior simulated using the PN approximation; contours show 68\% and 95\% CI. True values are denoted by the blue crosses.}
    \label{fig:simulated}
\end{figure}

\subsubsection{Post-Newtonian Approximation}
\label{sec:PN}

We follow the method outlined in \cite{2020PhRvD.102f4063C} to simulate realistic parameter estimation samples from mock GW NSBH detections. 
\cite{2020PhRvD.102f4063C} use the post-Newtonian (PN) description of the GW inspiral, with PN coefficients $\psi_0, \psi_2$, and $\psi_3$ that depend on the masses and spins.

\begin{equation}
    \psi_0(\mathcal{M}) = \frac{3}{128\mathcal{M}^{5/3}\pi^{5/3}}
\end{equation}

\begin{equation}
    \psi_2(\mathcal{M}, \nu) = \frac{5}{96\mathcal{M}\pi\nu^{2/5}}(\frac{743}{336} + \frac{11\nu}{4})
\end{equation}

\begin{equation}
    \beta = \frac{1}{3}(\frac{113-76\nu}{4}\chi_\text{eff} + \frac{76}{4}\delta m \nu \chi_{a})
\end{equation}

\begin{equation}
    \psi_3(\mathcal{M}, \nu, \beta) = \frac{3(4\beta-16\pi)}{128\mathcal{M}^{2/3}\pi^{2/3}\nu^{3/5}},
\end{equation}
where the mass difference $\delta m = (m_1-m_2)/(m_1+m_2)$ and the spin difference $\chi_a = (a_{1,z}-a_{2,z})/2$. The third coefficient $\psi_3$ encodes the spin-orbit degeneracy as $\beta$ includes the spins and $\nu$ is the mass ratio. In our case, unlike in \cite{2020PhRvD.102f4063C}, the $\chi_{a}$ term is not negligible. For NSBH systems, especially under the assumption of a spinning secondary and nonspinning primary, both the mass difference $\delta m$ and spin difference $\chi_a$ are significant.
For our mock events, we approximate the measured PN coefficients $\psi_i$ as independent Gaussian distributions with standard deviations $\sigma_i$. As in \citet{2020PhRvD.102f4063C}, we adopt $\sigma_0 = 0.0046\psi_0/\rho$, $\sigma_2 = 0.2341\psi_2/\rho$, and $\sigma_3 = -0.1293\psi_3/\rho$, where we draw the SNR $\rho$ according to $p(\rho) \propto \rho^{-4}$, an approximation to the SNR distribution of a uniform-in-comoving-volume distribution of sources~\citep{2014arXiv1409.0522C}. We then sample $m_1$, $m_2$, $a_{1,z}$, and $a_{2,z}$ from the $\psi_0$, $\psi_2$, $\psi_3$ likelihoods, accounting for the priors induced by the change of variables by calculating the appropriate Jacobian transformations. 

An example NSBH parameter estimation posterior generated according to this procedure is shown in Fig. \ref{fig:simulated}. We see that the masses and spins are highly correlated. In particular, the anti-correlation between the secondary mass and spin increases the uncertainty on $M_\mathrm{TOV}$ and the spin--maximum mass relationship.

\section{Application to LIGO--Virgo NSBH Detections}  \label{ligodata}
\subsection{Data and Event Selection}
\begin{figure}
    \includegraphics[width=0.5\textwidth]{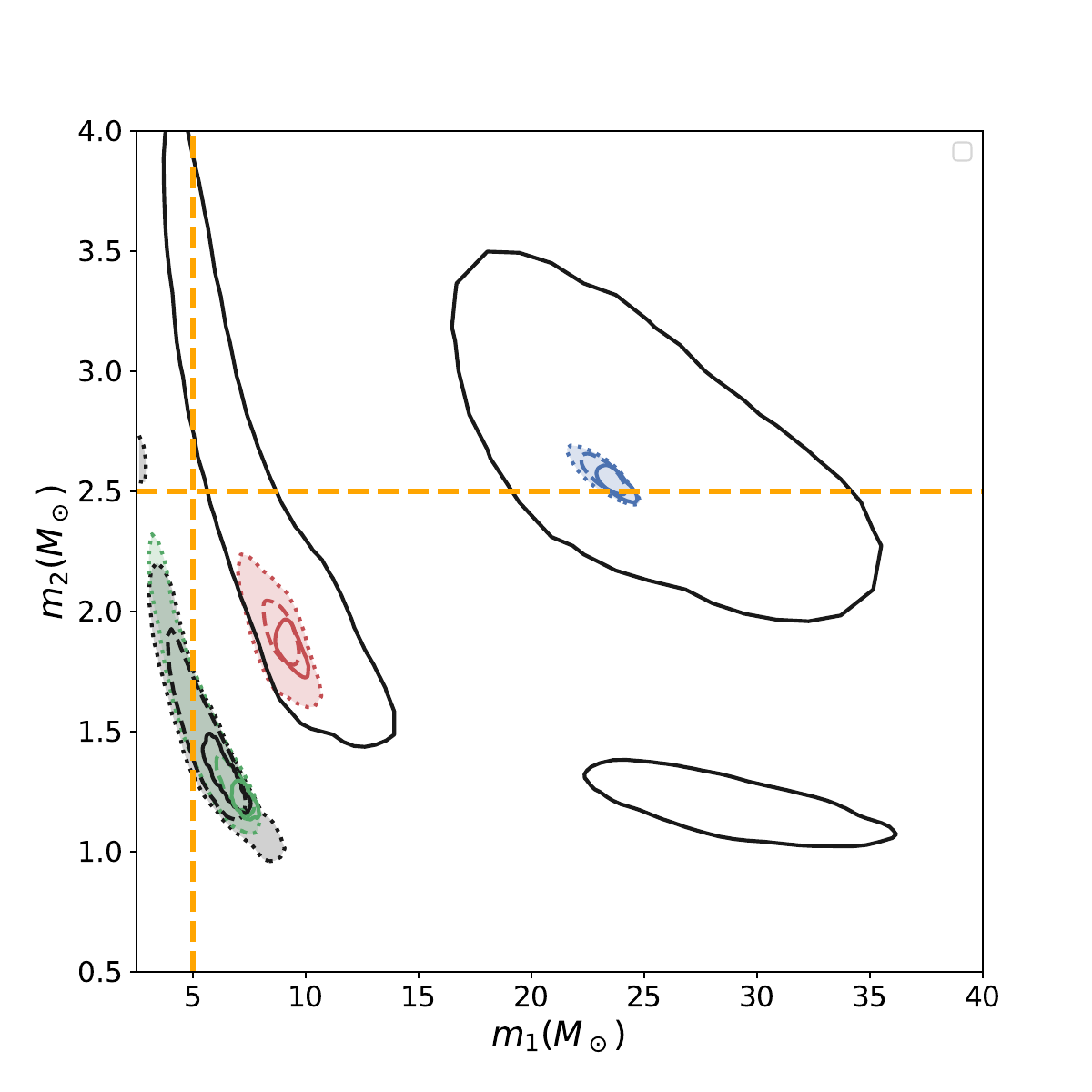}
    \caption{90\% contours on $m_1$ and $m_2$ from the four NSBH events: GW190426 (grey), GW190814 (blue), GW200105 (green), GW200115 (red). Plot features $m_1, m_2$ under three spin priors ``U+I" in dotted lines; ``ZS" in dashed lines; ``ZS + AS" solid (all 90\% contours). The spin priors correspond to the models from Sec.~\ref{spin models} with $\beta_s = 0$. \reply{Three} additional GWTC-2 and GWTC-3 events GW190917\_114630, GW191219\_163120, GW200210\_092254 are shown in black, using default LVK spin priors.}
    \label{fig:ligom1m2}
\end{figure}
In our population inference, we consider up to four LIGO--Virgo triggers as NSBH detections:\\

\begin{enumerate}
    \item GW200105 \citep{2021ApJ...915L...5A}; (all measurements quoted at 90\% confidence level) $m_1 = 8.9^{+ 1.2}_{-1.5} M_\odot$, $m_2 = 1.9^{+ 0.3}_{-0.2} M_\odot$ \\
    \item GW200115 \citep{2021ApJ...915L...5A}; $m_1 = 5.7^{+1.8}_{-2.1} M_\odot$, $m_2 = 1.5^{+0.7}_{-0.3} M_\odot$ \\
    \item GW190814 \citep{2020ApJ...896L..44A}; $m_1 = 23.2^{+1.1}_{-1.0} M_\odot$, $m_2 = 2.6^{+0.1}_{-0.1} M_\odot$. Because the secondary mass in GW190814 falls squarely into the putative lower mass gap, it is unclear whether GW190814 is a NSBH or BBH event. Accordingly, we do not include GW190814 in every analysis, but consider how it affects population estimates. \\
    \item GW190426\textunderscore 152155 (hereafter GW190426) \citep{2021PhRvX..11b1053A}; $m_1 = 5.7^{+3.9}_{-2.3} M_\odot$, $m_2 = 1.5^{+0.8}_{-0.5} M_\odot$. GW190426 is relatively low-significance with a network SNR of $\rho = 10.1$, and so may or may not be a real NSBH event. Accordingly, like with GW190814, we do not consider GW190426 in every analysis, but consider how it affects population estimates. \\
\end{enumerate}


\begin{figure*}
     \centering
     \begin{subfigure}[b]{0.23\textwidth}
         \centering
         \includegraphics[width=\textwidth]{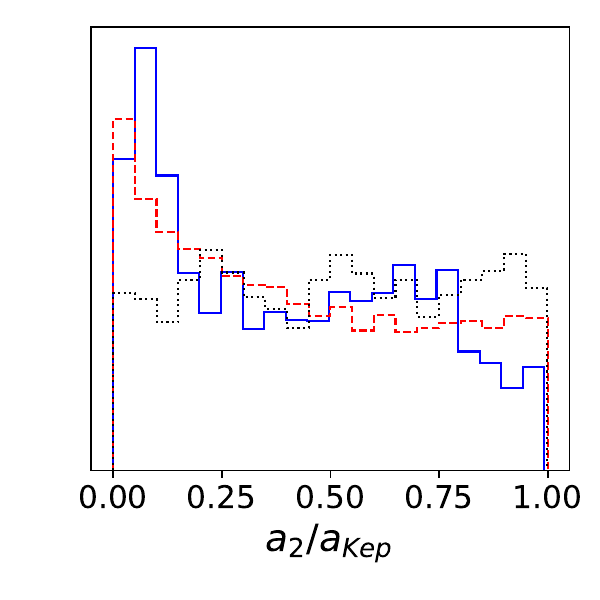}
         \caption{GW190426}
     \end{subfigure}
     \hfill
     \begin{subfigure}[b]{0.23\textwidth}
         \centering
         \includegraphics[width=\textwidth]{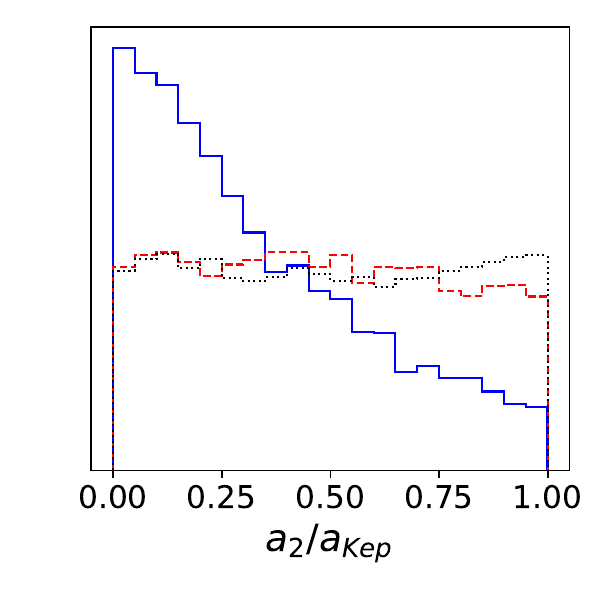}
         \caption{GW190814}
     \end{subfigure}
     \hfill
     \begin{subfigure}[b]{0.23\textwidth}
         \centering
         \includegraphics[width=\textwidth]{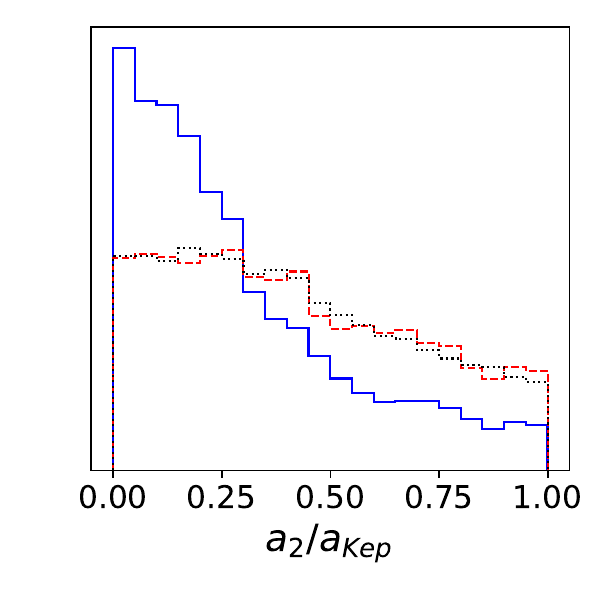}
         \caption{GW200105}
     \end{subfigure}
     \hfill
     \begin{subfigure}[b]{0.23\textwidth}
         \centering
         \includegraphics[width=\textwidth]{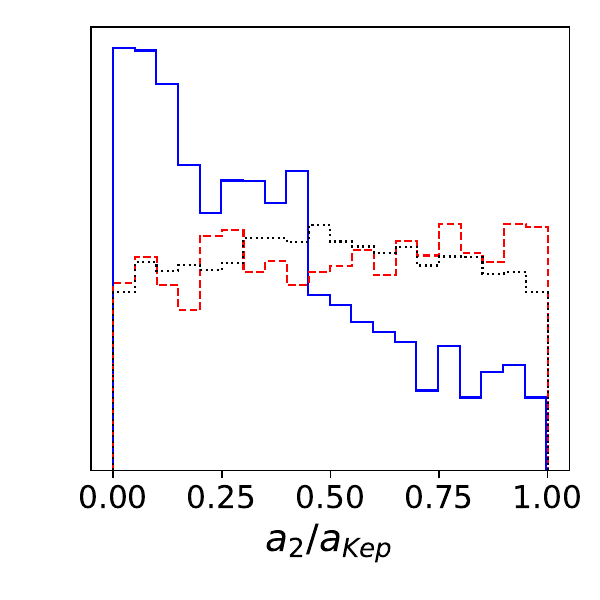}
         \caption{GW200115}
     \end{subfigure}
     \hfill
     \caption{Posteriors on NS spin $a_2/a_\text{Kep}$ from the NSBH events: GW190426, GW190814, GW200105, GW200115, inferred under different spin priors. ``U+I" $a_2$ samples in dotted black; ``ZS" in dashed red; ``ZS + AS" in solid blue.}
    \label{fig:ligoa2}
\end{figure*}
For GW200105 and GW200115, we use the ``Combined\textunderscore PHM\textunderscore high\textunderscore spin" parameter estimation samples from~\citet{2021ApJ...915L...5A}. For GW190426, we use the ``IMRPhenomNSBH" samples from~\citet{2021PhRvX..11b1053A}, and for GW190814, we use ``IMRPhenomPv3HM" from~\citet{2020ApJ...896L..44A}\footnote{The parameter estimation samples are available on the Gravitational Wave Open Science Center~\citep{GWOSC}}.
The default LIGO parameter estimation prior $\pi_\mathrm{PE}(\theta)$ is flat in component spin magnitudes and isotropic in spin tilts, following the ``U + I" spin prior. Meanwhile, the spin models ``ZS" and ``ZS + AS" described in Section~\ref{spin models} assume that the BH is nonspinning ($a_1 = 0$), and ``ZS + AS" further assumes that the NS spin is perfectly aligned.
In these models, we follow \citet{2020ApJ...895L..28M} and estimate $a_2 = |a_{2,z}/\text{cos(tilt$_2$)}|$ using the $\chi_\text{eff}$ posterior, accounting for the original $\chi_\text{eff}$ prior~\citep{2021arXiv210409508C}.
To illustrate the effect of the different spin assumptions on the inferred parameters of each NSBH event, we reweight the original parameter estimation posteriors by the three spin priors (the default ``ZS", as well as ``ZS + AS" and ``U + I") with $\beta_s = 0$. 
The $m_1, m_2$ posteriors for the four NSBH events under these three spin models are shown in Figure \ref{fig:ligom1m2}.
\reply{Analyses were performed on an initial set of 4 GW NSBH events from~\citet{2021PhRvX..11b1053A} and \citet{2021ApJ...915L...5A}, which were available at the start of this work}. During the course of this work, the latest LIGO--Virgo catalog GWTC-3 was released, which also includes the low-significance NSBH candidates GW190917\_114630, GW191219\_163120, GW200210\_092254~\citep{2021arXiv211103606T,ligo_scientific_collaboration_and_virgo_2021_5546663}; the inferred masses of these sources under the default priors are also shown in Figure \ref{fig:ligom1m2}. \reply{A similar full analysis could be applied to this larger sample of NSBH events, but we find only a slight shift in inferred values of $M_\text{TOV}$ and $M_\text{BH}$ with the addition of the 3 GWTC-3 events.}
In general, the ``U+I" model produces the broadest posteriors, while ``ZS + AS" provides the tightest constraints and the default ``ZS" model is in the middle. In the ``ZS" and ``ZS + AS" model, we see that fixing the BH spin to zero tends to increase the support for lower $m_2$ and higher $m_1$ because of the anti-correlation between $q = m_2/m_1$ and $\chi_\mathrm{eff}$, bringing both components out of the putative mass gap~\citep{2021arXiv210914759M}. Because the secondary spin is poorly measured, $a_2$ is poorly constrained and essentially recovers the broad prior (Figure~\ref{fig:ligoa2}). 

When fitting the population models, we divide the NSBH events into four different sets: ``confident'', with just GW200115 and GW200105; ``all'', with all four potential NSBH triggers; and excluding GW190814 and GW190426 one at a time each. For each event set, we repeat the population inference using the three different spin priors -- ``U+I", ``ZS", and ``ZS + AS" -- and three different NS mass models in Section \ref{ns models} -- uniform, 1-component (1C), and 2-component (2C). Finally, we also vary the pairing function between $\beta=3$ (preference for equal masses) and $\beta=0$ (random pairing). In total, we consider 72 model/dataset variations. Unless stated otherwise, results refer to the ``ZS" spin prior, a 1-component mass function, and random pairing ($\beta = 0$).


\subsection{Population Properties}

\begin{figure*}
     \centering
         \includegraphics[width=0.9\textwidth]{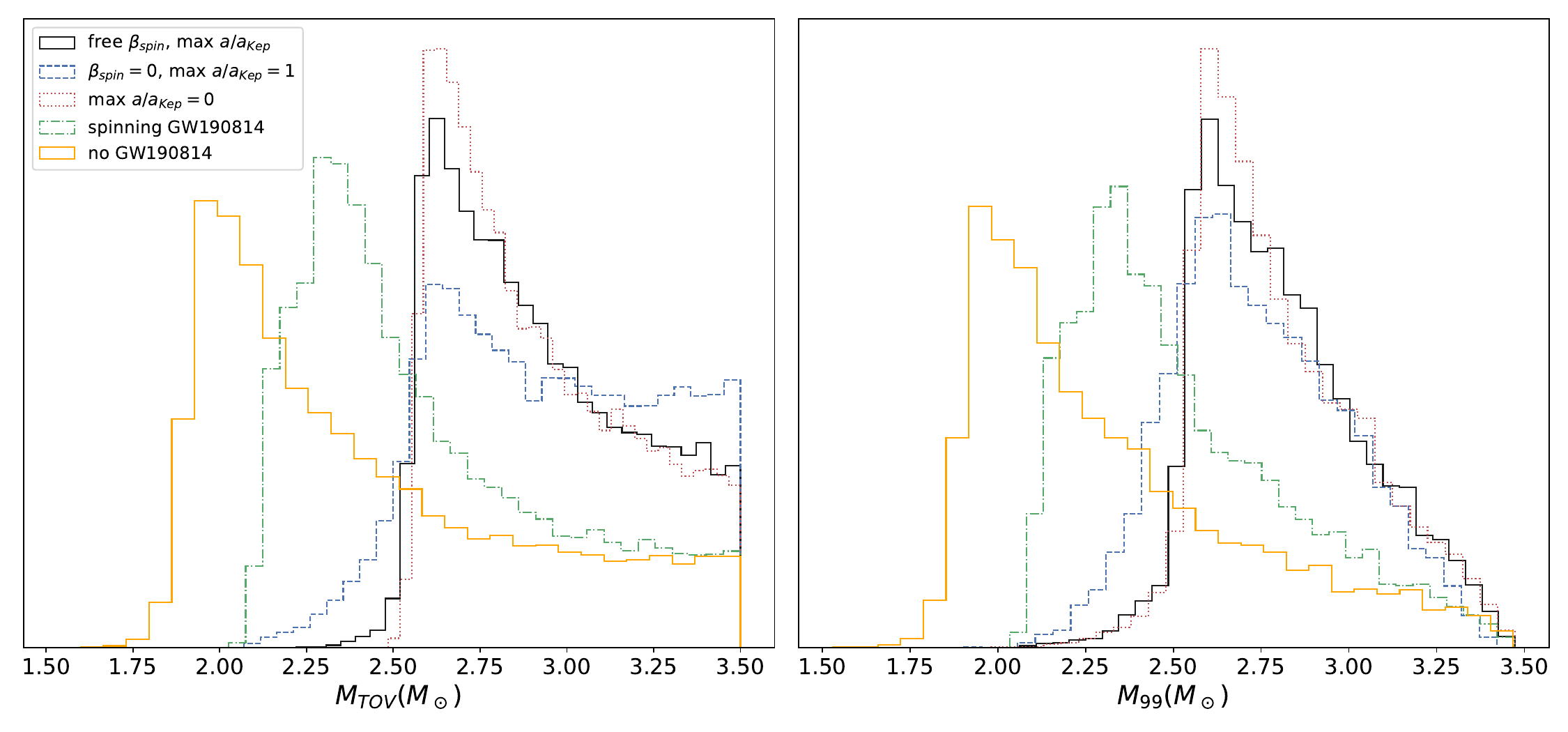}
    \caption{Estimates of $M_\text{TOV}$ and $M_{99}$ for different priors on the population $a_2$ distribution parameters $\beta_\text{s}$ and $a_\text{max}/a_\text{Kep}$, using the ``ZS" spin prior: free $\beta_\text{s}$ and $a_\text{max}/a_\text{Kep}$; fixed $\beta_\text{s} = 0$ fixed $a_\text{max}/a_\text{Kep}$; and $a_\text{max}/a_\text{Kep}$ (nonspinning); and fixed $a_\text{min}/a_\text{Kep} = 0.9$ (requiring maximally spinning GW190814). All events fit with a 1-component NS mass model and random pairing ($\beta = 0$).}
    \label{fig:tov_spin}
\end{figure*}

\subsubsection{$M_\text{TOV}$, $M_\text{BH}$, and the Mass Gap}
\begin{figure}
    \centering
     \begin{subfigure}[b]{0.4\textwidth}
         
         \includegraphics[width=\textwidth]{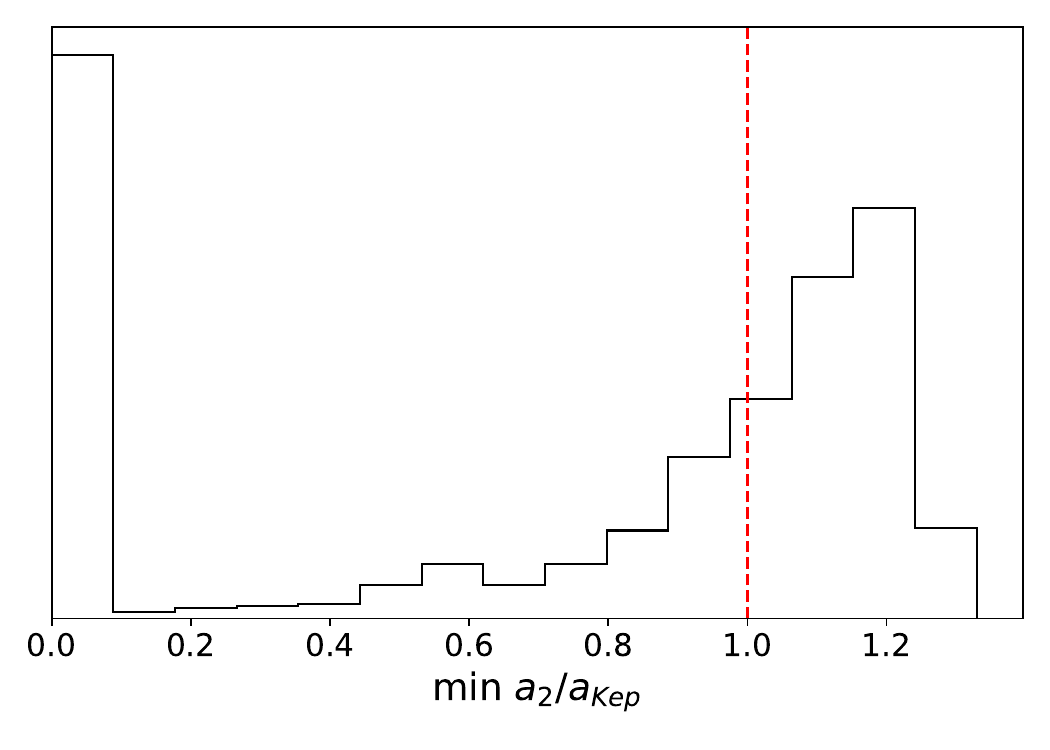}
         \caption{$M_\text{TOV}$ inferred from all other NSBH events.}
     \end{subfigure}
     \hfill
     \bigskip
     \begin{subfigure}[b]{0.4\textwidth}
         
         \includegraphics[width=\textwidth]{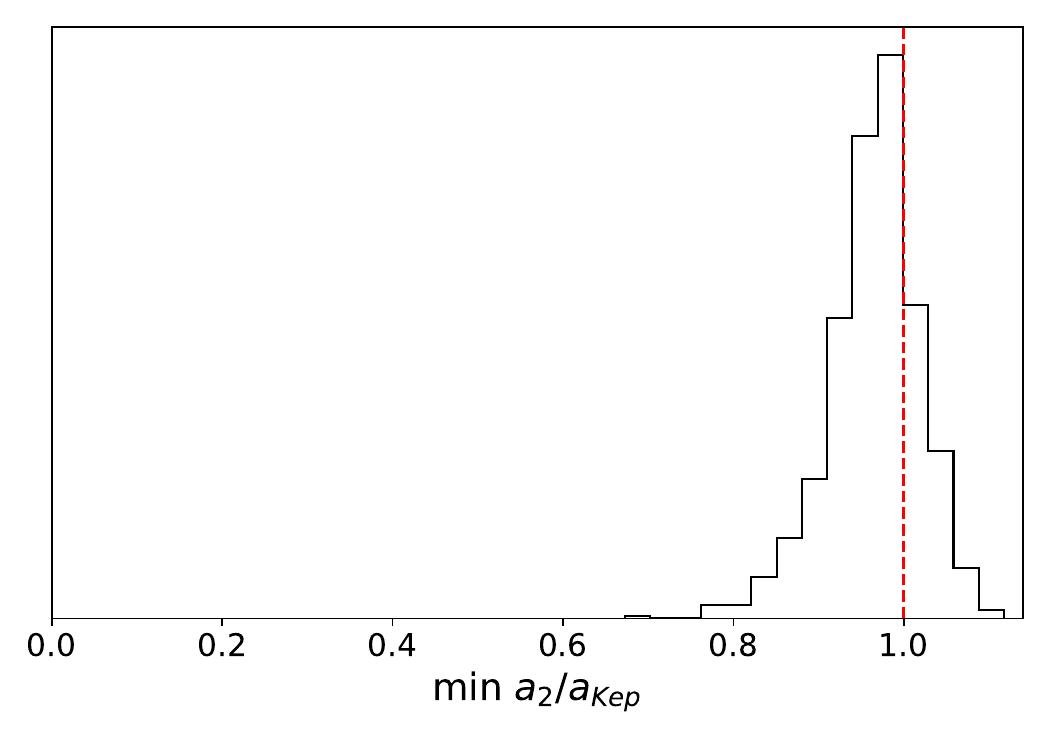}
         \caption{$M_\text{TOV}$ inferred from 150 future observations.}
     \end{subfigure}
     
    \caption{Constraints on the minimum spin of GW190814's secondary given its mass (using the ``ZS" spin model), assuming it is a massive rotation-supported neutron star, from population estimates of $M_\text{TOV}$. $a_2/a_\text{Kep} = 0$ means no spin support is required to make GW190814's mass consistent with the NSBH population inferred from the other events.}
    \label{fig:gw190814_spin}
\end{figure}

\begin{figure}
     \centering
         \includegraphics[width=0.5\textwidth]{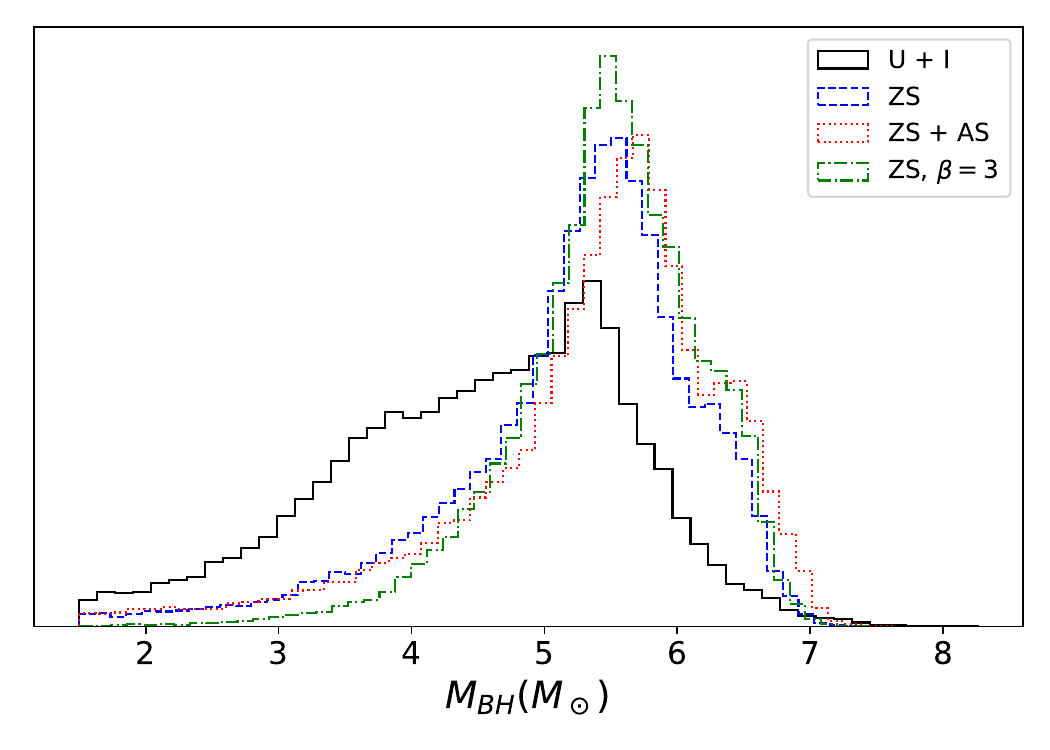}
    \caption{Estimates of $M_\text{BH}$ for different $a_2$ priors with pairing function $\beta = 0$: uniform and isotropic (``U + I", black solid), non-spinning BH (default ``ZS", blue dashed), and non-spinning BH + aligned spin NS (``ZS + AS", red dotted). Posterior on $M_\text{BH}$ for the ``ZS" spin prior with $\beta = 3$ is shown (green dash-dot). All events fit with a 1-component NS mass model.}
    \label{fig:mbh}
\end{figure}

\begin{figure*}
     \centering
     \begin{subfigure}[b]{0.49\textwidth}
         \centering
         \includegraphics[width=\textwidth]{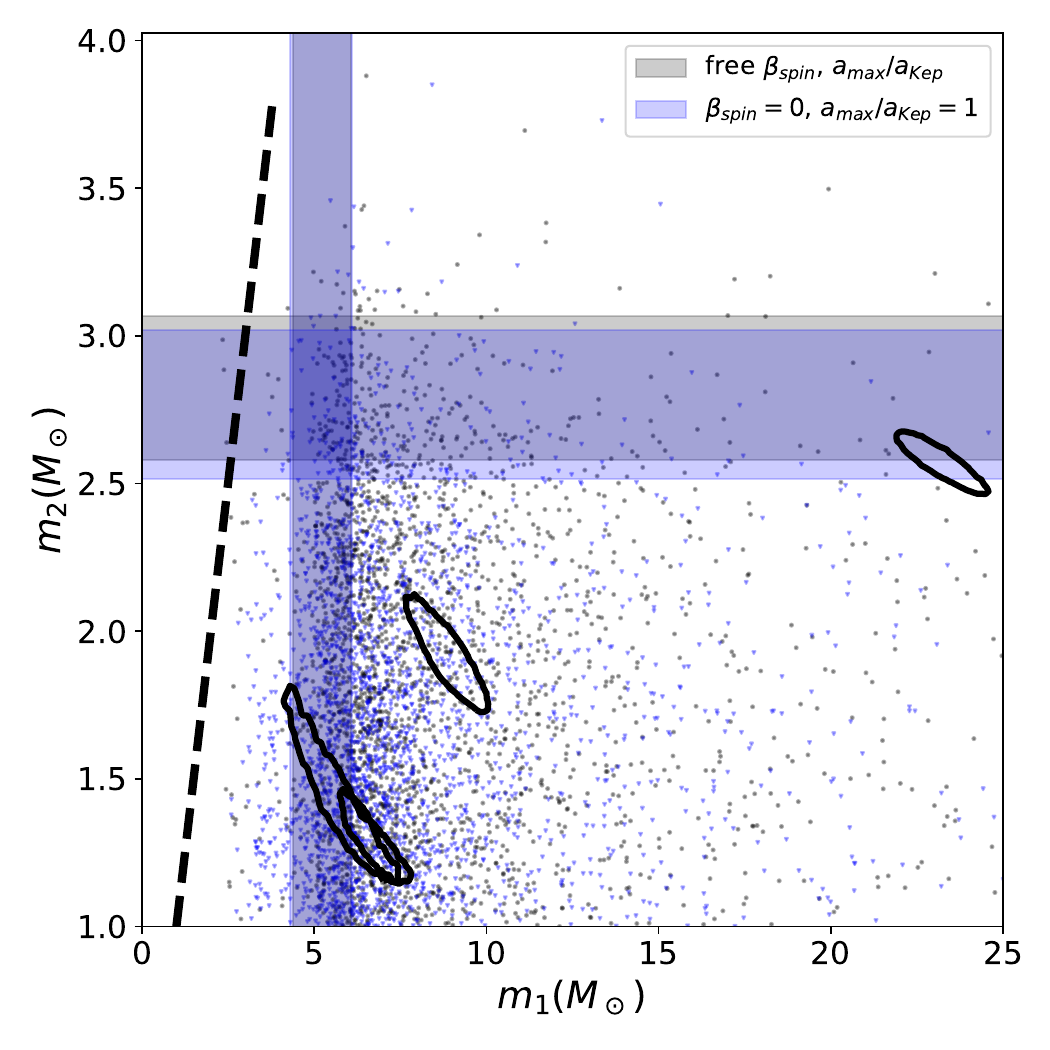}
         \caption{all events}
     \end{subfigure}
     \hfill
     \begin{subfigure}[b]{0.49\textwidth}
         \centering
         \includegraphics[width=\textwidth]{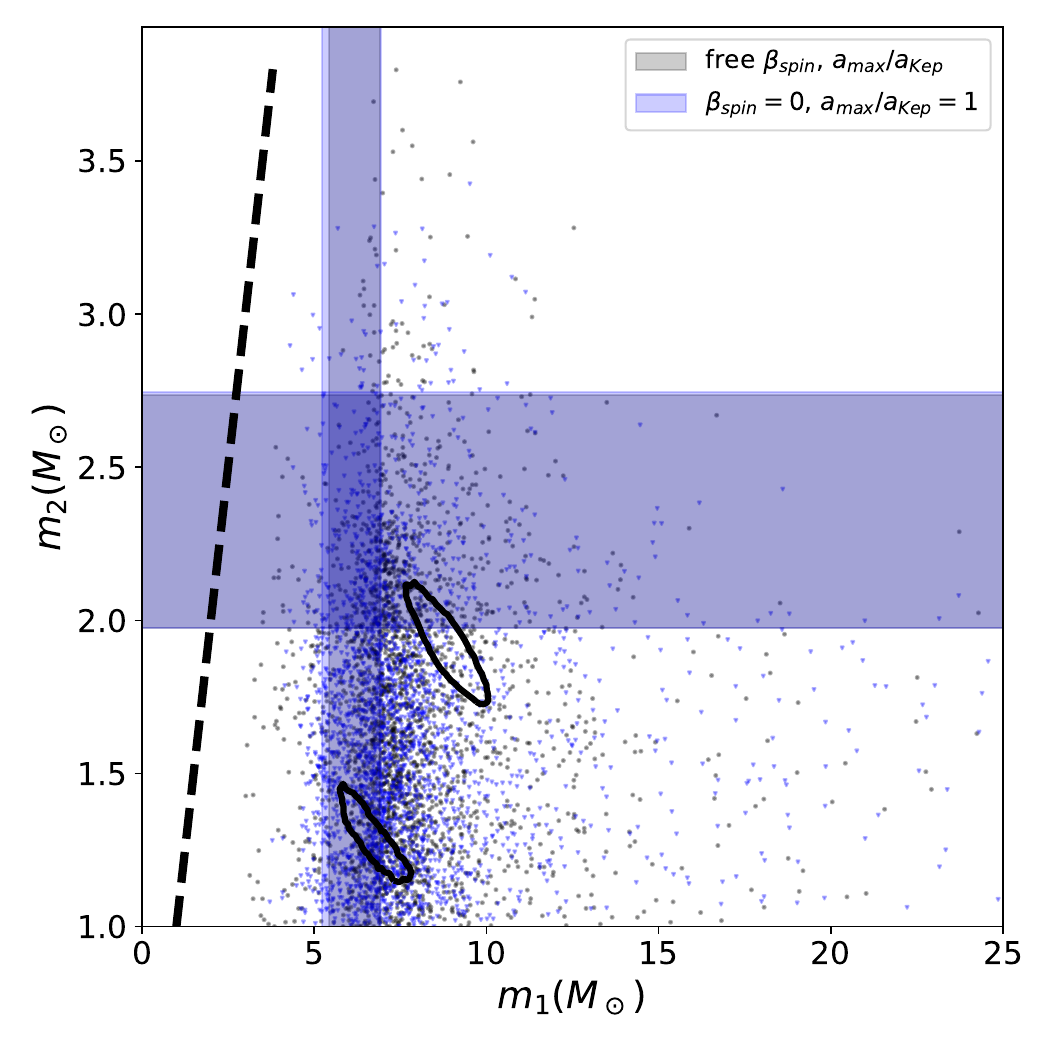}
         \caption{confident only}
     \end{subfigure}
     \bigskip{}
     \begin{subfigure}[b]{0.48\textwidth}
         \centering
         \includegraphics[width=\textwidth]{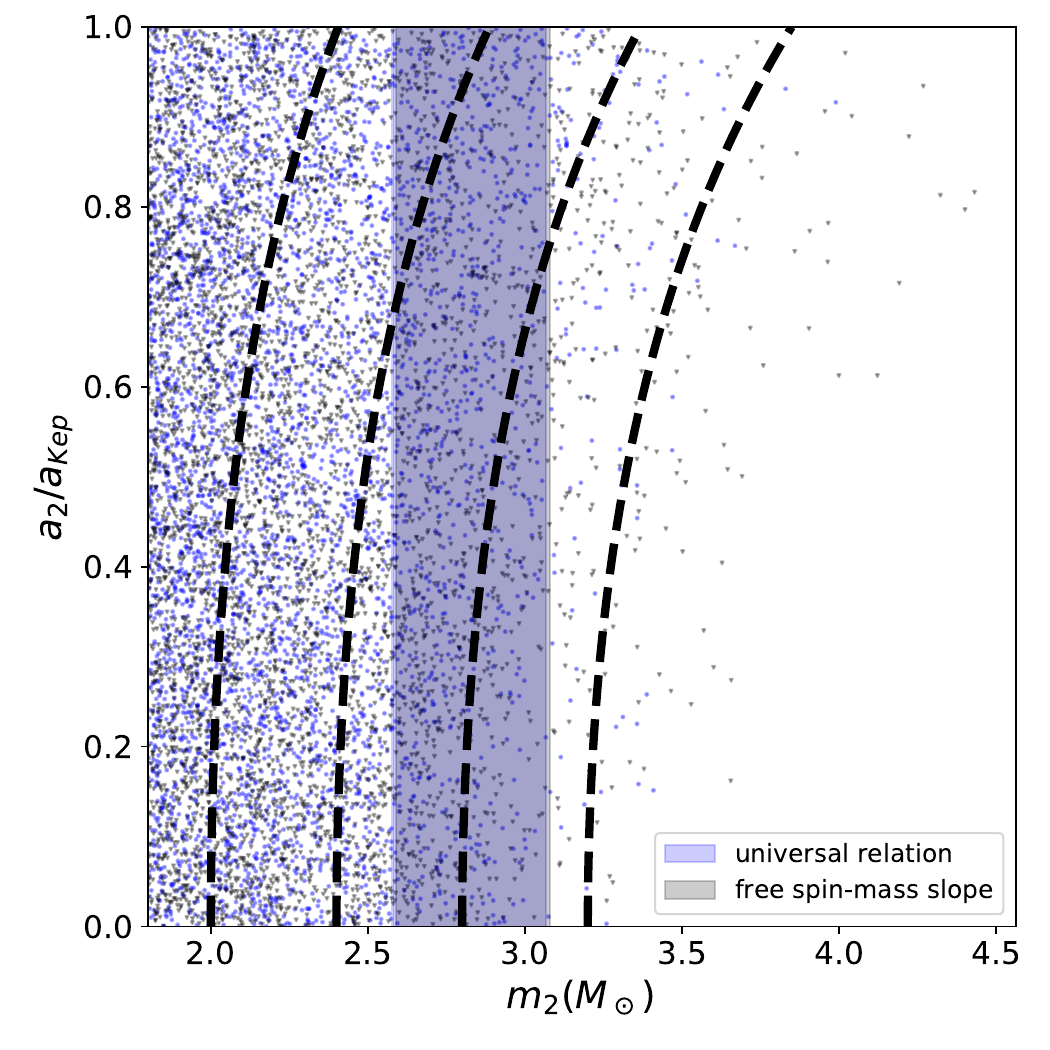}
         \caption{all events, $\beta=0$}
     \end{subfigure}
     \hfill
     \begin{subfigure}[b]{0.48\textwidth}
         \centering
         \includegraphics[width=\textwidth]{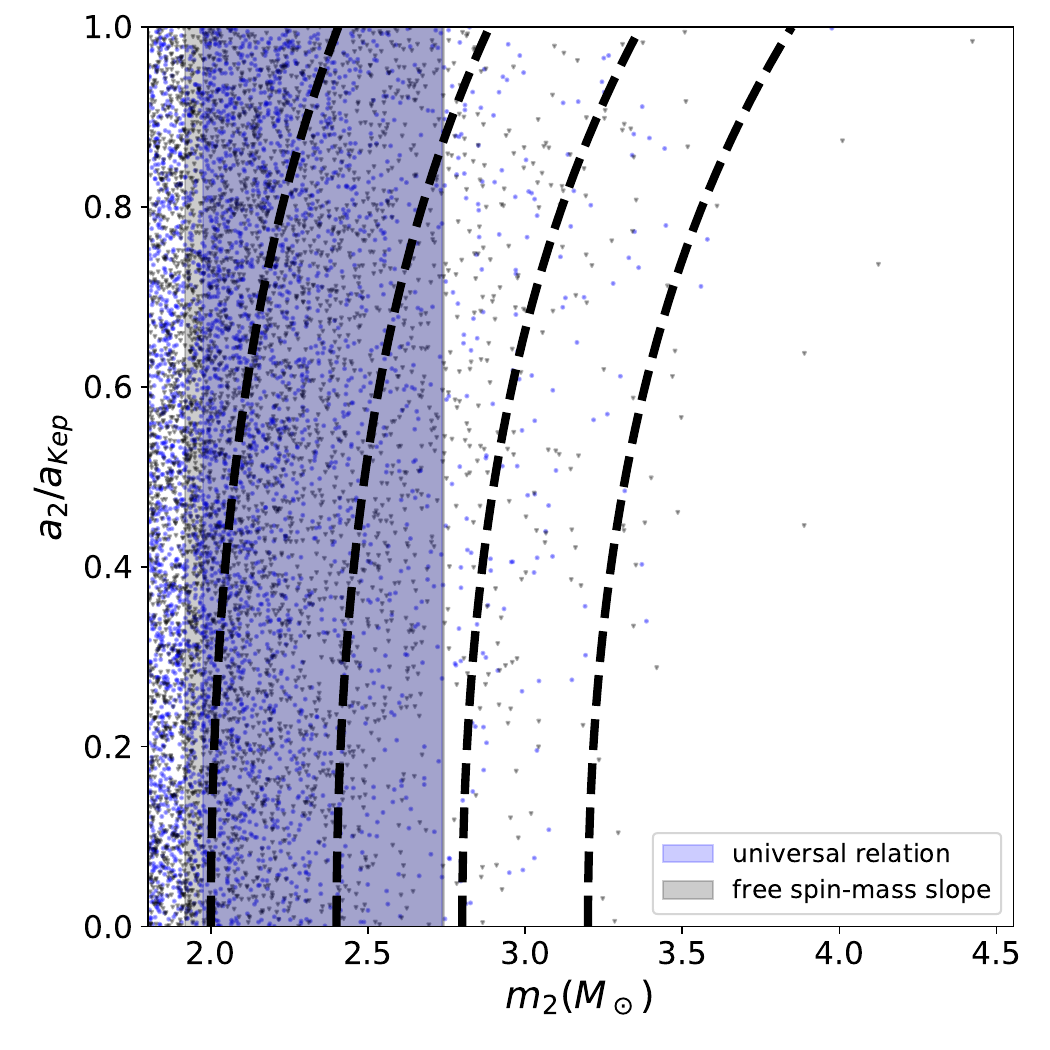}
         \caption{confident only, $\beta=0$}
     \end{subfigure}
    \caption{Posterior predictive distributions of NSBH events (conditioned on detection), as inferred under the different models, using the ``ZS" spin prior and pairing function $\beta = 0$. 68\% credible intervals on $M_\text{TOV}$ and $M_\text{BH}$ are shown. 90\% contours for the LVK NSBH events are overplotted (horizontal and vertical bars). In (a) and (b), the black dotted line shows equal NS and BH mass; we define a mass gap as $M_\text{TOV}  < M_\text{BH}$. In (c) and (d), black dotted lines show $M_\text{crit}$ for a given $M_\text{TOV}$ and spin $a_2/a_\text{Kep}$. (a) and (b) show 2500 draws each; (c) and (d) show events over $2 M_\odot$ from 10,000 draws.}
    \label{fig:ppd}
\end{figure*}

For each model and dataset variation, we infer the minimum BH mass $M_\text{BH}$, the NS $M_\text{TOV}$, and their difference (representing the width of the mass gap), marginalizing over all other parameters of the mass and spin distribution. Results for our \emph{Default} model are shown in Figs.~\ref{fig:tov_spin}-\ref{fig:fullcorner}, with Fig.-\ref{fig:fullcorner} showing a corner plot over all model parameters.\\

{\bf Maximum (spin-dependent) NS mass}: As discussed in \ref{ns models}, at a given secondary spin, we model a hard cut-off in the NS mass distribution $p(m_2)$. However, in the 1-component and 2-component models, some values of $\mu$ and $\sigma$ taper off the mass distribution between 2--3 $M_\odot$, making it difficult to discern a sharp truncation mass $M_\text{TOV}$ from the function's normal behavior. This results in long, flat tails to large posterior values of $M_\text{TOV}$ (see panel (a) of Fig.~\ref{fig:tov_spin}), \reply{reaching the prior bounds even if priors on $M_\text{TOV}$ are widened}. A better measured parameter is the 99th percentile of the nonspinning NS mass distribution, $M_{99}$ (panel b of Fig.~\ref{fig:tov_spin}). For models where the $M_\text{TOV}$ cutoff is significant, the 99th percentile is essentially identical to $M_\text{TOV}$. For models producing a softer cutoff without significant $M_\text{TOV}$ truncation, the 99th percentile still captures the largest NS we expect to observe, and, unlike $M_\text{TOV}$, the inference of $M_{99}$ is consistent between the three NS mass models.

For models including GW190814, we generally infer $M_{99}$ between $2.6-2.8 M_\odot$, with lower limits (95\% credibility) of 2.4-2.5 $M_\odot$. Our default model (all 4 events, $\beta = 0$, ``ZS" spin prior) measures $M_{99} = 2.8^{+0.3}_{-0.2}\,M_\odot$ (68\% credibility); \reply{the inclusion of 3 additional GWTC-3 events shifts $M_\text{99}$ to $2.9^{+0.2}_{-0.2} M_\odot$}. The cutoff mass is set by GW190814, where $m_2$ is extremely well-constrained. Without GW190814, we estimate $M_\text{TOV}$ between $2.0$--$2.3\,M_\odot$, with lower limit (95\% credibility) of 1.8-1.9 $M_\text{TOV}$. Without GW190814, our estimates are consistent with other estimates of $M_\text{TOV}$ from gravitational-wave NS observations that do not consider spin.

The spin distribution affects the inferred value of $M_\text{TOV}$ and $M_\text{99}$. For all four events, $m_2$ is consistent with being both non-spinning ($a_2 = 0$) or maximally spinning ($a_2 = 0.7$, $a/a_\text{Kep} = 1$). When the spin distribution allows or favors maximally spinning NS, lower values of $M_\text{TOV}$ are allowed and can still account for GW190814, the most massive secondary. When the spin distribution disfavors high spins, the spin-dependent maximum mass is lower and $M_\text{TOV}$ must be higher in order to accommodate GW190814.

This is shown in Figure \ref{fig:tov_spin}; the posterior on $M_\text{TOV}$ inferred under a uniform spin distribution ($\beta_\text{s} = 0$, $a_\text{max}/a_\text{Kep} = 1$), which has support at high NS spins, has a significant tail to lower values below $2.5\,M_\odot$ (dashed blue curve).  
A prior that requires GW190814 to be maximally spinning ($a_\text{min}/a_\text{Kep} = 0.9$) brings $M_\text{TOV}$ estimates even lower, to $\sim2.4$ $M_\odot$, with support below $2.2\,M_\odot$ (green dashed curve in Fig. \ref{fig:tov_spin}). Meanwhile, requiring all NSs to be nonspinning ($a_\text{max}/a_\text{Kep} = 0$) means that GW190814's secondary (if it is a NS) sets the non-spinning maximum mass for the population, and results in a narrower posterior preferring larger values. The difference between posteriors on $M_\text{TOV}$ and $M_\text{99}$ modeled with GW190814 (black solid, red dotted, blue dashed) and without GW90814 (solid yellow curve) is bridged partially by models assuming GW190814's spin is near-maximal. 
This effect is also visible in Fig. \ref{fig:traces}; if GW190814 is assumed spinning, the upper end of $p(m_2)$ visibly shifts to lower masses, and zero-spin NS mass functions truncating below GW190814's secondary's mass are allowed (see overplotted credible interval).  We see that even in the absence of well-constrained $a_2$, modeling a spin-dependent maximum mass has significant effects on the inferred NS mass distribution.


\textbf{GW190814's secondary spin}: Using the posterior on $M_\text{TOV}$ (Fig. \ref{fig:tov_spin}) inferred from the population of NSBHs excluding GW190814, we can infer the minimum secondary spin of GW190814 required for it to be consistent with the NSBH population. Results are shown in panel (a) of Fig. \ref{fig:gw190814_spin}. For our sample of NSBH events excluding GW190814, the results are inconclusive: because the posterior on $M_\text{TOV}$ is broad, GW190814 is consistent even if non-spinning (with the minimum required $a_2/a_\text{Kep} = 0$), but it may also be maximally spinning with $a_2/a_\mathrm{Kep} = 1$. GW190814 may also be an outlier from the NSBH population, even if it is maximally spinning: for this figure, we allow min $a_2/a_\text{Kep} > 1$, but $a_2/a_\text{Kep} > 1$ would imply inconsistency with the rest of the population as $a_\text{max}/a_\text{Kep} = 1$. Future GW observations of a larger population of NSBH events (see panel (b) of Fig. \ref{fig:gw190814_spin}) could allow a much tighter measurement of GW190814's secondary spin.

{\bf Minimum BH mass}: Across all models, the inferred BH minimum mass $M_\text{BH}$ is between 4--7 $M_\odot$ with typical uncertainties of $\pm 1 M_\odot$. Our default model using the ``ZS" spin model, all 4 NSBH events (GW190814, GW190426, GW200105, GW200115), and random pairing ($\beta = 0$) results in $M_\mathrm{BH} = 5.4^{+ 0.7}_{- 1.0}\,M_\odot$ (68\% credibility). At the low end, we infer $M_\text{BH} = 4.2^{+ 1.1}_{-1.0}\,M_\odot$ using all 4 NSBH events, a uniform NS mass distribution, pairing function $\beta=3$, and the ``U+I" spin model. At the high end, we infer $M_\text{BH} = 6.7^{+ 0.4}_{-0.8}\,M_\odot$ (68\% credibility) using only the confident NSBH events and the  spin model. The effect of the $m_1, m_2$ pairing function $\beta$ is minimal, but assuming equal-mass pairings further reduces posterior support for low $M_\text{BH}$ (see Figure \ref{fig:mbh}). \\

{\bf Mass gap}: We estimate the inferred width of the lower mass gap as the difference between the minimum BH mass, $M_\text{BH}$, and the maximum nonspinning NS mass, $M_\text{TOV}$ or $M_{99}$. \reply{The mass gap's width may range from 0 to a few $M_\odot$, while the mass gap's position may range from 2-7 $M_\odot$}. As seen in Figures \ref{fig:tov_spin} and \ref{fig:mbh}; the overlap between the posteriors on $M_\text{BH}$ and $M_\text{99}$ is low, suggesting the existence of a mass gap. Similarly, panels (a) and (b) in Figure \ref{fig:ppd} show inferred ($m_1, m_2$) posterior predictive distributions, overplotted with the LVK $m_1, m_2$ posteriors. As Fig. \ref{fig:ppd} illustrates, {for all model variations we find evidence for a separation between the upper end of the NS mass distribution and the lower end of the BH mass distribution}. 

For our default model, we measure a mass gap of $2.5^{+0.8}_{-1.0} M_\odot$ \reply{($2.3^{+0.7}_{-1.0} M_\odot$ with 3 additional GWTC-3 events)}, wider than $0\,M_\odot$ with 97\% credibility and $1\,M_\odot$ with 90\% credibility. The inferred mass gap is widest when only using the confident NSBH events, between $3.0-4.5 M_\odot$, and narrowest when using all 4 NSBH events, between $1.5-3.0 M_\odot$. This is because the mass gap is narrowed from the NS side by the inclusion of GW190814, and from the BH side by the inclusion of GW190426 (see Figure \ref{fig:ligom1m2}. All model variations (spin prior, $\beta$, events) support for the existence of a mass gap: $>0\,M_\odot$ with 92\% or higher (up to $>99.9$\%) credibility, and $>1 M_\odot$ with 68\% or higher (up to $>99.9$\%) credibility.  

As seen in Fig. \ref{fig:ligom1m2}, additional spin assumptions (namely assuming that the BH is nonspinning and/or the NS spin is aligned) tend to prefer lower $m_2$ and higher $m_1$, which widens the inferred mass gap. 
When using spin priors in which the BH is assumed to be nonspinning, even when modeling all 4 events (including GW190814) we infer a mass gap exists with $>96\%$ credibility and that it is wider than $1\,M_\odot$ with $>90\%$ credibility.

\subsubsection{Mass and spin distributions}
\begin{figure}
    \centering
     \begin{subfigure}[b]{0.45\textwidth}
         \centering
         \includegraphics[width=\textwidth]{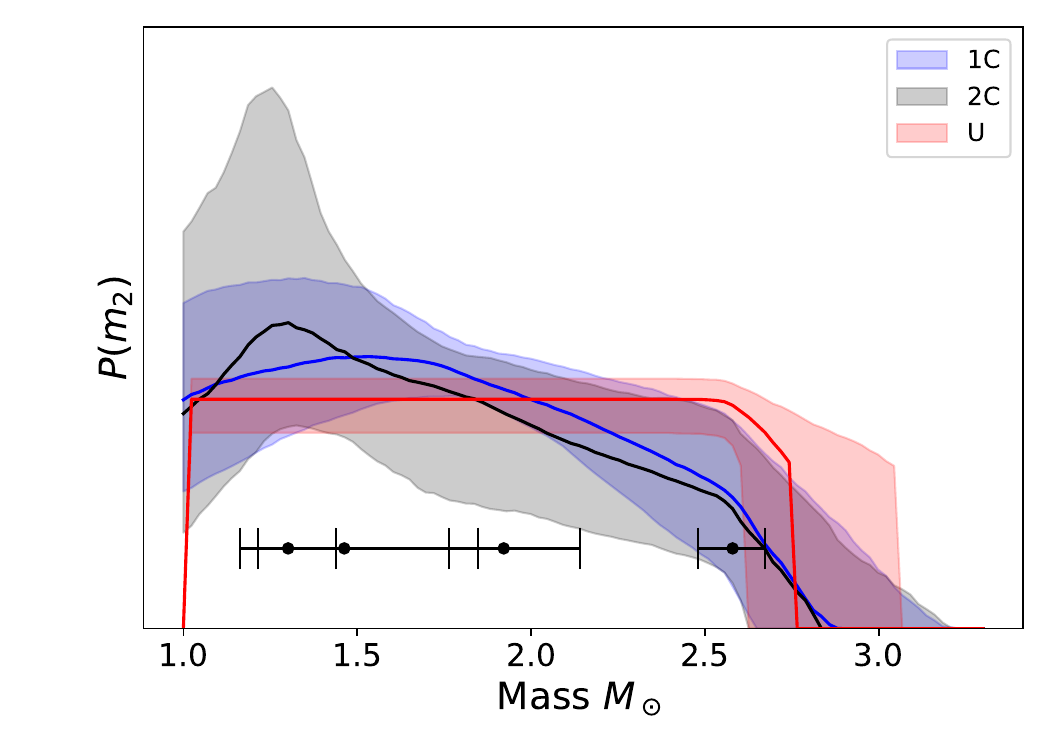}
         \caption{varying NS model}
     \end{subfigure}
     \hfill
     \centering
     \begin{subfigure}[b]{0.45\textwidth}
         \centering
         \includegraphics[width=\textwidth]{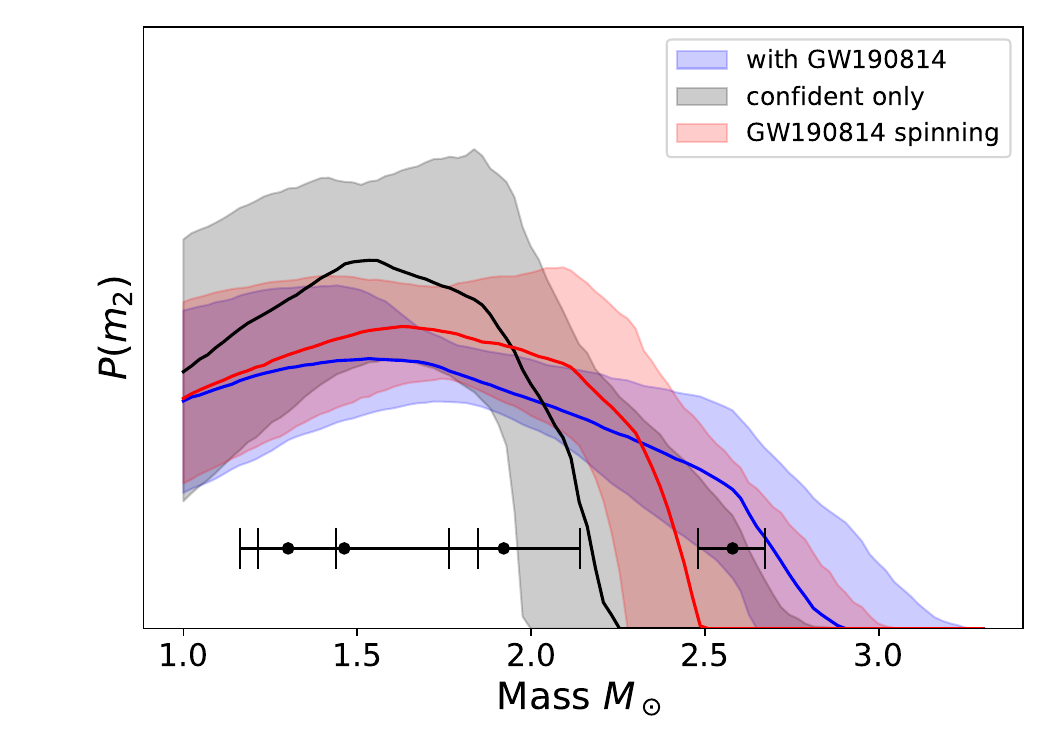}
         \caption{varying event selection}
     \end{subfigure}
     \hfill
    \caption{Median and 68\% credible interval of the non-spinning NS mass distribution $p(m_2)$ as inferred from variations on our fiducial model: all 4 NSBH events with a 1-component NS mass function, the ``ZS" spin prior, and $\beta = 0$. Each panel shows a different set of variations. The 95\% credible interval for $m_2$ is shown for each NSBH event.}
    \label{fig:traces}
\end{figure}

In addition to the most astrophysically relevant parameters -- $M_\text{BH}$, $M_\text{TOV}$, and the width of the mass gap -- we also constrain other parameters of the primary and secondary mass functions. In this section, we discuss general trends in the mass distribution shape, as inferred from posterior traces (Figure \ref{fig:traces}).

We first consider the NS mass distribution, $p(m_2)$, which differs slightly depending on the mass model used. 
For the 1-component model, we generally infer a broad distribution ($\sigma \simeq 0.5$) with mean $\mu$ between $1.2$ and $1.6 M_\odot$. A broad distribution is especially necessary to explain the large secondary mass of GW190814. The 2-component model generally agrees well with the 1-component model, although additional substructure (see panel (a) of Fig. \ref{fig:traces}), particularly a narrower peak at around 1.3 $M_\odot$ and a longer tail to high NS masses (above $2 M_\odot$) is possible. The only free parameter in the uniform model is the cutoff mass $M_\text{TOV}$. Though the flatness of the uniform model means we necessarily infer higher probability at masses near $M_\text{TOV}$, $M_\text{TOV}$ is generally consistent with the upper limit (99th percentile $M_{99}$) inferred from other mass models.  

The BH mass function is consistent between the three NS mass models. The most significant influence is the pairing function ($\beta = 0$ for random or $\beta = 3$ for equal-mass preference). For example, under our default model \reply{(4 events)}, which includes random pairing ($\beta = 0$), we infer a distribution with power-law slope $\alpha_\text{BH} = 3.4^{+1.4}_{-0.9}$ \reply{($\alpha_\text{BH} = 2.3^{+7}_{-1.0}$ with all 7 events)}. Under the same assumptions but preferring equal masses, $\beta = 3$, the inferred distribution shifts to significantly shallower slopes, $\alpha_\text{BH} = 0.9^{+1.1}_{-0.6}$. This is because the preference for equal-mass pairing requires a shallower slope in order to account for higher-mass black holes, especially the primary of GW190814. 

\begin{figure*}
    \centering
    \includegraphics[width=0.9\textwidth]{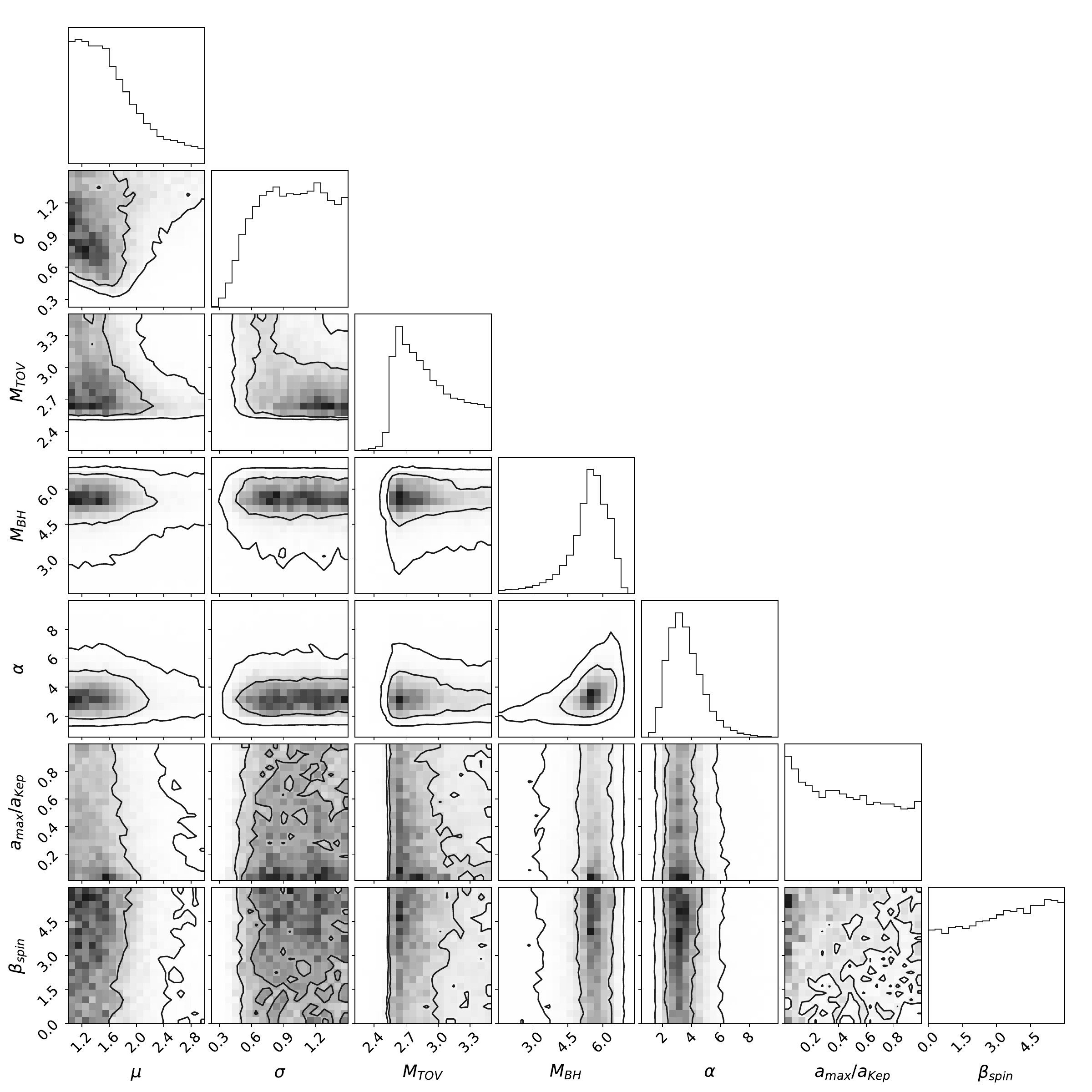}
     \hfill
     \caption{Joint posterior (68\% and 95\% contours) based on data from all NSBH events, the default ``ZS" spin prior, and a 1-component NS mass model. }
    \label{fig:fullcorner}
\end{figure*}

As seen in Fig. \ref{fig:fullcorner}, the joint posterior on $\beta_\text{spin}$ and $a_\text{max}/a_\text{Kep}$ prefers low $a_\text{max}/a_\text{Kep}$ and high $\beta_\text{spin}$, but mainly recovers the flat prior, which inherently prefers steeper and smaller spin distributions. Thus our measurement of the NS spin distribution is mostly uninformative, with a very mild preference for small spins. 


\section{Projections for aLIGO and A+} \label{Projections}

\begin{figure*}
     \centering
     \begin{subfigure}[b]{0.49\textwidth}
         \centering
         \includegraphics[width=\textwidth]{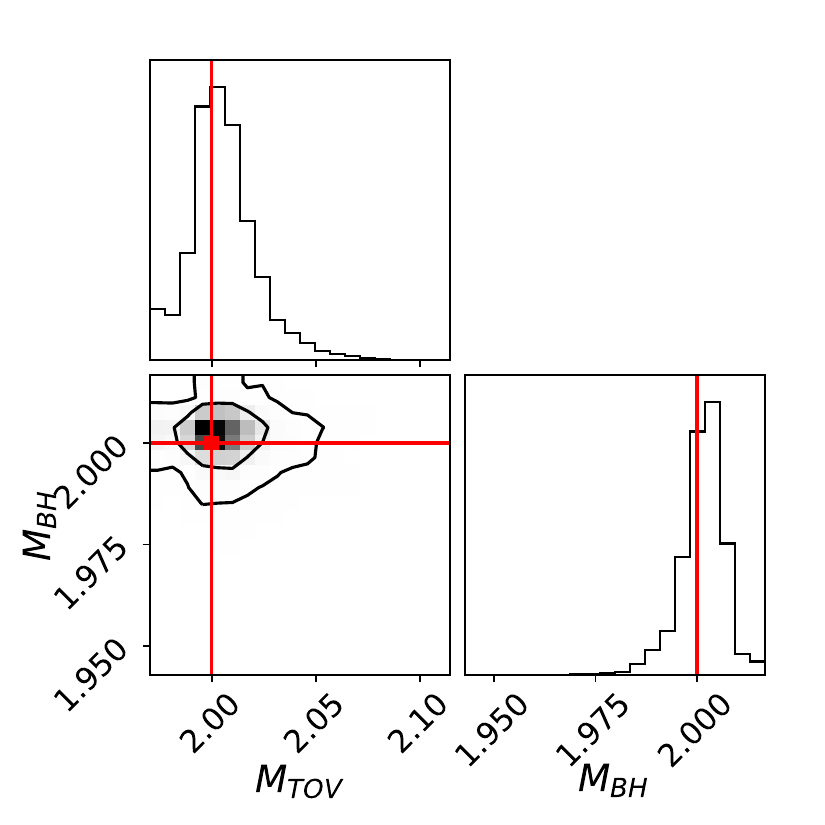}
         \caption{$M_{\text{TOV}} = 2.0$, $M_{BH} = 2.0$}
     \end{subfigure}
     \hfill
     \begin{subfigure}[b]{0.49\textwidth}
         \centering
         \includegraphics[width=\textwidth]{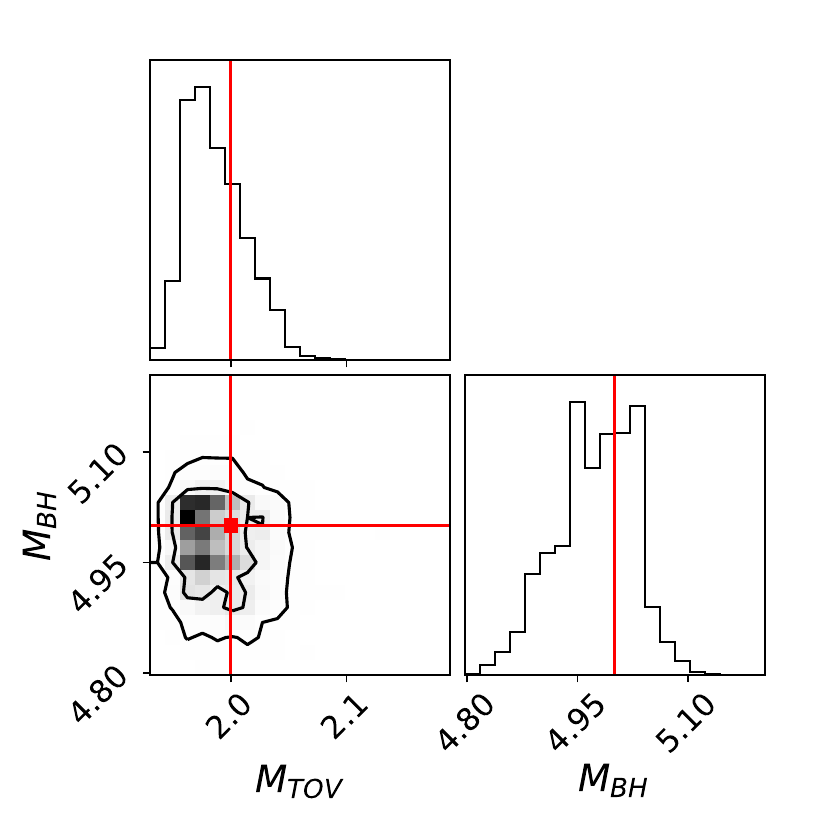}
        \caption{$M_{\text{TOV}} = 2.0$, $M_{BH} = 5.0$}
     \end{subfigure}
     \hfill
    \caption{Simulated posteriors on $M_{\text{TOV}}$ and $M_{BH}$ from 150 NSBH events at LIGO A+ sensitivity. Contours enclose 68\% and 95\% of the posterior probability.}
    \label{fig:simulated_massgap}
\end{figure*}

\subsection{Simulations}

In this section, we study measurements of NS and BH population properties from future observations. For our simulations, we use a fiducial set of parameters. We consider the three NS mass models. For the uniform NS mass distribution, we take $M_\text{TOV} = 2\,M_\odot$ or $2.2\,M_\odot$. For the 1-component distribution, we take $\mu = 1.5$ and $\sigma = 0.5$. For the 2-component distribution, based on \cite{2020PhRvD.102f4063C}, we take $ \mathcal{A}=0.63$, $\mu_1 = 1.35$, $\sigma_1 = 0.07$, $\mu_2 = 1.85$, and $\sigma_2 = 0.35$. We truncate the 1- and 2-component mass distributions at the maximum NS mass given by $M_\mathrm{TOV}$ and the NS spin. For the BH distribution, we take $\alpha = 2$, and consider three examples of a lower mass gap for each $M_\text{TOV}$ value: no mass gap ($M_\text{BH} = M_\text{TOV}$); a narrow mass gap where $M_\text{BH} = M_\text{crit}(a/a_\text{Kep} = 1)$ ($2.41\,M_\odot$ for $M_\text{TOV} = 2\,M_\odot$, $2.65\,M_\odot$ for $M_\text{TOV} = 2.2\,M_\odot$); and a wide mass gap with $M_\text{BH} = 5\,M_\odot$. For the pairing function, we take $\beta = 3$. We use the ``ZS + AS" spin model and work with three different values of $\beta_s$ and $a_\mathrm{max}$: a uniform distribution with $\beta_s = 0$ and $a_\text{max}/a_\text{Kep} = 1$ (``uniform" spin) or $a_\text{max}/a_\text{Kep} = 0.5$ (``medium" spin), and $\beta_s = 2$ with $a_\text{max}/a_\text{Kep} = 1$ (``low" spin). 
We simulate observations for LIGO at Design and A+ sensitivity. In total, we consider 3 NS models x 2 $M_\text{TOV}$ values x 3 spin models x 2 detector sensitivities = 36 variations. 

Assuming GW200105 and GW200115 are representative of the NSBH population, NSBH are expected to merge at a rate of $45^{+75}_{-33} \text{Gpc}^{-3} \text{yr}^{-1}$ (90\% credibility) \citep{2021ApJ...915L...5A}, resulting in between 2-20 NSBH/year at Design sensitivity and 8-80 NSBH/year during A+. Assuming a broader component mass distribution produces rate estimates from LVK observations of $130^{+112}_{-69} \text{Gpc}^{-3} \text{yr}^{-1}$, for detection rates of 8-30 NSBH/year at Design sensitivity and 40-160 NSBH/year during A+. Accordingly, we simulate constraints for future datasets of 10, 20, 30, 40, 50, 60, 90, 120, and 150 NSBH detections, and explore how key parameters converge. 

\subsection{Maximum Mass Constraints}

For the 1-component population model and $M_\text{TOV} = 2\,M_\odot$, marginalizing over uncertainty in the underlying spin distribution ($\beta_\text{s}$ and $a_\text{max}/a_\text{Kep}$), 10 NSBH detections allow $M_\text{TOV}$ to be constrained to $2.0^{+0.15}_{-0.08}\,M_\odot$, or $2.2^{+0.19}_{-0.07}\,M_\odot$ for $M_\text{TOV} = 2.2$, with the lower limit on $M_\text{TOV}$ generally much tighter than the upper limit. In our models, 50 NSBH detections allows constraints of $\pm 0.05$, and determining $M_\text{TOV}$ within $\pm 0.02$ is achievable with 150 events. $M_\text{TOV}$ is also slightly better measured for distributions favoring lower spin; the ``medium" and ``low" spin distributions allow constraints down to $\pm 0.03$ for 50 events and $\pm 0.01$ for 150. 
Constraints on $M_\text{TOV}$ generally scale as $N^{-0.5}$; the exact convergence depends on how well the drop-off in events can be resolved given the mass function and $M_\text{TOV}$ value. Compared to constraints from a 1-component population, $M_\text{TOV}$ converges fastest for lower values of $M_\text{TOV}$. Convergence is also fastest for a uniform mass distribution. This is expected, as both of these variations produce the most events close to $M_\text{TOV}$. 

\subsection{Lower Mass Gap}
 We find that $M_\text{TOV}$ and $M_\text{BH}$ can be measured virtually independently, under the optimistic assumption that all BH and NS can be confidently identified (see Section \ref{sec:conclusion}). As a result, all three mass gap widths (wide, $M_\text{BH} = 5 M_\odot$; narrow, $M_\text{BH} = M_\text{crit}(M_\text{TOV}, a/a_\text{Kep} = 1)$; none, $M_\text{BH} = M_\text{TOV}$) can be resolved by modeling a population of spinning NSBH binaries. 
 
For the ``no mass gap" case of $M_\text{BH} = M_\text{TOV} = 2\,M_\odot$, 10 events constrain the mass gap width to $0.0^{+0.07}_{-0.15} M_\odot$. In general, the lower bound on the mass gap width is  more uncertain given the extended tails to high $M_\text{TOV}$ and low $M_\text{BH}$ seen on posteriors (see Figs. \ref{fig:tov_spin}, \ref{fig:mbh}, \ref{fig:fullcorner}). 50 events allow measurements within $0.00 \pm 0.02  M_\odot$, and 150 events can measure the width of the mass gap as precisely as $\pm 0.01  M_\odot$.  
For a wider mass gap, with $M_\text{TOV} = 2\,M_\odot$ and $M_\text{BH} = 5\,M_\odot$, 50 NSBH events can measure the mass gap width to $3.00 \pm 0.08 M_\odot$, and $\pm 0.05  M_\odot$ can be achieved with 150 events. \reply{This is primarily because a wider mass gap is achieved with a larger value of $M_\text{BH}$, which thus has a proportionally higher uncertainty, leading to wider credible intervals for wider mass gaps.} In general, assuming sharp gap edges, the width of the mass gap converges as $N^{-1}$. Factors that lead to sharper constraints on $M_\text{TOV}$ or $M_\text{BH}$, such as a smaller value of $M_\text{TOV}$, a spin distribution favoring low $a_2$, or a steeper BH slope $\alpha$, unsurprisingly also result in faster convergence for the mass gap width. 
Example posteriors (for multiple input parameter variations) on $M_\text{BH}$ and $M_\text{TOV}$, from which the mass gap width is calculated, are shown in Fig.~\ref{fig:simulated_massgap}.

\subsection{Bias from Assuming Neutron Stars Are Non-Spinning}
A handful of events are still expected above the nonspinning maximum NS mass thanks to the effects of rotation support. For a ``uniform" spin distribution, allowing maximally spinning NS, and $M_\text{TOV} = 2\,M_\odot$, around 5\% of our simulated 2-component mass function will have rotation support $M_\text{TOV}$. 6\% of the 1-component mass function, and up to 10\% of the uniform mass function, will show evidence of rotation support above the maximum mass. For $M_\text{TOV} = 2.2\,M_\odot$, this drops to around 2\%, 3\%, and 8\% respectively. For $M_\text{TOV} = 2\,M_\odot$ and the ``low" spin distribution, which strongly disfavors maximally spinning NS, just 1\%, 2\%, and 3\% of the population show this behavior. These events can be seen in Fig. \ref{fig:distributions}, with masses greater than the red line marking $M_\text{TOV}$. If a population contains these events, where the most massive neutron star is measured above the true nonspinning $M_\text{TOV}$, then in order to accurately estimate $M_\text{TOV}$ this rotation support must be properly modeled. If NSs are wrongly assumed to be nonspinning, estimates of $M_\text{TOV}$ will be biased. 

\begin{figure}
     \centering
     \begin{subfigure}[b]{0.45\textwidth}
         \centering
         \includegraphics[width=\textwidth]{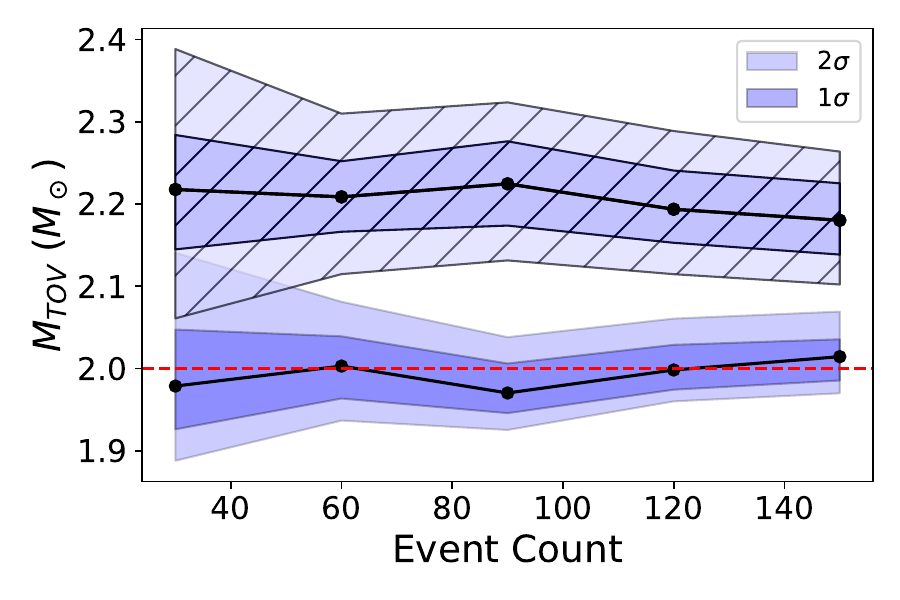}
         \caption{$M_{\text{TOV}} = 2\,M_\odot$, uniform spin, $a_\mathrm{max}/a_{\text{Kep}}$ = 1}
     \end{subfigure}
     \hfill
     \begin{subfigure}[b]{0.45\textwidth}
         \centering
         \includegraphics[width=\textwidth]{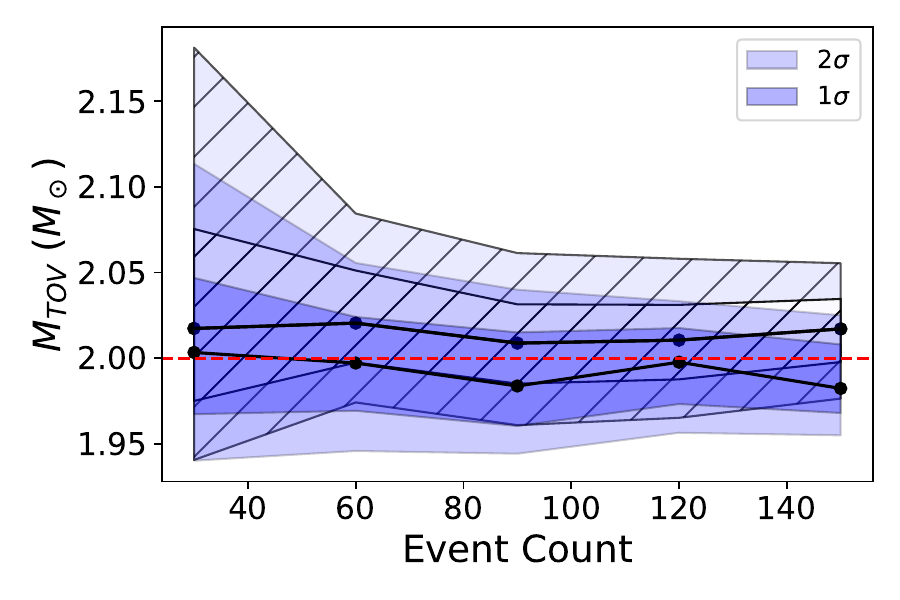}
        \caption{$M_{\text{TOV}} = 2\,M_\odot$, ``low spin" with $p(a_2) \propto (1-a_2)^{2}$}
     \end{subfigure}
     \hfill
    \caption{Inferred $M_\mathrm{TOV}$  when ignoring (hashed pattern) versus properly accounting for a spin-dependent maximum mass. Median, 68\% and 95\% credibility intervals are shown.} 
    \label{fig:bias}
\end{figure}

For an underlying ``uniform" spin distribution, if all NSs are assumed to be nonspinning, it can take as few as 10--20 events to wrongly exclude the true value of $M_\text{TOV}$ with 99.7\% credibility. At 50--150 events, the lower bound of the 99.7\% credibility interval can be as much as 0.2-0.3 $M_\odot$ above $M_\text{TOV}$, with the true value excluded entirely. On the other hand, if spins are relatively low, the bias from neglecting the spin-dependent maximum mass is smaller, but still often present. For the ``low" ($\beta_\text{spin} = 2, a_\text{max}/a_\text{Kep} = 1)$ and ``medium" spin distributions ($\beta_\text{spin} = 0, a_\text{max}/a_\text{Kep} = 0.5)$, which disfavor and disallow large spins, respectively, it usually takes 30--90 events to exclude the correct $M_\text{TOV}$ at 99.7\% credibility. This is partially because even substantial NS spins may have a relatively small effect on $M_\text{crit}$; for a NS with $a_2/a_\text{Kep} = 0.5$, $M_\text{crit}$ is just $1.037 M_\text{TOV}$, a change of less than 4\%. If spins and masses are low enough compared to $M_\text{TOV}$, it is possible to reach ~hundreds of NSBH detections without seeing substantial bias. However, the exact amount of bias depends heavily on the number of massive spinning neutron stars in the observed population, which is unknown. The difference in convergence between spin distributions for a specific realization of events is shown in Fig~\ref{fig:bias}. 

\begin{figure*}
     \centering
     \begin{subfigure}[b]{0.49\textwidth}
         \centering
         \includegraphics[width=\textwidth]{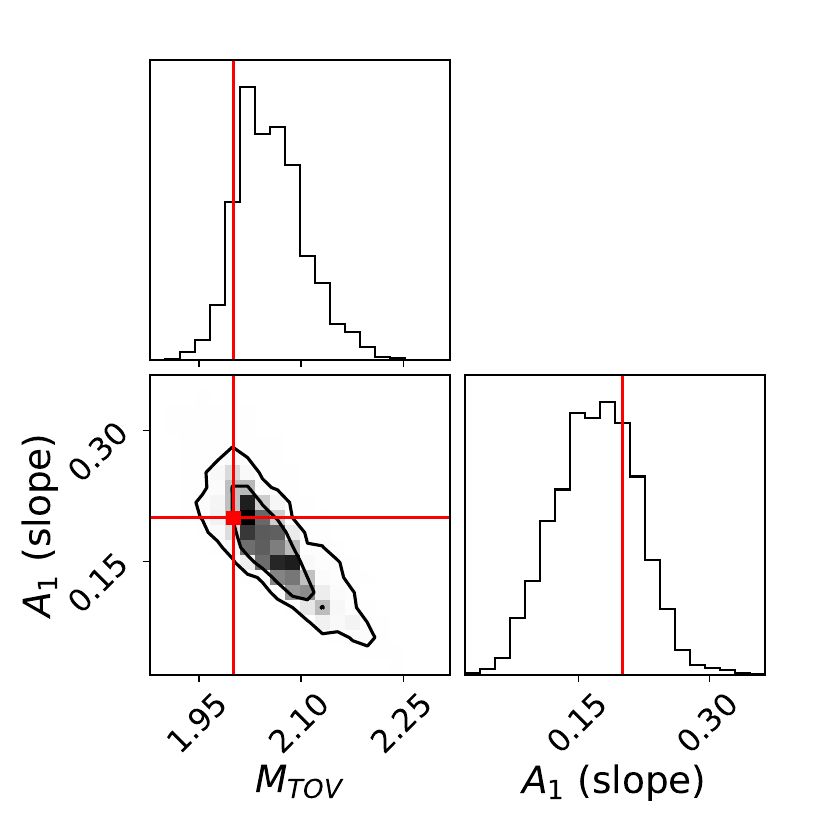}
         \caption{Assume slope $A_1 = 0.2$, uniform NS spin distribution with $a_\mathrm{max}/a_{\text{Kep}} = 1$}
     \end{subfigure}
     \hfill
     \begin{subfigure}[b]{0.49\textwidth}
         \centering
         \includegraphics[width=\textwidth]{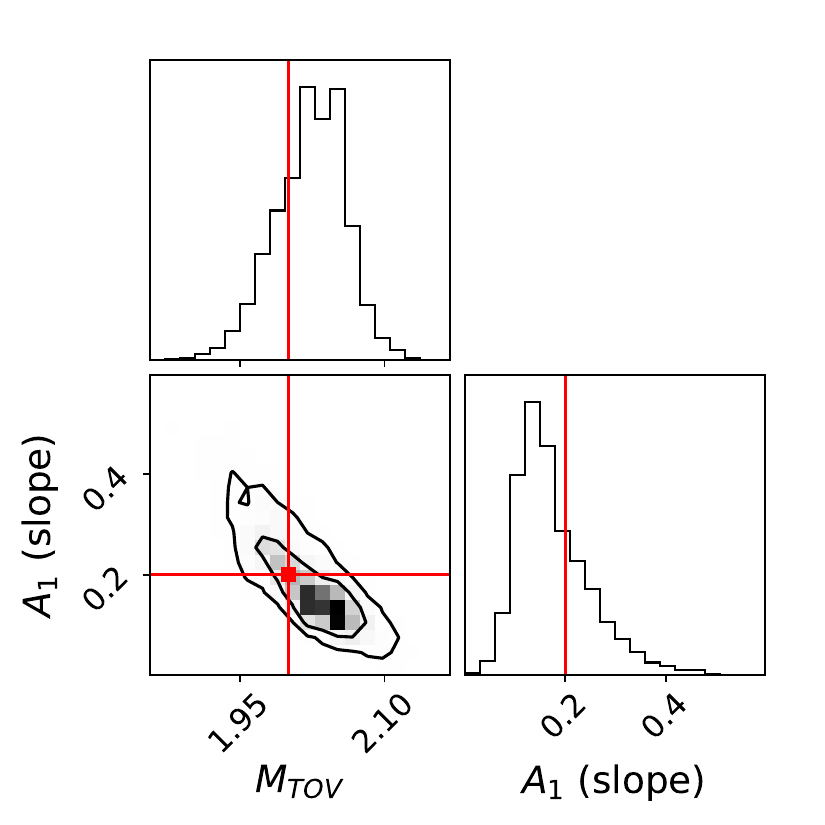}
        \caption{Assumed slope $A_1 = 0.2$, NS spin distribution follows $p(a_2) \propto (1-a_2)^{2}$}
     \end{subfigure}
     \hfill
    \caption{Posteriors on $M_{\text{TOV}}$ and the slope $A_1$ the governs the maximum NS mass as a function of NS spin, inferred from 150 mock events at LIGO A+ sensitivity. Contours enclose 68\% and 95\% of the posterior probability.}
    \label{fig:slope}
\end{figure*}
\subsection{Inferring the Relation Between Maximum NS Mass and Spin} \label{sec:spinmass}
In previous sections, we consider the ``universal relation" between the spin and critical mass as reported by \citet{2020MNRAS.499L..82M}. However, this may only hold for certain families of equations of state. As a result, measuring the relationship between $M_{\text{crit}}$ and $a/a_\text{Kep}$ as a high-degree polynomial may provide insights into the nuclear physics that informs $M_\text{TOV}$ and rotation-supported neutron stars. We consider the simplest case, a linear dependence between spin and maximum mass, with first-order coefficient $A_1$:
\begin{equation}
    M_{\text{crit}}(a_2, a_{\text{Kep}}, M_{\text{TOV}}) = M_{\text{TOV}} A_1(\frac{a_2}{a_{\text{Kep}}})
\end{equation}
and infer $A_1$ jointly with other population parameters.

We consider models with $A_1 = 0.2$ and $0.4$. For a population with a uniform NS spin distribution up to $a_\mathrm{Kep}$ and $A_1 = 0.2$, 10 events can constrain $A_1$ to around $0.2^{+0.2}_{-0.1}$, around $\pm 0.07$ for 50 events, and around $\pm 0.04$ for 150 events, assuming a known spin distribution. Generally, posteriors on $A_1$ are better constrained at low values, as a minimum amount of rotation support above $M_\text{TOV}$ is necessary to explain observations of extra-massive neutron stars. Constraints on $A_1$ converge as $N^{-0.5}$. Given that constraining $A_1$ requires measuring a number of NS with mass greater than $M_\text{TOV}$, populations with ``medium" or ``low" spin distributions constrain $A_1$ much more weakly, as do populations with fewer events close to $M_\text{TOV}$ (i.e. for larger values of $M_\text{TOV}$). For both the ``medium" and ``low" spin distributions, 50 events can constrain $A_1$ to $\pm 0.1$, or $\pm 0.06$ for 150 events. $A_1$ is also covariant with $M_\text{TOV}$, as illustrated in Figure \ref{fig:slope}. A lower value of $M_\text{TOV}$ with a higher $A_1$, and a higher value of $M_\text{TOV}$ with a lower $A_1$, can account for the high masses of rotation-supported neutron stars equally well. 

\section{Conclusion}
\label{sec:conclusion}
We considered the impact of a spin-dependent maximum NS mass on measurements of the mass gap and maximum NS mass from NSBH observations. Our main conclusions are as follows:
\begin{itemize}
     \item The existing NSBH observations prefer a maximum non-spinning NS mass $\sim2.6\,M_\odot$ (including GW190814, the event with the ``mass gap" secondary), or $\sim2.2\,M_\odot$ (excluding GW190814). {\bf Allowing for spin distributions with a broad range of NS spins up to the maximal value $a_
     \mathrm{Kep} \sim 0.7$ allows the inferred $M_\mathrm{TOV}$ to be as low as $\sim 2.3\,M_\odot$, even when including GW190814.} Future GW observations may constrain $M_\text{99}$ and $M_\text{TOV}$ to $\pm 0.02\,M_\odot$ with 150 events by LIGO at A+ sensitivity.
    \item The current NSBH observations support a mass gap between NSs and BHs with width $1.5-4.5 M_\odot$, with typical uncertainties (68\% credibility) of $\pm 1.0$. Exact values depend on event selection, pairing $\beta$, spin prior, and NS mass model; in particular, the mass gap is widened by assuming the BH is non-spinning. \textbf{Regardless of model variation, we infer the presence of a mass gap $> 0 M_\odot$ with high confidence (between $92\%$ and $>99.9\%$), and a mass gap $> 1 M_\odot$ with moderate confidence (between $75\%$ and $>99.9\%$)}. Future observations may constrain this value to $\pm 0.02$ with 150 events by LIGO at A+ sensitivity.
    \item \textbf{If massive, fast-spinning, rotation-supported NS exist, they must be modeled in order to not bias the NS mass function and $M_\text{TOV}$}. If they are common in the astrophysical population, the relationship between spin and maximum mass ($M_\text{crit}$) can be inferred directly from the data. Even without detecting confidently rotation-supported NS, the assumed spin distribution affects the inferred $M_\text{TOV}$ posterior, and spins of individual NS can be constrained simultaneously with the population inference of $M_\text{TOV}$. 
\end{itemize}

In our analysis and projections for the future, we have made several simplifying assumptions. 
In order to focus only on the NSBH section of the compact binary mass distribution, we have assumed that NSBH systems can be confidently distinguished from BBH systems, \reply{and implemented models using definite source classifications for events}. In reality, the classification of events is uncertain, especially without prior knowledge of the mass distribution. Future population analyses should jointly model the entire compact binary mass distribution as in\reply{~\citet{2017MNRAS.465.3254M,2020ApJ...899L...8F,2021arXiv211103498F},
 and \citet{2019MNRAS.488.3810P}, as well as the compact binary spin distribution and neutron star matter effects, while simultaneously inferring source classification}. \reply{In this work, rather than marginalizing over the uncertain source classification, we analyze all events with $m_2 < 3\,M_\odot$ and $m_1 > 3\,M_\odot$ as NSBHs, and illustrate the effect of different assumptions about source identities by repeating the inference with and without GW190814}. Since NSs are expected to follow a different spin distribution from BHs, the population-level spin distributions may provide another clue to distinguish NSs and BHs in merging binaries, in addition to masses and any tidal information~\citep{2020arXiv200101747W,2021arXiv210615745G}. We have also assumed that the astrophysical NS mass distribution cuts off at the maximum possible mass set by nuclear physics. In reality, even if there is a mass gap between NS and BH, the lower edge of the mass gap may be either above or below the non-spinning NS maximum mass $M$. In the future, it would be useful to incorporate external knowledge of the NS EOS, particularly to compare the inferred location of the lower mass gap edge against external $M_\mathrm{TOV}$ constraints. 

\acknowledgments
We thank Phil Landry for helpful comments on the manuscript. This material is based upon work supported by NSF's LIGO Laboratory which is a major facility fully funded by the National Science Foundation. The authors are grateful for computational resources provided by the LIGO Laboratory and supported by National Science Foundation Grants PHY-0757058 and PHY-0823459. M.F. is supported by NASA through NASA Hubble Fellowship grant HST-HF2-51455.001-A awarded by the Space Telescope Science Institute, which is operated by the Association of Universities for
Research in Astronomy, Incorporated, under NASA contract NAS5-26555.
M.F. is grateful for the hospitality of Perimeter Institute where part of this work was carried out.
Research at Perimeter Institute is supported in part by the Government of Canada through the
Department of Innovation, Science and Economic Development Canada and by the Province of
Ontario through the Ministry of Economic Development, Job Creation and Trade.

\bibliographystyle{apsrev}
\bibliography{references}

\begin{thebibliography}{73}
\expandafter\ifx\csname natexlab\endcsname\relax\def\natexlab#1{#1}\fi
\expandafter\ifx\csname bibnamefont\endcsname\relax
  \def\bibnamefont#1{#1}\fi
\expandafter\ifx\csname bibfnamefont\endcsname\relax
  \def\bibfnamefont#1{#1}\fi
\expandafter\ifx\csname citenamefont\endcsname\relax
  \def\citenamefont#1{#1}\fi
\expandafter\ifx\csname url\endcsname\relax
  \def\url#1{\texttt{#1}}\fi
\expandafter\ifx\csname urlprefix\endcsname\relax\def\urlprefix{URL }\fi
\providecommand{\bibinfo}[2]{#2}
\providecommand{\eprint}[2][]{\url{#2}}

\bibitem[{\citenamefont{{Bombaci}}(1996)}]{1996A&A...305..871B}
\bibinfo{author}{\bibfnamefont{I.}~\bibnamefont{{Bombaci}}},
  \bibinfo{journal}{\aap} \textbf{\bibinfo{volume}{305}}, \bibinfo{pages}{871}
  (\bibinfo{year}{1996}).

\bibitem[{\citenamefont{{Kalogera} and {Baym}}(1996)}]{1996ApJ...470L..61K}
\bibinfo{author}{\bibfnamefont{V.}~\bibnamefont{{Kalogera}}} \bibnamefont{and}
  \bibinfo{author}{\bibfnamefont{G.}~\bibnamefont{{Baym}}},
  \bibinfo{journal}{\apjl} \textbf{\bibinfo{volume}{470}}, \bibinfo{pages}{L61}
  (\bibinfo{year}{1996}), \eprint{astro-ph/9608059}.

\bibitem[{\citenamefont{{Lattimer}}(2012)}]{2012ARNPS..62..485L}
\bibinfo{author}{\bibfnamefont{J.~M.} \bibnamefont{{Lattimer}}},
  \bibinfo{journal}{Annual Review of Nuclear and Particle Science}
  \textbf{\bibinfo{volume}{62}}, \bibinfo{pages}{485} (\bibinfo{year}{2012}),
  \eprint{1305.3510}.

\bibitem[{\citenamefont{{Bogdanov} et~al.}(2019)\citenamefont{{Bogdanov},
  {Guillot}, {Ray}, {Wolff}, {Chakrabarty}, {Ho}, {Kerr}, {Lamb}, {Lommen},
  {Ludlam} et~al.}}]{2019ApJ...887L..25B}
\bibinfo{author}{\bibfnamefont{S.}~\bibnamefont{{Bogdanov}}},
  \bibinfo{author}{\bibfnamefont{S.}~\bibnamefont{{Guillot}}},
  \bibinfo{author}{\bibfnamefont{P.~S.} \bibnamefont{{Ray}}},
  \bibinfo{author}{\bibfnamefont{M.~T.} \bibnamefont{{Wolff}}},
  \bibinfo{author}{\bibfnamefont{D.}~\bibnamefont{{Chakrabarty}}},
  \bibinfo{author}{\bibfnamefont{W.~C.~G.} \bibnamefont{{Ho}}},
  \bibinfo{author}{\bibfnamefont{M.}~\bibnamefont{{Kerr}}},
  \bibinfo{author}{\bibfnamefont{F.~K.} \bibnamefont{{Lamb}}},
  \bibinfo{author}{\bibfnamefont{A.}~\bibnamefont{{Lommen}}},
  \bibinfo{author}{\bibfnamefont{R.~M.} \bibnamefont{{Ludlam}}},
  \bibnamefont{et~al.}, \bibinfo{journal}{\apjl}
  \textbf{\bibinfo{volume}{887}}, \bibinfo{eid}{L25} (\bibinfo{year}{2019}),
  \eprint{1912.05706}.

\bibitem[{\citenamefont{{Abbott} et~al.}(2018)\citenamefont{{Abbott}, {Abbott},
  {Abbott}, {Acernese}, {Ackley}, {Adams}, {Adams}, {Addesso}, {Adhikari},
  {Adya} et~al.}}]{2018PhRvL.121p1101A}
\bibinfo{author}{\bibfnamefont{B.~P.} \bibnamefont{{Abbott}}},
  \bibinfo{author}{\bibfnamefont{R.}~\bibnamefont{{Abbott}}},
  \bibinfo{author}{\bibfnamefont{T.~D.} \bibnamefont{{Abbott}}},
  \bibinfo{author}{\bibfnamefont{F.}~\bibnamefont{{Acernese}}},
  \bibinfo{author}{\bibfnamefont{K.}~\bibnamefont{{Ackley}}},
  \bibinfo{author}{\bibfnamefont{C.}~\bibnamefont{{Adams}}},
  \bibinfo{author}{\bibfnamefont{T.}~\bibnamefont{{Adams}}},
  \bibinfo{author}{\bibfnamefont{P.}~\bibnamefont{{Addesso}}},
  \bibinfo{author}{\bibfnamefont{R.~X.} \bibnamefont{{Adhikari}}},
  \bibinfo{author}{\bibfnamefont{V.~B.} \bibnamefont{{Adya}}},
  \bibnamefont{et~al.}, \bibinfo{journal}{\prl} \textbf{\bibinfo{volume}{121}},
  \bibinfo{eid}{161101} (\bibinfo{year}{2018}), \eprint{1805.11581}.

\bibitem[{\citenamefont{{Lim} and {Holt}}(2019)}]{2019EPJA...55..209L}
\bibinfo{author}{\bibfnamefont{Y.}~\bibnamefont{{Lim}}} \bibnamefont{and}
  \bibinfo{author}{\bibfnamefont{J.~W.} \bibnamefont{{Holt}}},
  \bibinfo{journal}{European Physical Journal A} \textbf{\bibinfo{volume}{55}},
  \bibinfo{eid}{209} (\bibinfo{year}{2019}), \eprint{1902.05502}.

\bibitem[{\citenamefont{{Landry} et~al.}(2020)\citenamefont{{Landry}, {Essick},
  and {Chatziioannou}}}]{2020PhRvD.101l3007L}
\bibinfo{author}{\bibfnamefont{P.}~\bibnamefont{{Landry}}},
  \bibinfo{author}{\bibfnamefont{R.}~\bibnamefont{{Essick}}}, \bibnamefont{and}
  \bibinfo{author}{\bibfnamefont{K.}~\bibnamefont{{Chatziioannou}}},
  \bibinfo{journal}{\prd} \textbf{\bibinfo{volume}{101}}, \bibinfo{eid}{123007}
  (\bibinfo{year}{2020}), \eprint{2003.04880}.

\bibitem[{\citenamefont{{Dietrich} et~al.}(2020)\citenamefont{{Dietrich},
  {Coughlin}, {Pang}, {Bulla}, {Heinzel}, {Issa}, {Tews}, and
  {Antier}}}]{2020Sci...370.1450D}
\bibinfo{author}{\bibfnamefont{T.}~\bibnamefont{{Dietrich}}},
  \bibinfo{author}{\bibfnamefont{M.~W.} \bibnamefont{{Coughlin}}},
  \bibinfo{author}{\bibfnamefont{P.~T.~H.} \bibnamefont{{Pang}}},
  \bibinfo{author}{\bibfnamefont{M.}~\bibnamefont{{Bulla}}},
  \bibinfo{author}{\bibfnamefont{J.}~\bibnamefont{{Heinzel}}},
  \bibinfo{author}{\bibfnamefont{L.}~\bibnamefont{{Issa}}},
  \bibinfo{author}{\bibfnamefont{I.}~\bibnamefont{{Tews}}}, \bibnamefont{and}
  \bibinfo{author}{\bibfnamefont{S.}~\bibnamefont{{Antier}}},
  \bibinfo{journal}{Science} \textbf{\bibinfo{volume}{370}},
  \bibinfo{pages}{1450} (\bibinfo{year}{2020}), \eprint{2002.11355}.

\bibitem[{\citenamefont{{Margalit} and {Metzger}}(2017)}]{2017ApJ...850L..19M}
\bibinfo{author}{\bibfnamefont{B.}~\bibnamefont{{Margalit}}} \bibnamefont{and}
  \bibinfo{author}{\bibfnamefont{B.~D.} \bibnamefont{{Metzger}}},
  \bibinfo{journal}{\apjl} \textbf{\bibinfo{volume}{850}}, \bibinfo{eid}{L19}
  (\bibinfo{year}{2017}), \eprint{1710.05938}.

\bibitem[{\citenamefont{{Rezzolla} et~al.}(2018)\citenamefont{{Rezzolla},
  {Most}, and {Weih}}}]{2018ApJ...852L..25R}
\bibinfo{author}{\bibfnamefont{L.}~\bibnamefont{{Rezzolla}}},
  \bibinfo{author}{\bibfnamefont{E.~R.} \bibnamefont{{Most}}},
  \bibnamefont{and} \bibinfo{author}{\bibfnamefont{L.~R.}
  \bibnamefont{{Weih}}}, \bibinfo{journal}{\apjl}
  \textbf{\bibinfo{volume}{852}}, \bibinfo{eid}{L25} (\bibinfo{year}{2018}),
  \eprint{1711.00314}.

\bibitem[{\citenamefont{{Adhikari} et~al.}(2021)\citenamefont{{Adhikari},
  {Albataineh}, {Androic}, {Aniol}, {Armstrong}, {Averett}, {Ayerbe Gayoso},
  {Barcus}, {Bellini}, {Beminiwattha} et~al.}}]{2021PhRvL.126q2502A}
\bibinfo{author}{\bibfnamefont{D.}~\bibnamefont{{Adhikari}}},
  \bibinfo{author}{\bibfnamefont{H.}~\bibnamefont{{Albataineh}}},
  \bibinfo{author}{\bibfnamefont{D.}~\bibnamefont{{Androic}}},
  \bibinfo{author}{\bibfnamefont{K.}~\bibnamefont{{Aniol}}},
  \bibinfo{author}{\bibfnamefont{D.~S.} \bibnamefont{{Armstrong}}},
  \bibinfo{author}{\bibfnamefont{T.}~\bibnamefont{{Averett}}},
  \bibinfo{author}{\bibfnamefont{C.}~\bibnamefont{{Ayerbe Gayoso}}},
  \bibinfo{author}{\bibfnamefont{S.}~\bibnamefont{{Barcus}}},
  \bibinfo{author}{\bibfnamefont{V.}~\bibnamefont{{Bellini}}},
  \bibinfo{author}{\bibfnamefont{R.~S.} \bibnamefont{{Beminiwattha}}},
  \bibnamefont{et~al.}, \bibinfo{journal}{\prl} \textbf{\bibinfo{volume}{126}},
  \bibinfo{eid}{172502} (\bibinfo{year}{2021}), \eprint{2102.10767}.

\bibitem[{\citenamefont{{Legred} et~al.}(2021)\citenamefont{{Legred},
  {Chatziioannou}, {Essick}, {Han}, and {Landry}}}]{2021PhRvD.104f3003L}
\bibinfo{author}{\bibfnamefont{I.}~\bibnamefont{{Legred}}},
  \bibinfo{author}{\bibfnamefont{K.}~\bibnamefont{{Chatziioannou}}},
  \bibinfo{author}{\bibfnamefont{R.}~\bibnamefont{{Essick}}},
  \bibinfo{author}{\bibfnamefont{S.}~\bibnamefont{{Han}}}, \bibnamefont{and}
  \bibinfo{author}{\bibfnamefont{P.}~\bibnamefont{{Landry}}},
  \bibinfo{journal}{\prd} \textbf{\bibinfo{volume}{104}}, \bibinfo{eid}{063003}
  (\bibinfo{year}{2021}), \eprint{2106.05313}.

\bibitem[{\citenamefont{{Valentim} et~al.}(2011)\citenamefont{{Valentim},
  {Rangel}, and {Horvath}}}]{2011MNRAS.414.1427V}
\bibinfo{author}{\bibfnamefont{R.}~\bibnamefont{{Valentim}}},
  \bibinfo{author}{\bibfnamefont{E.}~\bibnamefont{{Rangel}}}, \bibnamefont{and}
  \bibinfo{author}{\bibfnamefont{J.~E.} \bibnamefont{{Horvath}}},
  \bibinfo{journal}{\mnras} \textbf{\bibinfo{volume}{414}},
  \bibinfo{pages}{1427} (\bibinfo{year}{2011}), \eprint{1101.4872}.

\bibitem[{\citenamefont{{{\"O}zel} et~al.}(2012)\citenamefont{{{\"O}zel},
  {Psaltis}, {Narayan}, and {Santos Villarreal}}}]{2012ApJ...757...55O}
\bibinfo{author}{\bibfnamefont{F.}~\bibnamefont{{{\"O}zel}}},
  \bibinfo{author}{\bibfnamefont{D.}~\bibnamefont{{Psaltis}}},
  \bibinfo{author}{\bibfnamefont{R.}~\bibnamefont{{Narayan}}},
  \bibnamefont{and} \bibinfo{author}{\bibfnamefont{A.}~\bibnamefont{{Santos
  Villarreal}}}, \bibinfo{journal}{\apj} \textbf{\bibinfo{volume}{757}},
  \bibinfo{eid}{55} (\bibinfo{year}{2012}), \eprint{1201.1006}.

\bibitem[{\citenamefont{{Alsing} et~al.}(2018)\citenamefont{{Alsing}, {Silva},
  and {Berti}}}]{2018MNRAS.478.1377A}
\bibinfo{author}{\bibfnamefont{J.}~\bibnamefont{{Alsing}}},
  \bibinfo{author}{\bibfnamefont{H.~O.} \bibnamefont{{Silva}}},
  \bibnamefont{and} \bibinfo{author}{\bibfnamefont{E.}~\bibnamefont{{Berti}}},
  \bibinfo{journal}{\mnras} \textbf{\bibinfo{volume}{478}},
  \bibinfo{pages}{1377} (\bibinfo{year}{2018}), \eprint{1709.07889}.

\bibitem[{\citenamefont{{Farrow} et~al.}(2019)\citenamefont{{Farrow}, {Zhu},
  and {Thrane}}}]{2019ApJ...876...18F}
\bibinfo{author}{\bibfnamefont{N.}~\bibnamefont{{Farrow}}},
  \bibinfo{author}{\bibfnamefont{X.-J.} \bibnamefont{{Zhu}}}, \bibnamefont{and}
  \bibinfo{author}{\bibfnamefont{E.}~\bibnamefont{{Thrane}}},
  \bibinfo{journal}{\apj} \textbf{\bibinfo{volume}{876}}, \bibinfo{eid}{18}
  (\bibinfo{year}{2019}), \eprint{1902.03300}.

\bibitem[{\citenamefont{{Farr} and
  {Chatziioannou}}(2020)}]{2020RNAAS...4...65F}
\bibinfo{author}{\bibfnamefont{W.~M.} \bibnamefont{{Farr}}} \bibnamefont{and}
  \bibinfo{author}{\bibfnamefont{K.}~\bibnamefont{{Chatziioannou}}},
  \bibinfo{journal}{Research Notes of the American Astronomical Society}
  \textbf{\bibinfo{volume}{4}}, \bibinfo{eid}{65} (\bibinfo{year}{2020}),
  \eprint{2005.00032}.

\bibitem[{\citenamefont{{Antoniadis} et~al.}(2013)\citenamefont{{Antoniadis},
  {Freire}, {Wex}, {Tauris}, {Lynch}, {van Kerkwijk}, {Kramer}, {Bassa},
  {Dhillon}, {Driebe} et~al.}}]{2013Sci...340..448A}
\bibinfo{author}{\bibfnamefont{J.}~\bibnamefont{{Antoniadis}}},
  \bibinfo{author}{\bibfnamefont{P.~C.~C.} \bibnamefont{{Freire}}},
  \bibinfo{author}{\bibfnamefont{N.}~\bibnamefont{{Wex}}},
  \bibinfo{author}{\bibfnamefont{T.~M.} \bibnamefont{{Tauris}}},
  \bibinfo{author}{\bibfnamefont{R.~S.} \bibnamefont{{Lynch}}},
  \bibinfo{author}{\bibfnamefont{M.~H.} \bibnamefont{{van Kerkwijk}}},
  \bibinfo{author}{\bibfnamefont{M.}~\bibnamefont{{Kramer}}},
  \bibinfo{author}{\bibfnamefont{C.}~\bibnamefont{{Bassa}}},
  \bibinfo{author}{\bibfnamefont{V.~S.} \bibnamefont{{Dhillon}}},
  \bibinfo{author}{\bibfnamefont{T.}~\bibnamefont{{Driebe}}},
  \bibnamefont{et~al.}, \bibinfo{journal}{Science}
  \textbf{\bibinfo{volume}{340}}, \bibinfo{pages}{448} (\bibinfo{year}{2013}),
  \eprint{1304.6875}.

\bibitem[{\citenamefont{{Cromartie} et~al.}(2020)\citenamefont{{Cromartie},
  {Fonseca}, {Ransom}, {Demorest}, {Arzoumanian}, {Blumer}, {Brook}, {DeCesar},
  {Dolch}, {Ellis} et~al.}}]{2020NatAs...4...72C}
\bibinfo{author}{\bibfnamefont{H.~T.} \bibnamefont{{Cromartie}}},
  \bibinfo{author}{\bibfnamefont{E.}~\bibnamefont{{Fonseca}}},
  \bibinfo{author}{\bibfnamefont{S.~M.} \bibnamefont{{Ransom}}},
  \bibinfo{author}{\bibfnamefont{P.~B.} \bibnamefont{{Demorest}}},
  \bibinfo{author}{\bibfnamefont{Z.}~\bibnamefont{{Arzoumanian}}},
  \bibinfo{author}{\bibfnamefont{H.}~\bibnamefont{{Blumer}}},
  \bibinfo{author}{\bibfnamefont{P.~R.} \bibnamefont{{Brook}}},
  \bibinfo{author}{\bibfnamefont{M.~E.} \bibnamefont{{DeCesar}}},
  \bibinfo{author}{\bibfnamefont{T.}~\bibnamefont{{Dolch}}},
  \bibinfo{author}{\bibfnamefont{J.~A.} \bibnamefont{{Ellis}}},
  \bibnamefont{et~al.}, \bibinfo{journal}{Nature Astronomy}
  \textbf{\bibinfo{volume}{4}}, \bibinfo{pages}{72} (\bibinfo{year}{2020}),
  \eprint{1904.06759}.

\bibitem[{\citenamefont{{Fryer} and {Kalogera}}(2001)}]{2001ApJ...554..548F}
\bibinfo{author}{\bibfnamefont{C.~L.} \bibnamefont{{Fryer}}} \bibnamefont{and}
  \bibinfo{author}{\bibfnamefont{V.}~\bibnamefont{{Kalogera}}},
  \bibinfo{journal}{\apj} \textbf{\bibinfo{volume}{554}}, \bibinfo{pages}{548}
  (\bibinfo{year}{2001}), \eprint{astro-ph/9911312}.

\bibitem[{\citenamefont{{Fryer} et~al.}(2012)\citenamefont{{Fryer},
  {Belczynski}, {Wiktorowicz}, {Dominik}, {Kalogera}, and
  {Holz}}}]{2012ApJ...749...91F}
\bibinfo{author}{\bibfnamefont{C.~L.} \bibnamefont{{Fryer}}},
  \bibinfo{author}{\bibfnamefont{K.}~\bibnamefont{{Belczynski}}},
  \bibinfo{author}{\bibfnamefont{G.}~\bibnamefont{{Wiktorowicz}}},
  \bibinfo{author}{\bibfnamefont{M.}~\bibnamefont{{Dominik}}},
  \bibinfo{author}{\bibfnamefont{V.}~\bibnamefont{{Kalogera}}},
  \bibnamefont{and} \bibinfo{author}{\bibfnamefont{D.~E.}
  \bibnamefont{{Holz}}}, \bibinfo{journal}{\apj}
  \textbf{\bibinfo{volume}{749}}, \bibinfo{eid}{91} (\bibinfo{year}{2012}),
  \eprint{1110.1726}.

\bibitem[{\citenamefont{{Belczynski} et~al.}(2012)\citenamefont{{Belczynski},
  {Wiktorowicz}, {Fryer}, {Holz}, and {Kalogera}}}]{2012ApJ...757...91B}
\bibinfo{author}{\bibfnamefont{K.}~\bibnamefont{{Belczynski}}},
  \bibinfo{author}{\bibfnamefont{G.}~\bibnamefont{{Wiktorowicz}}},
  \bibinfo{author}{\bibfnamefont{C.~L.} \bibnamefont{{Fryer}}},
  \bibinfo{author}{\bibfnamefont{D.~E.} \bibnamefont{{Holz}}},
  \bibnamefont{and}
  \bibinfo{author}{\bibfnamefont{V.}~\bibnamefont{{Kalogera}}},
  \bibinfo{journal}{\apj} \textbf{\bibinfo{volume}{757}}, \bibinfo{eid}{91}
  (\bibinfo{year}{2012}), \eprint{1110.1635}.

\bibitem[{\citenamefont{{Liu} et~al.}(2021)\citenamefont{{Liu}, {Wei}, {Xue},
  and {Sun}}}]{2021ApJ...908..106L}
\bibinfo{author}{\bibfnamefont{T.}~\bibnamefont{{Liu}}},
  \bibinfo{author}{\bibfnamefont{Y.-F.} \bibnamefont{{Wei}}},
  \bibinfo{author}{\bibfnamefont{L.}~\bibnamefont{{Xue}}}, \bibnamefont{and}
  \bibinfo{author}{\bibfnamefont{M.-Y.} \bibnamefont{{Sun}}},
  \bibinfo{journal}{\apj} \textbf{\bibinfo{volume}{908}}, \bibinfo{eid}{106}
  (\bibinfo{year}{2021}), \eprint{2011.14361}.

\bibitem[{\citenamefont{{{\"O}zel} et~al.}(2010)\citenamefont{{{\"O}zel},
  {Psaltis}, {Narayan}, and {McClintock}}}]{2010ApJ...725.1918O}
\bibinfo{author}{\bibfnamefont{F.}~\bibnamefont{{{\"O}zel}}},
  \bibinfo{author}{\bibfnamefont{D.}~\bibnamefont{{Psaltis}}},
  \bibinfo{author}{\bibfnamefont{R.}~\bibnamefont{{Narayan}}},
  \bibnamefont{and} \bibinfo{author}{\bibfnamefont{J.~E.}
  \bibnamefont{{McClintock}}}, \bibinfo{journal}{\apj}
  \textbf{\bibinfo{volume}{725}}, \bibinfo{pages}{1918} (\bibinfo{year}{2010}),
  \eprint{1006.2834}.

\bibitem[{\citenamefont{{Farr} et~al.}(2011)\citenamefont{{Farr}, {Sravan},
  {Cantrell}, {Kreidberg}, {Bailyn}, {Mandel}, and
  {Kalogera}}}]{2011ApJ...741..103F}
\bibinfo{author}{\bibfnamefont{W.~M.} \bibnamefont{{Farr}}},
  \bibinfo{author}{\bibfnamefont{N.}~\bibnamefont{{Sravan}}},
  \bibinfo{author}{\bibfnamefont{A.}~\bibnamefont{{Cantrell}}},
  \bibinfo{author}{\bibfnamefont{L.}~\bibnamefont{{Kreidberg}}},
  \bibinfo{author}{\bibfnamefont{C.~D.} \bibnamefont{{Bailyn}}},
  \bibinfo{author}{\bibfnamefont{I.}~\bibnamefont{{Mandel}}}, \bibnamefont{and}
  \bibinfo{author}{\bibfnamefont{V.}~\bibnamefont{{Kalogera}}},
  \bibinfo{journal}{\apj} \textbf{\bibinfo{volume}{741}}, \bibinfo{eid}{103}
  (\bibinfo{year}{2011}), \eprint{1011.1459}.

\bibitem[{\citenamefont{{Thompson} et~al.}(2019)\citenamefont{{Thompson},
  {Kochanek}, {Stanek}, {Badenes}, {Post}, {Jayasinghe}, {Latham}, {Bieryla},
  {Esquerdo}, {Berlind} et~al.}}]{2019Sci...366..637T}
\bibinfo{author}{\bibfnamefont{T.~A.} \bibnamefont{{Thompson}}},
  \bibinfo{author}{\bibfnamefont{C.~S.} \bibnamefont{{Kochanek}}},
  \bibinfo{author}{\bibfnamefont{K.~Z.} \bibnamefont{{Stanek}}},
  \bibinfo{author}{\bibfnamefont{C.}~\bibnamefont{{Badenes}}},
  \bibinfo{author}{\bibfnamefont{R.~S.} \bibnamefont{{Post}}},
  \bibinfo{author}{\bibfnamefont{T.}~\bibnamefont{{Jayasinghe}}},
  \bibinfo{author}{\bibfnamefont{D.~W.} \bibnamefont{{Latham}}},
  \bibinfo{author}{\bibfnamefont{A.}~\bibnamefont{{Bieryla}}},
  \bibinfo{author}{\bibfnamefont{G.~A.} \bibnamefont{{Esquerdo}}},
  \bibinfo{author}{\bibfnamefont{P.}~\bibnamefont{{Berlind}}},
  \bibnamefont{et~al.}, \bibinfo{journal}{Science}
  \textbf{\bibinfo{volume}{366}}, \bibinfo{pages}{637} (\bibinfo{year}{2019}),
  \eprint{1806.02751}.

\bibitem[{\citenamefont{{Abbott}
  et~al.}(2020{\natexlab{a}})\citenamefont{{Abbott}, {Abbott}, {Abraham},
  {Acernese}, {Ackley}, {Adams}, {Adhikari}, {Adya}, {Affeldt}, {Agathos}
  et~al.}}]{2020ApJ...896L..44A}
\bibinfo{author}{\bibfnamefont{R.}~\bibnamefont{{Abbott}}},
  \bibinfo{author}{\bibfnamefont{T.~D.} \bibnamefont{{Abbott}}},
  \bibinfo{author}{\bibfnamefont{S.}~\bibnamefont{{Abraham}}},
  \bibinfo{author}{\bibfnamefont{F.}~\bibnamefont{{Acernese}}},
  \bibinfo{author}{\bibfnamefont{K.}~\bibnamefont{{Ackley}}},
  \bibinfo{author}{\bibfnamefont{C.}~\bibnamefont{{Adams}}},
  \bibinfo{author}{\bibfnamefont{R.~X.} \bibnamefont{{Adhikari}}},
  \bibinfo{author}{\bibfnamefont{V.~B.} \bibnamefont{{Adya}}},
  \bibinfo{author}{\bibfnamefont{C.}~\bibnamefont{{Affeldt}}},
  \bibinfo{author}{\bibfnamefont{M.}~\bibnamefont{{Agathos}}},
  \bibnamefont{et~al.}, \bibinfo{journal}{\apjl}
  \textbf{\bibinfo{volume}{896}}, \bibinfo{eid}{L44}
  (\bibinfo{year}{2020}{\natexlab{a}}), \eprint{2006.12611}.

\bibitem[{\citenamefont{{Aasi} et~al.}(2015)\citenamefont{{Aasi}, {Abbott},
  {Abbott}, {Abbott}, {Abernathy}, {Ackley}, {Adams}, {Adams}, {Addesso}, and
  et~al.}}]{2015CQGra..32g4001L}
\bibinfo{author}{\bibfnamefont{J.}~\bibnamefont{{Aasi}}},
  \bibinfo{author}{\bibfnamefont{B.~P.} \bibnamefont{{Abbott}}},
  \bibinfo{author}{\bibfnamefont{R.}~\bibnamefont{{Abbott}}},
  \bibinfo{author}{\bibfnamefont{T.}~\bibnamefont{{Abbott}}},
  \bibinfo{author}{\bibfnamefont{M.~R.} \bibnamefont{{Abernathy}}},
  \bibinfo{author}{\bibfnamefont{K.}~\bibnamefont{{Ackley}}},
  \bibinfo{author}{\bibfnamefont{C.}~\bibnamefont{{Adams}}},
  \bibinfo{author}{\bibfnamefont{T.}~\bibnamefont{{Adams}}},
  \bibinfo{author}{\bibfnamefont{P.}~\bibnamefont{{Addesso}}},
  \bibnamefont{and} \bibinfo{author}{\bibnamefont{et~al.}},
  \bibinfo{journal}{Classical and Quantum Gravity}
  \textbf{\bibinfo{volume}{32}}, \bibinfo{eid}{074001} (\bibinfo{year}{2015}),
  \eprint{1411.4547}.

\bibitem[{\citenamefont{{Acernese} et~al.}(2015)\citenamefont{{Acernese},
  {Agathos}, {Agatsuma}, {Aisa}, {Allemandou}, {Allocca}, {Amarni}, {Astone},
  {Balestri}, {Ballardin} et~al.}}]{2015CQGra..32b4001A}
\bibinfo{author}{\bibfnamefont{F.}~\bibnamefont{{Acernese}}},
  \bibinfo{author}{\bibfnamefont{M.}~\bibnamefont{{Agathos}}},
  \bibinfo{author}{\bibfnamefont{K.}~\bibnamefont{{Agatsuma}}},
  \bibinfo{author}{\bibfnamefont{D.}~\bibnamefont{{Aisa}}},
  \bibinfo{author}{\bibfnamefont{N.}~\bibnamefont{{Allemandou}}},
  \bibinfo{author}{\bibfnamefont{A.}~\bibnamefont{{Allocca}}},
  \bibinfo{author}{\bibfnamefont{J.}~\bibnamefont{{Amarni}}},
  \bibinfo{author}{\bibfnamefont{P.}~\bibnamefont{{Astone}}},
  \bibinfo{author}{\bibfnamefont{G.}~\bibnamefont{{Balestri}}},
  \bibinfo{author}{\bibfnamefont{G.}~\bibnamefont{{Ballardin}}},
  \bibnamefont{et~al.}, \bibinfo{journal}{Classical and Quantum Gravity}
  \textbf{\bibinfo{volume}{32}}, \bibinfo{eid}{024001} (\bibinfo{year}{2015}),
  \eprint{1408.3978}.

\bibitem[{\citenamefont{{Abbott} et~al.}(2016)\citenamefont{{Abbott}, {Abbott},
  {Abbott}, {Abernathy}, {Acernese}, {Ackley}, {Adams}, {Adams}, {Addesso},
  {Adhikari} et~al.}}]{2016PhRvL.116f1102A}
\bibinfo{author}{\bibfnamefont{B.~P.} \bibnamefont{{Abbott}}},
  \bibinfo{author}{\bibfnamefont{R.}~\bibnamefont{{Abbott}}},
  \bibinfo{author}{\bibfnamefont{T.~D.} \bibnamefont{{Abbott}}},
  \bibinfo{author}{\bibfnamefont{M.~R.} \bibnamefont{{Abernathy}}},
  \bibinfo{author}{\bibfnamefont{F.}~\bibnamefont{{Acernese}}},
  \bibinfo{author}{\bibfnamefont{K.}~\bibnamefont{{Ackley}}},
  \bibinfo{author}{\bibfnamefont{C.}~\bibnamefont{{Adams}}},
  \bibinfo{author}{\bibfnamefont{T.}~\bibnamefont{{Adams}}},
  \bibinfo{author}{\bibfnamefont{P.}~\bibnamefont{{Addesso}}},
  \bibinfo{author}{\bibfnamefont{R.~X.} \bibnamefont{{Adhikari}}},
  \bibnamefont{et~al.}, \bibinfo{journal}{\prl} \textbf{\bibinfo{volume}{116}},
  \bibinfo{eid}{061102} (\bibinfo{year}{2016}), \eprint{1602.03837}.

\bibitem[{\citenamefont{{Abbott} et~al.}(2017)\citenamefont{{Abbott}, {Abbott},
  {Abbott}, {Acernese}, {Ackley}, {Adams}, {Adams}, {Addesso}, {Adhikari},
  {Adya} et~al.}}]{2017PhRvL.119p1101A}
\bibinfo{author}{\bibfnamefont{B.~P.} \bibnamefont{{Abbott}}},
  \bibinfo{author}{\bibfnamefont{R.}~\bibnamefont{{Abbott}}},
  \bibinfo{author}{\bibfnamefont{T.~D.} \bibnamefont{{Abbott}}},
  \bibinfo{author}{\bibfnamefont{F.}~\bibnamefont{{Acernese}}},
  \bibinfo{author}{\bibfnamefont{K.}~\bibnamefont{{Ackley}}},
  \bibinfo{author}{\bibfnamefont{C.}~\bibnamefont{{Adams}}},
  \bibinfo{author}{\bibfnamefont{T.}~\bibnamefont{{Adams}}},
  \bibinfo{author}{\bibfnamefont{P.}~\bibnamefont{{Addesso}}},
  \bibinfo{author}{\bibfnamefont{R.~X.} \bibnamefont{{Adhikari}}},
  \bibinfo{author}{\bibfnamefont{V.~B.} \bibnamefont{{Adya}}},
  \bibnamefont{et~al.}, \bibinfo{journal}{\prl} \textbf{\bibinfo{volume}{119}},
  \bibinfo{eid}{161101} (\bibinfo{year}{2017}), \eprint{1710.05832}.

\bibitem[{\citenamefont{{Abbott}
  et~al.}(2020{\natexlab{b}})\citenamefont{{Abbott}, {Abbott}, {Abbott},
  {Abraham}, {Acernese}, {Ackley}, {Adams}, {Adhikari}, {Adya}, {Affeldt}
  et~al.}}]{2020ApJ...892L...3A}
\bibinfo{author}{\bibfnamefont{B.~P.} \bibnamefont{{Abbott}}},
  \bibinfo{author}{\bibfnamefont{R.}~\bibnamefont{{Abbott}}},
  \bibinfo{author}{\bibfnamefont{T.~D.} \bibnamefont{{Abbott}}},
  \bibinfo{author}{\bibfnamefont{S.}~\bibnamefont{{Abraham}}},
  \bibinfo{author}{\bibfnamefont{F.}~\bibnamefont{{Acernese}}},
  \bibinfo{author}{\bibfnamefont{K.}~\bibnamefont{{Ackley}}},
  \bibinfo{author}{\bibfnamefont{C.}~\bibnamefont{{Adams}}},
  \bibinfo{author}{\bibfnamefont{R.~X.} \bibnamefont{{Adhikari}}},
  \bibinfo{author}{\bibfnamefont{V.~B.} \bibnamefont{{Adya}}},
  \bibinfo{author}{\bibfnamefont{C.}~\bibnamefont{{Affeldt}}},
  \bibnamefont{et~al.}, \bibinfo{journal}{\apjl}
  \textbf{\bibinfo{volume}{892}}, \bibinfo{eid}{L3}
  (\bibinfo{year}{2020}{\natexlab{b}}), \eprint{2001.01761}.

\bibitem[{\citenamefont{{Abbott}
  et~al.}(2021{\natexlab{a}})\citenamefont{{Abbott}, {Abbott}, {Abraham},
  {Acernese}, {Ackley}, {Adams}, {Adams}, {Adhikari}, {Adya}, {Affeldt}
  et~al.}}]{2021ApJ...915L...5A}
\bibinfo{author}{\bibfnamefont{R.}~\bibnamefont{{Abbott}}},
  \bibinfo{author}{\bibfnamefont{T.~D.} \bibnamefont{{Abbott}}},
  \bibinfo{author}{\bibfnamefont{S.}~\bibnamefont{{Abraham}}},
  \bibinfo{author}{\bibfnamefont{F.}~\bibnamefont{{Acernese}}},
  \bibinfo{author}{\bibfnamefont{K.}~\bibnamefont{{Ackley}}},
  \bibinfo{author}{\bibfnamefont{A.}~\bibnamefont{{Adams}}},
  \bibinfo{author}{\bibfnamefont{C.}~\bibnamefont{{Adams}}},
  \bibinfo{author}{\bibfnamefont{R.~X.} \bibnamefont{{Adhikari}}},
  \bibinfo{author}{\bibfnamefont{V.~B.} \bibnamefont{{Adya}}},
  \bibinfo{author}{\bibfnamefont{C.}~\bibnamefont{{Affeldt}}},
  \bibnamefont{et~al.}, \bibinfo{journal}{\apjl}
  \textbf{\bibinfo{volume}{915}}, \bibinfo{eid}{L5}
  (\bibinfo{year}{2021}{\natexlab{a}}), \eprint{2106.15163}.

\bibitem[{\citenamefont{{Chatziioannou} and
  {Farr}}(2020)}]{2020PhRvD.102f4063C}
\bibinfo{author}{\bibfnamefont{K.}~\bibnamefont{{Chatziioannou}}}
  \bibnamefont{and} \bibinfo{author}{\bibfnamefont{W.~M.}
  \bibnamefont{{Farr}}}, \bibinfo{journal}{\prd}
  \textbf{\bibinfo{volume}{102}}, \bibinfo{eid}{064063} (\bibinfo{year}{2020}),
  \eprint{2005.00482}.

\bibitem[{\citenamefont{{Galaudage} et~al.}(2021)\citenamefont{{Galaudage},
  {Adamcewicz}, {Zhu}, {Stevenson}, and {Thrane}}}]{2021ApJ...909L..19G}
\bibinfo{author}{\bibfnamefont{S.}~\bibnamefont{{Galaudage}}},
  \bibinfo{author}{\bibfnamefont{C.}~\bibnamefont{{Adamcewicz}}},
  \bibinfo{author}{\bibfnamefont{X.-J.} \bibnamefont{{Zhu}}},
  \bibinfo{author}{\bibfnamefont{S.}~\bibnamefont{{Stevenson}}},
  \bibnamefont{and} \bibinfo{author}{\bibfnamefont{E.}~\bibnamefont{{Thrane}}},
  \bibinfo{journal}{\apjl} \textbf{\bibinfo{volume}{909}}, \bibinfo{eid}{L19}
  (\bibinfo{year}{2021}), \eprint{2011.01495}.

\bibitem[{\citenamefont{{Landry} and {Read}}(2021)}]{2021arXiv210704559L}
\bibinfo{author}{\bibfnamefont{P.}~\bibnamefont{{Landry}}} \bibnamefont{and}
  \bibinfo{author}{\bibfnamefont{J.~S.} \bibnamefont{{Read}}},
  \bibinfo{journal}{arXiv e-prints} \bibinfo{eid}{arXiv:2107.04559}
  (\bibinfo{year}{2021}), \eprint{2107.04559}.

\bibitem[{\citenamefont{{Li} et~al.}(2021)\citenamefont{{Li}, {Tang}, {Wang},
  {Yuan}, {Fan}, and {Wei}}}]{2021arXiv210806986L}
\bibinfo{author}{\bibfnamefont{Y.-J.} \bibnamefont{{Li}}},
  \bibinfo{author}{\bibfnamefont{S.-P.} \bibnamefont{{Tang}}},
  \bibinfo{author}{\bibfnamefont{Y.-Z.} \bibnamefont{{Wang}}},
  \bibinfo{author}{\bibfnamefont{Q.}~\bibnamefont{{Yuan}}},
  \bibinfo{author}{\bibfnamefont{Y.-Z.} \bibnamefont{{Fan}}}, \bibnamefont{and}
  \bibinfo{author}{\bibfnamefont{D.-M.} \bibnamefont{{Wei}}},
  \bibinfo{journal}{arXiv e-prints} \bibinfo{eid}{arXiv:2108.06986}
  (\bibinfo{year}{2021}), \eprint{2108.06986}.

\bibitem[{\citenamefont{{Zhu} et~al.}(2021)\citenamefont{{Zhu}, {Wu}, {Qin},
  {Zhang}, {Gao}, and {Cao}}}]{2021arXiv211202605Z}
\bibinfo{author}{\bibfnamefont{J.-P.} \bibnamefont{{Zhu}}},
  \bibinfo{author}{\bibfnamefont{S.}~\bibnamefont{{Wu}}},
  \bibinfo{author}{\bibfnamefont{Y.}~\bibnamefont{{Qin}}},
  \bibinfo{author}{\bibfnamefont{B.}~\bibnamefont{{Zhang}}},
  \bibinfo{author}{\bibfnamefont{H.}~\bibnamefont{{Gao}}}, \bibnamefont{and}
  \bibinfo{author}{\bibfnamefont{Z.}~\bibnamefont{{Cao}}},
  \bibinfo{journal}{arXiv e-prints} \bibinfo{eid}{arXiv:2112.02605}
  (\bibinfo{year}{2021}), \eprint{2112.02605}.

\bibitem[{\citenamefont{{The LIGO Scientific Collaboration}
  et~al.}(2021{\natexlab{a}})\citenamefont{{The LIGO Scientific Collaboration},
  {the Virgo Collaboration}, {the KAGRA Collaboration}, {Abbott}, {Abbott},
  {Acernese}, {Ackley}, {Adams}, {Adhikari}, {Adhikari}
  et~al.}}]{2021arXiv211103634T}
\bibinfo{author}{\bibnamefont{{The LIGO Scientific Collaboration}}},
  \bibinfo{author}{\bibnamefont{{the Virgo Collaboration}}},
  \bibinfo{author}{\bibnamefont{{the KAGRA Collaboration}}},
  \bibinfo{author}{\bibfnamefont{R.}~\bibnamefont{{Abbott}}},
  \bibinfo{author}{\bibfnamefont{T.~D.} \bibnamefont{{Abbott}}},
  \bibinfo{author}{\bibfnamefont{F.}~\bibnamefont{{Acernese}}},
  \bibinfo{author}{\bibfnamefont{K.}~\bibnamefont{{Ackley}}},
  \bibinfo{author}{\bibfnamefont{C.}~\bibnamefont{{Adams}}},
  \bibinfo{author}{\bibfnamefont{N.}~\bibnamefont{{Adhikari}}},
  \bibinfo{author}{\bibfnamefont{R.~X.} \bibnamefont{{Adhikari}}},
  \bibnamefont{et~al.}, \bibinfo{journal}{arXiv e-prints}
  \bibinfo{eid}{arXiv:2111.03634} (\bibinfo{year}{2021}{\natexlab{a}}),
  \eprint{2111.03634}.

\bibitem[{\citenamefont{{Mandel} et~al.}(2017)\citenamefont{{Mandel}, {Farr},
  {Colonna}, {Stevenson}, {Ti{\v{n}}o}, and {Veitch}}}]{2017MNRAS.465.3254M}
\bibinfo{author}{\bibfnamefont{I.}~\bibnamefont{{Mandel}}},
  \bibinfo{author}{\bibfnamefont{W.~M.} \bibnamefont{{Farr}}},
  \bibinfo{author}{\bibfnamefont{A.}~\bibnamefont{{Colonna}}},
  \bibinfo{author}{\bibfnamefont{S.}~\bibnamefont{{Stevenson}}},
  \bibinfo{author}{\bibfnamefont{P.}~\bibnamefont{{Ti{\v{n}}o}}},
  \bibnamefont{and} \bibinfo{author}{\bibfnamefont{J.}~\bibnamefont{{Veitch}}},
  \bibinfo{journal}{\mnras} \textbf{\bibinfo{volume}{465}},
  \bibinfo{pages}{3254} (\bibinfo{year}{2017}), \eprint{1608.08223}.

\bibitem[{\citenamefont{{Fishbach} et~al.}(2020)\citenamefont{{Fishbach},
  {Essick}, and {Holz}}}]{2020ApJ...899L...8F}
\bibinfo{author}{\bibfnamefont{M.}~\bibnamefont{{Fishbach}}},
  \bibinfo{author}{\bibfnamefont{R.}~\bibnamefont{{Essick}}}, \bibnamefont{and}
  \bibinfo{author}{\bibfnamefont{D.~E.} \bibnamefont{{Holz}}},
  \bibinfo{journal}{\apjl} \textbf{\bibinfo{volume}{899}}, \bibinfo{eid}{L8}
  (\bibinfo{year}{2020}), \eprint{2006.13178}.

\bibitem[{\citenamefont{{Farah} et~al.}(2021)\citenamefont{{Farah}, {Fishbach},
  {Essick}, {Holz}, and {Galaudage}}}]{2021arXiv211103498F}
\bibinfo{author}{\bibfnamefont{A.~M.} \bibnamefont{{Farah}}},
  \bibinfo{author}{\bibfnamefont{M.}~\bibnamefont{{Fishbach}}},
  \bibinfo{author}{\bibfnamefont{R.}~\bibnamefont{{Essick}}},
  \bibinfo{author}{\bibfnamefont{D.~E.} \bibnamefont{{Holz}}},
  \bibnamefont{and}
  \bibinfo{author}{\bibfnamefont{S.}~\bibnamefont{{Galaudage}}},
  \bibinfo{journal}{arXiv e-prints} \bibinfo{eid}{arXiv:2111.03498}
  (\bibinfo{year}{2021}), \eprint{2111.03498}.

\bibitem[{\citenamefont{{Abbott}
  et~al.}(2021{\natexlab{b}})\citenamefont{{Abbott}, {Abbott}, {Abraham},
  {Acernese}, {Ackley}, {Adams}, {Adams}, {Adhikari}, {Adya}, {Affeldt}
  et~al.}}]{2021ApJ...913L...7A}
\bibinfo{author}{\bibfnamefont{R.}~\bibnamefont{{Abbott}}},
  \bibinfo{author}{\bibfnamefont{T.~D.} \bibnamefont{{Abbott}}},
  \bibinfo{author}{\bibfnamefont{S.}~\bibnamefont{{Abraham}}},
  \bibinfo{author}{\bibfnamefont{F.}~\bibnamefont{{Acernese}}},
  \bibinfo{author}{\bibfnamefont{K.}~\bibnamefont{{Ackley}}},
  \bibinfo{author}{\bibfnamefont{A.}~\bibnamefont{{Adams}}},
  \bibinfo{author}{\bibfnamefont{C.}~\bibnamefont{{Adams}}},
  \bibinfo{author}{\bibfnamefont{R.~X.} \bibnamefont{{Adhikari}}},
  \bibinfo{author}{\bibfnamefont{V.~B.} \bibnamefont{{Adya}}},
  \bibinfo{author}{\bibfnamefont{C.}~\bibnamefont{{Affeldt}}},
  \bibnamefont{et~al.}, \bibinfo{journal}{\apjl}
  \textbf{\bibinfo{volume}{913}}, \bibinfo{eid}{L7}
  (\bibinfo{year}{2021}{\natexlab{b}}), \eprint{2010.14533}.

\bibitem[{\citenamefont{{Mandel} and {Smith}}(2021)}]{2021arXiv210914759M}
\bibinfo{author}{\bibfnamefont{I.}~\bibnamefont{{Mandel}}} \bibnamefont{and}
  \bibinfo{author}{\bibfnamefont{R.~J.~E.} \bibnamefont{{Smith}}},
  \bibinfo{journal}{arXiv e-prints} \bibinfo{eid}{arXiv:2109.14759}
  (\bibinfo{year}{2021}), \eprint{2109.14759}.

\bibitem[{\citenamefont{{Essick} et~al.}(2021)\citenamefont{{Essick}, {Farah},
  {Galaudage}, {Talbot}, {Fishbach}, {Thrane}, and
  {Holz}}}]{2021arXiv210900418E}
\bibinfo{author}{\bibfnamefont{R.}~\bibnamefont{{Essick}}},
  \bibinfo{author}{\bibfnamefont{A.}~\bibnamefont{{Farah}}},
  \bibinfo{author}{\bibfnamefont{S.}~\bibnamefont{{Galaudage}}},
  \bibinfo{author}{\bibfnamefont{C.}~\bibnamefont{{Talbot}}},
  \bibinfo{author}{\bibfnamefont{M.}~\bibnamefont{{Fishbach}}},
  \bibinfo{author}{\bibfnamefont{E.}~\bibnamefont{{Thrane}}}, \bibnamefont{and}
  \bibinfo{author}{\bibfnamefont{D.~E.} \bibnamefont{{Holz}}},
  \bibinfo{journal}{arXiv e-prints} \bibinfo{eid}{arXiv:2109.00418}
  (\bibinfo{year}{2021}), \eprint{2109.00418}.

\bibitem[{\citenamefont{{Essick} and {Landry}}(2020)}]{2020ApJ...904...80E}
\bibinfo{author}{\bibfnamefont{R.}~\bibnamefont{{Essick}}} \bibnamefont{and}
  \bibinfo{author}{\bibfnamefont{P.}~\bibnamefont{{Landry}}},
  \bibinfo{journal}{\apj} \textbf{\bibinfo{volume}{904}}, \bibinfo{eid}{80}
  (\bibinfo{year}{2020}), \eprint{2007.01372}.

\bibitem[{\citenamefont{{Most} et~al.}(2020)\citenamefont{{Most}, {Papenfort},
  {Weih}, and {Rezzolla}}}]{2020MNRAS.499L..82M}
\bibinfo{author}{\bibfnamefont{E.~R.} \bibnamefont{{Most}}},
  \bibinfo{author}{\bibfnamefont{L.~J.} \bibnamefont{{Papenfort}}},
  \bibinfo{author}{\bibfnamefont{L.~R.} \bibnamefont{{Weih}}},
  \bibnamefont{and}
  \bibinfo{author}{\bibfnamefont{L.}~\bibnamefont{{Rezzolla}}},
  \bibinfo{journal}{\mnras} \textbf{\bibinfo{volume}{499}},
  \bibinfo{pages}{L82} (\bibinfo{year}{2020}), \eprint{2006.14601}.

\bibitem[{\citenamefont{{Cook} et~al.}(1994)\citenamefont{{Cook}, {Shapiro},
  and {Teukolsky}}}]{1994ApJ...424..823C}
\bibinfo{author}{\bibfnamefont{G.~B.} \bibnamefont{{Cook}}},
  \bibinfo{author}{\bibfnamefont{S.~L.} \bibnamefont{{Shapiro}}},
  \bibnamefont{and} \bibinfo{author}{\bibfnamefont{S.~A.}
  \bibnamefont{{Teukolsky}}}, \bibinfo{journal}{\apj}
  \textbf{\bibinfo{volume}{424}}, \bibinfo{pages}{823} (\bibinfo{year}{1994}).

\bibitem[{\citenamefont{{Biscoveanu} et~al.}(2022)\citenamefont{{Biscoveanu},
  {Talbot}, and {Vitale}}}]{2022MNRAS.511.4350B}
\bibinfo{author}{\bibfnamefont{S.}~\bibnamefont{{Biscoveanu}}},
  \bibinfo{author}{\bibfnamefont{C.}~\bibnamefont{{Talbot}}}, \bibnamefont{and}
  \bibinfo{author}{\bibfnamefont{S.}~\bibnamefont{{Vitale}}},
  \bibinfo{journal}{\mnras} \textbf{\bibinfo{volume}{511}},
  \bibinfo{pages}{4350} (\bibinfo{year}{2022}), \eprint{2111.13619}.

\bibitem[{\citenamefont{{Lynch} et~al.}(2012)\citenamefont{{Lynch}, {Freire},
  {Ransom}, and {Jacoby}}}]{2012ApJ...745..109L}
\bibinfo{author}{\bibfnamefont{R.~S.} \bibnamefont{{Lynch}}},
  \bibinfo{author}{\bibfnamefont{P.~C.~C.} \bibnamefont{{Freire}}},
  \bibinfo{author}{\bibfnamefont{S.~M.} \bibnamefont{{Ransom}}},
  \bibnamefont{and} \bibinfo{author}{\bibfnamefont{B.~A.}
  \bibnamefont{{Jacoby}}}, \bibinfo{journal}{\apj}
  \textbf{\bibinfo{volume}{745}}, \bibinfo{eid}{109} (\bibinfo{year}{2012}),
  \eprint{1112.2612}.

\bibitem[{\citenamefont{{Hessels} et~al.}(2006)\citenamefont{{Hessels},
  {Ransom}, {Stairs}, {Freire}, {Kaspi}, and {Camilo}}}]{2006Sci...311.1901H}
\bibinfo{author}{\bibfnamefont{J.~W.~T.} \bibnamefont{{Hessels}}},
  \bibinfo{author}{\bibfnamefont{S.~M.} \bibnamefont{{Ransom}}},
  \bibinfo{author}{\bibfnamefont{I.~H.} \bibnamefont{{Stairs}}},
  \bibinfo{author}{\bibfnamefont{P.~C.~C.} \bibnamefont{{Freire}}},
  \bibinfo{author}{\bibfnamefont{V.~M.} \bibnamefont{{Kaspi}}},
  \bibnamefont{and} \bibinfo{author}{\bibfnamefont{F.}~\bibnamefont{{Camilo}}},
  \bibinfo{journal}{Science} \textbf{\bibinfo{volume}{311}},
  \bibinfo{pages}{1901} (\bibinfo{year}{2006}), \eprint{astro-ph/0601337}.

\bibitem[{\citenamefont{{Chattopadhyay}
  et~al.}(2021)\citenamefont{{Chattopadhyay}, {Stevenson}, {Hurley}, {Bailes},
  and {Broekgaarden}}}]{2021MNRAS.504.3682C}
\bibinfo{author}{\bibfnamefont{D.}~\bibnamefont{{Chattopadhyay}}},
  \bibinfo{author}{\bibfnamefont{S.}~\bibnamefont{{Stevenson}}},
  \bibinfo{author}{\bibfnamefont{J.~R.} \bibnamefont{{Hurley}}},
  \bibinfo{author}{\bibfnamefont{M.}~\bibnamefont{{Bailes}}}, \bibnamefont{and}
  \bibinfo{author}{\bibfnamefont{F.}~\bibnamefont{{Broekgaarden}}},
  \bibinfo{journal}{\mnras} \textbf{\bibinfo{volume}{504}},
  \bibinfo{pages}{3682} (\bibinfo{year}{2021}), \eprint{2011.13503}.

\bibitem[{\citenamefont{{Qin} et~al.}(2018)\citenamefont{{Qin}, {Fragos},
  {Meynet}, {Andrews}, {S{\o}rensen}, and {Song}}}]{2018A&A...616A..28Q}
\bibinfo{author}{\bibfnamefont{Y.}~\bibnamefont{{Qin}}},
  \bibinfo{author}{\bibfnamefont{T.}~\bibnamefont{{Fragos}}},
  \bibinfo{author}{\bibfnamefont{G.}~\bibnamefont{{Meynet}}},
  \bibinfo{author}{\bibfnamefont{J.}~\bibnamefont{{Andrews}}},
  \bibinfo{author}{\bibfnamefont{M.}~\bibnamefont{{S{\o}rensen}}},
  \bibnamefont{and} \bibinfo{author}{\bibfnamefont{H.~F.}
  \bibnamefont{{Song}}}, \bibinfo{journal}{\aap}
  \textbf{\bibinfo{volume}{616}}, \bibinfo{eid}{A28} (\bibinfo{year}{2018}),
  \eprint{1802.05738}.

\bibitem[{\citenamefont{{Fuller} and {Ma}}(2019)}]{2019ApJ...881L...1F}
\bibinfo{author}{\bibfnamefont{J.}~\bibnamefont{{Fuller}}} \bibnamefont{and}
  \bibinfo{author}{\bibfnamefont{L.}~\bibnamefont{{Ma}}},
  \bibinfo{journal}{\apjl} \textbf{\bibinfo{volume}{881}}, \bibinfo{eid}{L1}
  (\bibinfo{year}{2019}), \eprint{1907.03714}.

\bibitem[{\citenamefont{{Fishbach} and {Holz}}(2020)}]{2020ApJ...891L..27F}
\bibinfo{author}{\bibfnamefont{M.}~\bibnamefont{{Fishbach}}} \bibnamefont{and}
  \bibinfo{author}{\bibfnamefont{D.~E.} \bibnamefont{{Holz}}},
  \bibinfo{journal}{\apjl} \textbf{\bibinfo{volume}{891}}, \bibinfo{eid}{L27}
  (\bibinfo{year}{2020}), \eprint{1905.12669}.

\bibitem[{\citenamefont{{Loredo}}(2004)}]{2004AIPC..735..195L}
\bibinfo{author}{\bibfnamefont{T.~J.} \bibnamefont{{Loredo}}}, in
  \emph{\bibinfo{booktitle}{Bayesian Inference and Maximum Entropy Methods in
  Science and Engineering: 24th International Workshop on Bayesian Inference
  and Maximum Entropy Methods in Science and Engineering}}, edited by
  \bibinfo{editor}{\bibfnamefont{R.}~\bibnamefont{{Fischer}}},
  \bibinfo{editor}{\bibfnamefont{R.}~\bibnamefont{{Preuss}}}, \bibnamefont{and}
  \bibinfo{editor}{\bibfnamefont{U.~V.} \bibnamefont{{Toussaint}}}
  (\bibinfo{year}{2004}), vol. \bibinfo{volume}{735} of
  \emph{\bibinfo{series}{American Institute of Physics Conference Series}}, pp.
  \bibinfo{pages}{195--206}, \eprint{astro-ph/0409387}.

\bibitem[{\citenamefont{{Mandel} et~al.}(2019)\citenamefont{{Mandel}, {Farr},
  and {Gair}}}]{2019MNRAS.486.1086M}
\bibinfo{author}{\bibfnamefont{I.}~\bibnamefont{{Mandel}}},
  \bibinfo{author}{\bibfnamefont{W.~M.} \bibnamefont{{Farr}}},
  \bibnamefont{and} \bibinfo{author}{\bibfnamefont{J.~R.}
  \bibnamefont{{Gair}}}, \bibinfo{journal}{\mnras}
  \textbf{\bibinfo{volume}{486}}, \bibinfo{pages}{1086} (\bibinfo{year}{2019}),
  \eprint{1809.02063}.

\bibitem[{\citenamefont{{Fishbach} et~al.}(2018)\citenamefont{{Fishbach},
  {Holz}, and {Farr}}}]{2018ApJ...863L..41F}
\bibinfo{author}{\bibfnamefont{M.}~\bibnamefont{{Fishbach}}},
  \bibinfo{author}{\bibfnamefont{D.~E.} \bibnamefont{{Holz}}},
  \bibnamefont{and} \bibinfo{author}{\bibfnamefont{W.~M.}
  \bibnamefont{{Farr}}}, \bibinfo{journal}{\apjl}
  \textbf{\bibinfo{volume}{863}}, \bibinfo{eid}{L41} (\bibinfo{year}{2018}),
  \eprint{1805.10270}.

\bibitem[{\citenamefont{{Fishbach} and {Holz}}(2017)}]{2017ApJ...851L..25F}
\bibinfo{author}{\bibfnamefont{M.}~\bibnamefont{{Fishbach}}} \bibnamefont{and}
  \bibinfo{author}{\bibfnamefont{D.~E.} \bibnamefont{{Holz}}},
  \bibinfo{journal}{\apjl} \textbf{\bibinfo{volume}{851}}, \bibinfo{eid}{L25}
  (\bibinfo{year}{2017}), \eprint{1709.08584}.

\bibitem[{\citenamefont{{Ng} et~al.}(2018)\citenamefont{{Ng}, {Vitale},
  {Zimmerman}, {Chatziioannou}, {Gerosa}, and {Haster}}}]{2018PhRvD..98h3007N}
\bibinfo{author}{\bibfnamefont{K.~K.~Y.} \bibnamefont{{Ng}}},
  \bibinfo{author}{\bibfnamefont{S.}~\bibnamefont{{Vitale}}},
  \bibinfo{author}{\bibfnamefont{A.}~\bibnamefont{{Zimmerman}}},
  \bibinfo{author}{\bibfnamefont{K.}~\bibnamefont{{Chatziioannou}}},
  \bibinfo{author}{\bibfnamefont{D.}~\bibnamefont{{Gerosa}}}, \bibnamefont{and}
  \bibinfo{author}{\bibfnamefont{C.-J.} \bibnamefont{{Haster}}},
  \bibinfo{journal}{\prd} \textbf{\bibinfo{volume}{98}}, \bibinfo{eid}{083007}
  (\bibinfo{year}{2018}), \eprint{1805.03046}.

\bibitem[{\citenamefont{{Abbott}
  et~al.}(2020{\natexlab{c}})\citenamefont{{Abbott}, {Abbott}, {Abbott},
  {Abraham}, {Acernese}, {Ackley}, {Adams}, {Adya}, {Affeldt}, {Agathos}
  et~al.}}]{2020LRR....23....3A}
\bibinfo{author}{\bibfnamefont{B.~P.} \bibnamefont{{Abbott}}},
  \bibinfo{author}{\bibfnamefont{R.}~\bibnamefont{{Abbott}}},
  \bibinfo{author}{\bibfnamefont{T.~D.} \bibnamefont{{Abbott}}},
  \bibinfo{author}{\bibfnamefont{S.}~\bibnamefont{{Abraham}}},
  \bibinfo{author}{\bibfnamefont{F.}~\bibnamefont{{Acernese}}},
  \bibinfo{author}{\bibfnamefont{K.}~\bibnamefont{{Ackley}}},
  \bibinfo{author}{\bibfnamefont{C.}~\bibnamefont{{Adams}}},
  \bibinfo{author}{\bibfnamefont{V.~B.} \bibnamefont{{Adya}}},
  \bibinfo{author}{\bibfnamefont{C.}~\bibnamefont{{Affeldt}}},
  \bibinfo{author}{\bibfnamefont{M.}~\bibnamefont{{Agathos}}},
  \bibnamefont{et~al.}, \bibinfo{journal}{Living Reviews in Relativity}
  \textbf{\bibinfo{volume}{23}}, \bibinfo{eid}{3}
  (\bibinfo{year}{2020}{\natexlab{c}}).

\bibitem[{\citenamefont{{Finn} and {Chernoff}}(1993)}]{1993PhRvD..47.2198F}
\bibinfo{author}{\bibfnamefont{L.~S.} \bibnamefont{{Finn}}} \bibnamefont{and}
  \bibinfo{author}{\bibfnamefont{D.~F.} \bibnamefont{{Chernoff}}},
  \bibinfo{journal}{\prd} \textbf{\bibinfo{volume}{47}}, \bibinfo{pages}{2198}
  (\bibinfo{year}{1993}), \eprint{gr-qc/9301003}.

\bibitem[{\citenamefont{{Chen} and {Holz}}(2014)}]{2014arXiv1409.0522C}
\bibinfo{author}{\bibfnamefont{H.-Y.} \bibnamefont{{Chen}}} \bibnamefont{and}
  \bibinfo{author}{\bibfnamefont{D.~E.} \bibnamefont{{Holz}}},
  \bibinfo{journal}{arXiv e-prints} \bibinfo{eid}{arXiv:1409.0522}
  (\bibinfo{year}{2014}), \eprint{1409.0522}.

\bibitem[{\citenamefont{{Abbott}
  et~al.}(2021{\natexlab{c}})\citenamefont{{Abbott}, {Abbott}, {Abraham},
  {Acernese}, {Ackley}, {Adams}, {Adams}, {Adhikari}, {Adya}, {Affeldt}
  et~al.}}]{2021PhRvX..11b1053A}
\bibinfo{author}{\bibfnamefont{R.}~\bibnamefont{{Abbott}}},
  \bibinfo{author}{\bibfnamefont{T.~D.} \bibnamefont{{Abbott}}},
  \bibinfo{author}{\bibfnamefont{S.}~\bibnamefont{{Abraham}}},
  \bibinfo{author}{\bibfnamefont{F.}~\bibnamefont{{Acernese}}},
  \bibinfo{author}{\bibfnamefont{K.}~\bibnamefont{{Ackley}}},
  \bibinfo{author}{\bibfnamefont{A.}~\bibnamefont{{Adams}}},
  \bibinfo{author}{\bibfnamefont{C.}~\bibnamefont{{Adams}}},
  \bibinfo{author}{\bibfnamefont{R.~X.} \bibnamefont{{Adhikari}}},
  \bibinfo{author}{\bibfnamefont{V.~B.} \bibnamefont{{Adya}}},
  \bibinfo{author}{\bibfnamefont{C.}~\bibnamefont{{Affeldt}}},
  \bibnamefont{et~al.}, \bibinfo{journal}{Physical Review X}
  \textbf{\bibinfo{volume}{11}}, \bibinfo{eid}{021053}
  (\bibinfo{year}{2021}{\natexlab{c}}), \eprint{2010.14527}.

\bibitem[{Note1()}]{Note1}
Note1, \bibinfo{note}{the parameter estimation samples are available on the
  Gravitational Wave Open Science Center~\protect \citep {GWOSC}}.

\bibitem[{\citenamefont{{Mandel} and {Fragos}}(2020)}]{2020ApJ...895L..28M}
\bibinfo{author}{\bibfnamefont{I.}~\bibnamefont{{Mandel}}} \bibnamefont{and}
  \bibinfo{author}{\bibfnamefont{T.}~\bibnamefont{{Fragos}}},
  \bibinfo{journal}{\apjl} \textbf{\bibinfo{volume}{895}}, \bibinfo{eid}{L28}
  (\bibinfo{year}{2020}), \eprint{2004.09288}.

\bibitem[{\citenamefont{{Callister}}(2021)}]{2021arXiv210409508C}
\bibinfo{author}{\bibfnamefont{T.~A.} \bibnamefont{{Callister}}},
  \bibinfo{journal}{arXiv e-prints} \bibinfo{eid}{arXiv:2104.09508}
  (\bibinfo{year}{2021}), \eprint{2104.09508}.

\bibitem[{\citenamefont{{The LIGO Scientific Collaboration}
  et~al.}(2021{\natexlab{b}})\citenamefont{{The LIGO Scientific Collaboration},
  {the Virgo Collaboration}, {the KAGRA Collaboration}, {Abbott}, {Abbott},
  {Acernese}, {Ackley}, {Adams}, {Adhikari}, {Adhikari}
  et~al.}}]{2021arXiv211103606T}
\bibinfo{author}{\bibnamefont{{The LIGO Scientific Collaboration}}},
  \bibinfo{author}{\bibnamefont{{the Virgo Collaboration}}},
  \bibinfo{author}{\bibnamefont{{the KAGRA Collaboration}}},
  \bibinfo{author}{\bibfnamefont{R.}~\bibnamefont{{Abbott}}},
  \bibinfo{author}{\bibfnamefont{T.~D.} \bibnamefont{{Abbott}}},
  \bibinfo{author}{\bibfnamefont{F.}~\bibnamefont{{Acernese}}},
  \bibinfo{author}{\bibfnamefont{K.}~\bibnamefont{{Ackley}}},
  \bibinfo{author}{\bibfnamefont{C.}~\bibnamefont{{Adams}}},
  \bibinfo{author}{\bibfnamefont{N.}~\bibnamefont{{Adhikari}}},
  \bibinfo{author}{\bibfnamefont{R.~X.} \bibnamefont{{Adhikari}}},
  \bibnamefont{et~al.}, \bibinfo{journal}{arXiv e-prints}
  \bibinfo{eid}{arXiv:2111.03606} (\bibinfo{year}{2021}{\natexlab{b}}),
  \eprint{2111.03606}.

\bibitem[{\citenamefont{Collaboration et~al.}(2021)\citenamefont{Collaboration,
  Collaboration, and
  Collaboration}}]{ligo_scientific_collaboration_and_virgo_2021_5546663}
\bibinfo{author}{\bibfnamefont{L.~S.} \bibnamefont{Collaboration}},
  \bibinfo{author}{\bibfnamefont{V.}~\bibnamefont{Collaboration}},
  \bibnamefont{and}
  \bibinfo{author}{\bibfnamefont{K.}~\bibnamefont{Collaboration}},
  \emph{\bibinfo{title}{{GWTC-3: Compact Binary Coalescences Observed by LIGO
  and Virgo During the Second Part of the Third Observing Run — Parameter
  estimation data release}}} (\bibinfo{year}{2021}),
  \urlprefix\url{https://doi.org/10.5281/zenodo.5546663}.

\bibitem[{\citenamefont{{Powell} et~al.}(2019)\citenamefont{{Powell},
  {Stevenson}, {Mandel}, and {Ti{\r{A}}o}}}]{2019MNRAS.488.3810P}
\bibinfo{author}{\bibfnamefont{J.}~\bibnamefont{{Powell}}},
  \bibinfo{author}{\bibfnamefont{S.}~\bibnamefont{{Stevenson}}},
  \bibinfo{author}{\bibfnamefont{I.}~\bibnamefont{{Mandel}}}, \bibnamefont{and}
  \bibinfo{author}{\bibfnamefont{P.}~\bibnamefont{{Ti{\r{A}}o}}},
  \bibinfo{journal}{\mnras} \textbf{\bibinfo{volume}{488}},
  \bibinfo{pages}{3810} (\bibinfo{year}{2019}), \eprint{1905.04825}.

\bibitem[{\citenamefont{{Wysocki} et~al.}(2020)\citenamefont{{Wysocki},
  {O'Shaughnessy}, {Wade}, and {Lange}}}]{2020arXiv200101747W}
\bibinfo{author}{\bibfnamefont{D.}~\bibnamefont{{Wysocki}}},
  \bibinfo{author}{\bibfnamefont{R.}~\bibnamefont{{O'Shaughnessy}}},
  \bibinfo{author}{\bibfnamefont{L.}~\bibnamefont{{Wade}}}, \bibnamefont{and}
  \bibinfo{author}{\bibfnamefont{J.}~\bibnamefont{{Lange}}},
  \bibinfo{journal}{arXiv e-prints} \bibinfo{eid}{arXiv:2001.01747}
  (\bibinfo{year}{2020}), \eprint{2001.01747}.

\bibitem[{\citenamefont{{Golomb} and {Talbot}}(2021)}]{2021arXiv210615745G}
\bibinfo{author}{\bibfnamefont{J.}~\bibnamefont{{Golomb}}} \bibnamefont{and}
  \bibinfo{author}{\bibfnamefont{C.}~\bibnamefont{{Talbot}}},
  \bibinfo{journal}{arXiv e-prints} \bibinfo{eid}{arXiv:2106.15745}
  (\bibinfo{year}{2021}), \eprint{2106.15745}.

\bibitem[{\citenamefont{Vallisneri et~al.}(2015)\citenamefont{Vallisneri,
  Kanner, Williams, Weinstein, and Stephens}}]{GWOSC}
\bibinfo{author}{\bibfnamefont{M.}~\bibnamefont{Vallisneri}},
  \bibinfo{author}{\bibfnamefont{J.}~\bibnamefont{Kanner}},
  \bibinfo{author}{\bibfnamefont{R.}~\bibnamefont{Williams}},
  \bibinfo{author}{\bibfnamefont{A.}~\bibnamefont{Weinstein}},
  \bibnamefont{and} \bibinfo{author}{\bibfnamefont{B.}~\bibnamefont{Stephens}},
  \bibinfo{journal}{Journal of Physics: Conference Series}
  \textbf{\bibinfo{volume}{610}}, \bibinfo{pages}{012021}
  (\bibinfo{year}{2015}), ISSN \bibinfo{issn}{1742-6596},
  \urlprefix\url{http://dx.doi.org/10.1088/1742-6596/610/1/012021}.

\end{thebibliography}
\end{document}